\documentclass[12pt, reqno]{amsart}
\usepackage{amsmath,amssymb,pdflscape,bm,xr}
\usepackage{amsfonts,natbib,color,xcolor}
\usepackage{graphicx,epsfig,setspace}
\usepackage[margin=1in]{geometry}

\numberwithin{equation}{section}
\begin{document}

\newcommand{\cov}{\textnormal{Cov}}
\newcommand{\var}{\textnormal{Var}}
\newcommand{\diag}{\textnormal{diag}}
\newcommand{\plim}{\textnormal{plim}_n}
\newcommand{\dum}{1\hspace{-2.5pt}\textnormal{l}}
\newcommand{\ind}{\bot\hspace{-6pt}\bot}
\newcommand{\co}{\textnormal{co}}
\newcommand{\fsgn}{\textnormal{\footnotesize sgn}}
\newcommand{\sgn}{\textnormal{sgn}}
\newcommand{\fatb}{\mathbf{b}}
\newcommand{\fatp}{\mathbf{p}}
\newcommand{\tr}{\textnormal{tr}}
\newcommand{\spn}{\textnormal{span}}
\newcommand{\im}{\textnormal{im}}
\newcommand{\id}{\textnormal{Id}}
\newcommand{\Eta}{\textnormal{H}}

\newtheorem{dfn}{Definition}[section]
\newtheorem{rem}{Remark}[section]
\newtheorem{cor}{Corollary}[section]
\newtheorem{thm}{Theorem}[section]
\newtheorem{lem}{Lemma}[section]
\newtheorem{notn}{Notation}[section]
\newtheorem{con}{Condition}[section]
\newtheorem{prp}{Proposition}[section]
\newtheorem{pty}{Property}[section]
\newtheorem{ass}{Assumption}[section]
\newtheorem{ex}{Example}[section]
\newtheorem*{cst1}{Constraint S}
\newtheorem*{cst2}{Constraint U}
\newtheorem{qn}{Question}[section]

\onehalfspacing

\title[Fixed-Population Causal Inference for Models of Equilibrium]{Fixed-Population Causal Inference for Models of Equilibrium}
\author[Menzel]{Konrad Menzel\\New York University}
\date{January, 2025, this version: August 2026.  Author email: \textsf{km125@nyu.edu}. I am grateful to seminar and conference participants at various institutions for valuable feedback. All remaining errors are my own.}

\begin{abstract} We consider the problem of how to interpret a randomized experiment on a network when outcomes are determined in equilibrium. Based on a single network experiment, a class of inverse probability weighting (IPW) estimators consistently estimates certain averages of direct and spillover causal effects, which we term \emph{contingent average exposure effects}. Identification and estimation of these effects require only minimal assumptions on the structure of interference and the extent of spillovers arising from equilibrium interactions. Unlike in the conventional design-based literature, however, contingent average exposure effects are not fixed features of the population: since they may vary with global features of the realized experimental assignment, they are themselves stochastic ex ante. Under a fixed-population, design-based approach, we establish unbiasedness, consistency, and asymptotic normality for IPW estimators of these effects. Because their asymptotic variance is generally not point-identified, we also propose a conservative variance estimator together with asymptotically valid confidence intervals. To investigate to what extent these reduced-form estimands can help anticipate policy effects and uncover structural properties of the equilibrium model, we analyze in detail the case of an equilibrium model with smooth response functions. We show that contingent average exposure effects can be combined, under additional structural assumptions, to recover LATE-type weighted averages of structural derivatives of unit-level responses and to predict policy effects, while retaining their less restrictive causal interpretation when those assumptions fail. At the same time, several stylized examples show that some policy-relevant global features of the assignment can remain irreducibly unidentified from a single experiment, regardless of the estimation approach.\\[8pt]
\end{abstract}

\maketitle

\section{Introduction}

\label{sec:introduction}

We consider causal inference for models of equilibrium, where the outcome for a unit is determined not only by the direct impact of a policy variable or economic shock (``treatment"), but also by other units' endogenous response (``outcome") to that intervention. For example, the labor market in a U.S. metropolitan area may not only be affected directly by rising import competition from a foreign trading partner, but also by reduced demand for intermediate goods from other domestic trading partners facing the same external competition. Similarly, for an immunization against a communicable disease, an individual may be less likely to be infected not only as a result of receiving the immunization herself, but also as a result of her peers receiving the same immunization, thereby disrupting potential transmission chains.

Spillover effects of this kind may not only confound estimates of the direct effect of a unit's treatment status on its outcome, but also capture welfare-relevant externalities that should be accounted for when evaluating a potential intervention. In particular, when scaling up a pilot or otherwise limited experimental implementation of an intervention, the policymaker would need to anticipate not only direct effects, but also indirect general equilibrium effects resulting from interference.

Empirical practice traditionally relies on restrictive (usually parametric) structural models to account for and predict equilibrium responses. In contrast, this paper first investigates which aspects of an equilibrium model can be estimated by imposing only minimal structure on the problem. We then return to the question of what additional structure is needed to characterize structural mechanisms and policy effects on the basis of those robustly identified causal effects.

We analyze the problem using a potential outcomes framework for a population of units connected via a spillover network, allowing for strong dependence in outcomes. We first consider estimation of ``reduced-form" response to assigned treatment, i.e. changes in observed outcomes that are due to certain changes in the allocation of unit-specific treatments across the population. In the now well-studied problem of interference in unit-specific (exogenous) treatments, estimation of spillovers is often operationalized using exposure mappings. Exposure mappings aggregate the features of the social assignment relevant to the outcome of a target unit using known functions, and such a mapping may often depend on the treatment status of a small number of units at a short network distance from that unit.

However, with interference in (endogenous) outcomes, this is often no longer plausible: for one, the structural mechanism now acts through the (unknown and heterogeneous) potential outcomes of neighboring units, so that there is no known exposure mapping summarizing the impact of the social treatment on the outcome of a reference unit. Also, even if the structural mechanism linking individual outcomes operates only at a short network distance and aggregates peer outcomes according to a known ``structural" mapping, outcomes may depend on other units' treatment status at much longer network distances, and the corresponding reduced-form mapping from the social allocation to outcomes can in general also not be easily derived from that structural mapping.

\subsection{Contribution}

We show how to interpret a large networked experiment as a collection of many overlapping ``local" experiments which allow us to estimate certain averages of direct and spillover effects from equilibrium outcomes.  We propose an interpretation of such average exposure effects under minimal assumptions, most importantly without the presumption that the specified exposures are sufficient for the combined effect of treatment assignments in the population on outcomes.

We argue that without further assumptions, we can identify certain meaningful causal contrasts from observing a single networked population, but that these effects are generally not stable across experimental allocations and instead may vary with the realized assignment. We show that it is indeed natural in typical use scenarios to interpret causal effects as being contingent on the assignment; our definition of contingent average exposure effects is meant to make that dependence explicit. Partial effects of this kind are nevertheless informative about causal mechanisms of direct response and interference, especially if combined with additional structural assumptions. That said, we also illustrate using examples that in general there may be global features of the vector of assignments that are ``irreducible" in the sense that they have a causal effect on individual outcomes but do not vary across nodes or neighborhoods for any given single realization of the experiment. A principled application of the Neyman-Rubin potential outcomes framework with general interference can therefore help map out the possibilities and limitations of an ``agnostic" approach to policy analysis in the presence of equilibrium effects.

To understand the relationship between these ``agnostic" and more structural parameters better, we analyze the problem in greater detail for a model of an infinitesimal intervention in an equilibrium model with smooth responses in Section \ref{sec:infinitesimal_model}. For that model we show that we can recover Local Average Treatment Effect (LATE)-type weighted average derivatives by adapting the identification strategy for the linear-in-means model proposed by \cite{BDF09}. Under the more restrictive assumption of the linear-in-means model, estimated ``local" responses are in fact sufficient to extrapolate to the ``global" equilibrium response to a counterfactual intervention. On the other hand, our framework also provides a more robust, design-based causal interpretation for such an estimand that does not rely on a particular set of assumptions on the structure of interference. Our analysis also illustrates how identification of parameters in a structural version of the equilibrium model ultimately derives from the reduced form, that is weighted average responses to certain changes in the social treatment.

We show that a fixed-population, design-based approach makes it possible to address the central challenges posed by this setup in a clear and principled manner. In this statistical framework, we regard the experimental assignment as the principal source of estimation uncertainty. We give an asymptotic theory for the randomization distribution and show that the estimators we propose are unbiased, consistent, and asymptotically normal. As in the leading case of randomization inference in the absence of spillovers, the asymptotic variance of fixed-population average treatment effects is not identified even under standard conditions. We therefore provide a conservative estimator of their asymptotic variance and asymptotically conservative confidence intervals.

Any probabilistic statements are therefore defined with respect to randomization of the assignment of unit-specific treatments in the experiment. This perspective is well-developed in the literature on causal inference (see e.g. \cite{AAIW17}). However, a novel aspect of the problem with unconstrained interference is that classical estimands not only are contingent on the potential outcomes for the fixed study population, but also turn out to vary with the realized experimental assignment. Hence statements about bias, consistency, and asymptotic distributions evaluate estimation error with respect to an estimand that is also random ex ante under this design-based interpretation.

\subsection*{Organization}

The remainder of the paper is organized as follows: we first describe our general fixed-population framework with interference, where we define average exposure effects as ``model-free" causal estimands. We then analyze more closely the case of marginal interventions in a model of equilibrium with smooth response functions to illustrate the relation between ``agnostic" causal estimands and structural model parameters. Section \ref{sec:asy_prop} gives fixed-population, large-sample results that characterize the statistical performance of IPW estimators for conditional average exposure effects. We also derive asymptotic rates for design-based inference from primitive conditions in Appendix \ref{sec:design_rate_app}, which considers two stylized models for the structure of a spillover network.

\subsection*{Literature}

\textbf{Foundational work.} A general framework to analyze identification of peer and spillover effects, including interference in outcomes, was first put forward by \cite{Man93} in econometrics, whereas \cite{HSt95} defined the problem of interference in causal models in the biomedical literature. Prominent structural approaches to the problem were subsequently developed by \cite{BDu01} and \cite{BDF09}.

\textbf{Exposure mappings.} In classical causal inference, spillover effects constitute a failure of \cite{Rub80}'s Stable Unit Treatment Value Assumption (SUTVA). In that context, \emph{exposure mappings} have become a key tool in simplifying the structure of possible social treatments. As shown by \cite{Man11}, flexible identification of causal responses with interference in a potential outcomes framework typically requires a reduction in the complexity of unrestricted dependence of outcomes on unit-specific assignments in the relevant reference group. An important focus in the literature has therefore been on interference mechanisms that operate through an ``exposure" mapping that summarizes the relevant aspects of the social treatments (\cite{HHa08}, \cite{Man11}, and \cite{ASa17}, see also \cite{OSSvL24}).

\cite{WSCA24} and \cite{FNi26} provide a model-free theory for reweighting and linear estimators for average spatial spillovers and global effects, respectively, under weak spatial dependence. \cite{Gao24} showed how to extend that framework to endogenize the structure of the interference network to the intervention. A more recent paper by \cite{LZT25} considers an estimator similar to ours, but their asymptotic results require that there are no spillover effects outside of a fixed network neighborhood of the reference node, and that generalized propensity scores are bounded away from zero, both of which are the conditions we specifically mean to relax in this paper. \cite{AEI18} proposed tests for qualitative hypotheses regarding the presence of direct, first- and higher-order spillover effects that allow for other types of unmodeled interference; our approach aims to recover the magnitude of causal effects that are valid under similarly weak conditions, but which can also be interpreted structurally under more stringent assumptions.

\textbf{Misspecification.} We argue in this paper that for equilibrium models with interference in (endogenous) outcomes rather than only (exogenous) treatments, \emph{misspecification} is the default scenario for any conventional exposure model, since the mechanism linking exogenous treatments to unit-specific outcomes is itself mediated by the heterogeneous potential outcomes mapping, so that there is no known exposure mapping to summarize the reduced-form social treatment on a given unit. Furthermore, depending on the strength of spillovers and density of the network, potential values for unit $i$ may be sensitive to assignments $d_j$ for units at a majority of other nodes in the network.

The paper closest to our approach is \cite{Sav24}, who shows that for misspecified exposure models, meaningful causal parameters can still be identified, although the contribution of unmodeled interference to estimation error is generally dependent across units and complicates inference. A similar point was made by \cite{EKU17}, who proposed ways of reducing ``estimand bias" relative to an oftentimes unidentified policy parameter through experimental design and adjustments to default estimators. We propose instead estimands that are proper causal parameters and that are consistently estimable even when estimation of the global effect of an intervention remains elusive.

Our approach differs from \cite{Sav24} specifically in how it treats that dependence on the realized assignment: we propose a theory that allows for estimands that are contingent on the experimental assignment, whereas \cite{Sav24} instead articulates conditions on the interference mechanism under which that dependency is asymptotically negligible, and regards any remaining dependence on the observed assignment as contributing to estimation error relative to a stable, unconditional effect. Under our interpretation, we provide causal parameters that reveal information about causal mechanisms, but also acknowledge their potential fragility and make explicit their potential limitations for extrapolation to counterfactual policies or anticipation of general equilibrium effects. We provide several scenarios in which counterfactuals depend on global features of the assignment whose effect cannot be identified from a single experiment.

\textbf{Sign preservation and policy relevance.} \cite{Leu24} gives conditions for unweighted average exposure contrasts to preserve the sign of individual causal effects independent of model assumptions. The IPW estimators considered in this paper estimate average causal effects by construction, but assume knowledge of the assignment mechanism, and our focus is furthermore on asymptotic properties of those IPW estimators. \cite{AATM24} caution that under misspecification, exposure contrasts, while representing a weighted average of causal effects, are in general not immediately policy relevant. We look at the relationship between ``agnostic" reduced form estimands and policy counterfactuals in more detail in Section \ref{sec:infinitesimal_model} for the special case of marginal interventions with differentiable responses. One possible interpretation of our findings is that identification of the overall equilibrium effect of an intervention may require additional structural assumptions, but a design-based approach that allows for misspecification may give a more robust causal interpretation to estimands motivated by such a structural framework.

\textbf{Approximate neighborhood interference.} In models with \emph{endogenous interference}, i.e. interference mediated by peer outcomes, causal effects can be estimated under an Approximate Neighborhood Interference (ANI) condition, see \cite{SAi17}, \cite{Leu22}, see also \cite{Leu24}. Broadly speaking, ANI holds if the reduced form for the endogenous response of any unit is known to be responsive only to assignments at a short distance in the interference network, and we consider asymptotic sequences of networks whose radius grows large. Alternatively, \cite{ATM21} propose to approximate causal effects using exposures defined on rooted network types that are defined in terms of treatment assignments in a small but growing subnetwork around the reference unit. Causal inference with equilibrium effects in centralized markets was analyzed by \cite{MWX22}.

\textbf{Design-based inference.} A systematic \emph{design-based theory} for estimation of causal models with interference was developed by \cite{HSW22}, and \cite{Cha23} provided general results for variance estimation. Earlier work by \cite{LHu14} and \cite{LHB16} considered partial interference within an observed sample of distinct, non-interfering groups, where indirect effects are identified by cross-group variation in the exposure. We depart from this literature by defining treatment effects conditional on other components of the assignment, i.e. an estimand that is allowed to vary with the experimental assignment.

\textbf{Inverse probability weighting.} IPW estimation in the presence of interference has been considered, among others, by \cite{TvW10}, \cite{CEU22}, \cite{HLWa22}, and \cite{LWa22}. \cite{DGa23} propose the regression-based H\'ajek (fixed sum of weights) estimator as an improvement over the classical Hurvitz-Thompson estimator.

\section{Problem Description}

\label{sec:setup}

Suppose an intervention targets a population of $n$ units $\mathcal{N}=\{1,\dots,n\}$, and that this intervention takes the form of unit-specific assignments $D_1,\dots,D_n$ from some set $\mathcal{D}$ of values for the policy variable. For simplicity we focus on the case in which $D_i$ is binary, $\mathcal{D}=\{0,1\}$. The causal response of unit-specific outcomes $Y_1,\dots,Y_n\in\mathbb{R}$ to that intervention may exhibit spillover effects, where the outcome of unit $i$ may depend on assignments $D_j$ or outcomes $Y_j$ for other units $j$. To be specific, we assume that spillovers act on a network $(\mathcal{N},\mathbf{L})$ with vertex set $\mathcal{N}$ and edges represented by an adjacency matrix $\mathbf{L}=(L_{ij})_{ij}$. In this paper we restrict attention to the leading case of (potentially directed) unweighted graphs, $L_{ij}\in\{0,1\}$ and leave an extension to weighted graphs for future research. We also assume no self-links, $L_{ii}=0$ for all $i=1,\dots,n$. In a typical application, the researcher may in addition observe a vector of unit-specific attributes $X_i$; however, we will ignore that possibility for the moment.

\subsection{The Equilibrium Model}

Unit-specific outcomes $\mathbf{Y}\equiv (Y_i)_{i=1}^n$ are then assumed to be determined in equilibrium, satisfying the fixed-point conditions
\begin{equation}\label{equilibrium_condition}Y_i = h(\mathbf{D},\mathbf{L},\mathbf{Y},\mathbf{U};i) \end{equation}
for $i=1,\dots,n$, where $\mathbf{U}:=(U_i)_{i=1}^n$ are unobservables of arbitrary dimension. That is, for any social assignment of the policy variable $\mathbf{D}=(D_i)_{i=1}^n$, outcomes are determined in a system of $n$ simultaneous equations corresponding to the coordinates of the response mapping $h(\cdot)$.
Existence or uniqueness of a solution to  (\ref{equilibrium_condition}) is not guaranteed without additional assumptions and must be discussed separately. Assuming that a solution exists and a particular mechanism for selection among potentially multiple solutions for any assignment $\mathbf{D}$ has been fixed, we denote the corresponding \textbf{potential values} with
\begin{equation}\label{potential_outcomes_generic}Y_i(\mathbf{D}) = Y_i(D_i;\mathbf{D}_{-i})\end{equation}
where dependence of $Y_i(\mathbf{D})\equiv Y_i(\mathbf{D},\mathbf{L},\mathbf{U})$ on $\mathbf{L},\mathbf{U}$ is left implicit for the moment. We return later to the question how restrictions on the mapping $h(\cdot)$ can yield additional structure on the potential outcomes $Y_i(\mathbf{D})$.

Without further restrictions, the potential values are a mapping
\[\mathbf{Y}:\mathcal{D}^n\rightarrow\mathbb{R}^n\]
which we also denote by $\mathbf{Y}(\cdot)$. Following traditional econometric terminology, we also refer to (\ref{equilibrium_condition}) as the \textbf{structural form} of the equilibrium model, and to the solution (\ref{potential_outcomes_generic}) as its \textbf{reduced form}.

Endogenous interactions of the form (\ref{equilibrium_condition}) therefore introduce a number of additional challenges even if we are only interested in reduced-form causal effects in terms of $\mathbf{D}$:
\begin{itemize}
\item For one, without additional restrictions, potential outcomes $Y_i(D_i;\mathbf{D}_{-i})$ generally depend on the full set of unit-specific assignments among peers $\mathbf{D}_{-i}$, so identification of any causal objects from a single realization of this model is generally more challenging than for the case of exogenous spillovers.
\item Since each response function in (\ref{equilibrium_condition}) also depends on agent-specific unobserved heterogeneity $U_i$, equilibrium generally induces cross-sectional dependence in potential outcomes in addition to dependence in $(U_i)$.
\item Furthermore, in the case of multiple solutions to the system (\ref{equilibrium_condition}), potential values also depend on an equilibrium selection rule. We assume that agents coordinate on an equilibrium autonomously from the experimenter, so that the selection rule is subsumed under the causal mechanism, and any causal statements have to be interpreted as being conditional on that selection rule.\footnote{Estimation of models with multiple equilibria has been analyzed by an extensive literature in econometrics, see e.g. \cite{BVu84}, \cite{BRe90}, \cite{Tam03}, and \cite{deP13} for an overview.}
\end{itemize}

The primary focus in the literature on causal inference has been on the case of exogenous interference, where spillovers are mediated only through the treatment status of other units, rather than their outcomes.\footnote{See \cite{HHa08}, \cite{ASe96}, \cite{TvW10}, \cite{LWa22}, and \cite{HLWa22}.} The present setup differs qualitatively in two important aspects: for one, the realized outcome for a unit may depend on treatment assignments to units that are distant from that unit under any relevant metric even if structural interactions in the mapping (\ref{equilibrium_condition}) are only with respect to peer outcomes at short network distances.\footnote{\cite{Leu22} gave conditions for endogenous interaction effects to exhibit approximate neighborhood interference within a bounded network neighborhood around $i$.} Furthermore, even if the mapping $h(\cdot;i)$ depends on outcomes $\mathbf{Y}_{-i}:=(Y_j)_{j\neq i}$ only  through known summary statistics $T_i:=T_i(\mathbf{Y},\mathbf{L})$ (``exposures"), the resulting spillovers are mediated via units' heterogeneous responses to the causal variable. Hence, under interference in outcomes there is in general no known exposure mapping to summarize the effective social treatment with respect to a given unit.

\subsection{Network Experiment}

We consider the problem of estimating causal effects and counterfactuals for the experimental population, where $\mathbf{D}$ are assigned at random, and we regard unit-specific assignments as random, but the unknown potential outcomes as fixed. Since we regard the population and potential values as fixed, the causal parameters we consider in this paper are average differences in potential outcomes between different counterfactual assignments.

Estimation is based on a network experiment, where individualized treatments are assigned according to a known, stochastic mechanism:

\begin{ass}\textbf{(Experimental Assignment)}\label{ass:experimental_assg} Under the experimental assignment, the vector $\mathbf{D}$ is distributed according to the known distribution $\pi_0$ on $\mathcal{D}^n$ given any potential outcomes $\mathbf{Y}(\cdot)$,
\[(D_1,\dots,D_n)|\mathbf{Y}(\cdot)\sim \pi_0(d_1,\dots,d_n)\]
\end{ass}

It is generally necessary, or at least advantageous, to implement dependent assignments to estimate spillover effects (see e.g. \cite{HHa08}), so we do not want to restrict the experimental assignment rule further at this point. We analyze a class of hierarchical randomization designs and the resulting rates of consistency in Appendix \ref{sec:design_rate_app}. In what follows, we assume that the data available to the researcher consists of $\mathbf{Y} = \mathbf{Y}(\mathbf{D})$ given the realized experimental assignment $\mathbf{D}$, along with $\mathbf{D}$ and the adjacency matrix $\mathbf{L}$. In the following, we let $\mathbb{P}_{\pi_0}[\cdot]$, $\mathbb{E}_{\pi_0}[\cdot]$, $\var_{\pi_0}(\cdot)$, and $\cov_{\pi_0}(\cdot,\cdot)$ denote probabilities and moments under the randomization distribution given the assignment $\pi_0$, where the $\pi_0$-subscript is left implicit when there is no ambiguity.

\subsection{Leading Case: Direct Effect} To frame ideas, consider first the case of estimation of the average direct effect of the assignment $D_i\in\{0,1\}$ on the outcome for that same unit, $Y_i$. We then define the average direct effect given the realized assignment $\mathbf{D}=(D_1,\dots,D_n)$ as
\begin{equation}\label{intro_tau_direct}\tau^{dir}_n\equiv\tau_{dir,n}(\mathbf{D}):=\frac1n\sum_{i=1}^n\left(Y_i(1;\mathbf{D}_{-i}) - Y_i(0;\mathbf{D}_{-i})\right)\end{equation}
where we follow the notational convention $Y_i(\mathbf{D}) = Y_i(D_i;\mathbf{D}_{-i})$ from (\ref{potential_outcomes_generic}).

The parameter $\tau^{dir}_n$ gives the average ceteris paribus effect of an intervention that changes a single unit's assignment on the outcome of that particular unit. Such a contrast is a well-defined causal parameter even though the average is over treatment assignments to nodes $j\notin\mathcal{N}_T(i)$ that differ across units $i\in\mathcal{N}$ and cannot be realized simultaneously in the cross-section of nodes. Without further restrictions on potential values, unit-specific causal effects $Y_i(1;\mathbf{D}_{-i}) - Y_i(0;\mathbf{D}_{-i})$ do depend on assignments to other units, so that $\tau_{dir,n}(\mathbf{D})$ is generally not structural but may vary with the social assignment.

The direct effect does not internalize the spillovers across units and therefore does not immediately address a social planner's problem of maximizing welfare in the aggregate. Rather it captures the unilateral benefit (or private incentive) from receiving the treatment given the treatment status of other units. The policy-relevant causal parameter of interest therefore generally depends on the question at hand: A regulator deciding whether or not to approve a novel vaccine or therapy for a communicable disease (for purposes of marketing by the manufacturer and coverage by insurers) may base the decision primarily on the interest of the individual patient rather than social welfare. On the other hand, a decision whether to mandate regular testing or immunization for employees in an institution would also seek to account for externalities from transmission within the workplace. Furthermore, the average unilateral benefit of a ``physiologically effective" immunization to a patient may be greater in a more adverse disease environment, which may change if an experimental treatment is distributed at scale in the networked population. Hence, dependence of $\tau_{dir,n}(\mathbf{D})$ on $\mathbf{D}$ accurately captures the sensitivity of---measured and actual---average direct effects to the allocation in the remaining population via equilibrium spillover effects.

\begin{figure}\footnotesize\hspace{-1cm}
\begin{tabular}{lcccccccccccccccc}
&\hspace{0.3cm}&$(0,0,\dots,0)$&\dots&$(0,0;\mathbf{d}_{-12})$&
$\mathbf{(1,0;\mathbf{d}_{-12})}$&
$(1,1;\mathbf{d}_{-12})$&\dots&$(1,1,\dots,1)$\\
\hline\hline
$1$&\hspace{0.3cm}&$Y_1(0,\dots,0)$&\dots&$\textcolor{darkgray}{\mathbf{Y_1(0,1;\mathbf{d}_{-12})}}$&
$\mathbf{Y_1(1,0;\mathbf{d}_{-12})}$&
$Y_1(1,1;\mathbf{d}_{-12})$&\dots&$Y_1(1,\dots,1)$\\
$2$&&$Y_2(0,\dots,0)$&\dots&$Y_2(0,0;\mathbf{d}_{-12})$&$\mathbf{Y_2(0,1;\mathbf{d}_{-12})}$&
$\textcolor{darkgray}{\mathbf{Y_2(1,1;\mathbf{d}_{-12})}}$&\dots&$Y_2(1,\dots,1)$\\
\vdots&&\vdots&&\vdots&$\mathbf{\vdots}$&\vdots&&\vdots\\
$i$&&$Y_i(0,\dots,0)$&\dots&$Y_i(d_i,0,0;\mathbf{d}_{-12i})$&$\mathbf{Y_i(d_i,1,0;\mathbf{d}_{-12i})}$&
$Y_i(d_i,1,1;\mathbf{d}_{-12i})$&\dots&$Y_i(1,\dots,1)$\\
\vdots&&\vdots&&\vdots&$\mathbf{\vdots}$&\vdots&&\vdots\\
$n$&&$Y_n(0,\dots,0)$&\dots&$Y_n(d_n,0,0;\mathbf{d}_{-12n})$&$\mathbf{Y_n(d_n,1,0;\mathbf{d}_{-12n})}$&
$Y_n(d_n,1,1;\mathbf{d}_{-12n})$&\dots&$Y_n(1,\dots,1)$\\\hline\hline
\end{tabular}
\caption{Estimation of direct effects: potential outcomes, highlighting the experimental sample (boldface) and relevant counterfactuals (dark gray boldface), where $\mathbf{d}_{-i_1i_2\dots i_s}$ denotes the vector $\mathbf{d}$ after deleting the $i_1,\dots,i_s$th components.}
\label{fig:table_of_knowledge}
\end{figure}

To appreciate the nature of the missing data problem in recovering the direct effect from an experimental sample, consider the representation of potential outcomes in Table \ref{fig:table_of_knowledge}: The potential outcomes for the fixed population $\mathbf{Y}(\mathbf{D})$ are a mapping from $\{0,1\}^n$ to $\mathbb{R}^n$, which can be represented as an $n\times 2^n$ matrix whose columns correspond to social allocations $\mathbf{d}\equiv (d_1,d_2,\dots,d_n)\in\{0,1\}^n$. The realized experimental allocation, say $\mathbf{D}=(1,0,\dots,1)$ then corresponds to a particular column of the matrix $\mathbf{Y}(\cdot)$ (highlighted in boldface) which is then observed by the researcher, whereas the remaining entries of that matrix remain hidden. The network experiment therefore samples among those columns at random with probabilities given by the experimental distribution $\pi_0$. The direct treatment effect on unit $i$ given that realized allocation is $Y_i(1;\mathbf{d}_{-i})-Y_i(0;\mathbf{d}_{-i})$, where the potential values $\mathbf{Y_i(d_i;\mathbf{d}_{-i})}$ are observed, but the counterfactuals $\mathbf{Y_i(1-d_i;\mathbf{d}_{-i})}$ (highlighted in gray boldface) are missing data that has to be imputed in order to estimate $\tau_{dir,n}(d_i,\mathbf{d}_{-i})$.

For our discussion of estimation, we assume for simplicity that the experimental assignments $D_1,\dots,D_n$ are i.i.d. Bernoulli with success probability $\pi_0$. Then, a natural estimator for the treatment effect (\ref{intro_tau_direct}) is the Inverse Probability Weighting (Hurvitz-Thompson) estimator
\begin{equation}\label{hurvitz_thompson_direct}\hat{\tau}_{dir,n}:=\frac1n\sum_{i=1}^n\left(\frac{D_iY_i}{\pi_0}-\frac{(1-D_i)Y_i}{1-\pi_0}\right)
\end{equation}
We define more general average exposure effects from varying the distribution of treatments in network neighborhoods of node $i$ in the next section to measure indirect or spillover effects, also allowing for general assignment mechanisms.

We provide an inferential theory for IPW estimators for a broader class of average causal effects which specifically needs to account for the fact that estimands like $\tau^{dir}_n\equiv\tau_{dir,n}(\mathbf{D})$ vary with the experimental assignment $\mathbf{D}$ and is therefore stochastic ex ante under our design-based interpretation. In particular, we give conditions for such an estimator to be consistent in the sense that the estimation error $\hat{\tau}_{dir,n}-\tau_{dir,n}(\mathbf{D})$ converges to zero in probability. We also propose confidence intervals that guarantee asymptotic coverage of $\tau_{dir,n}(\mathbf{D})$ at the pre-specified nominal confidence rate, where probabilities are with respect to the experimental randomization distribution $\mathbf{D}\sim\pi_0$ but regard potential values $\mathbf{Y}(\cdot)$ as fixed.

\subsection{Baseline Cases} To fix ideas, we next present a number of illustrative examples that we will keep referring to as we discuss our framework below. First it is always instructive to compare scenarios with spillover effects to the baseline case of no interference, corresponding to \cite{Rub80}'s Stable Unit Treatment Value Assumption (SUTVA):

\begin{ex}\textbf{(SUTVA)} The Stable Unit Treatment Value Assumption corresponds to the model (\ref{equilibrium_condition}) with
\[h(\mathbf{D},\mathbf{L},\mathbf{Y},\mathbf{U};i)=\tilde{h}(D_i,U_i)\]
for some function $\tilde{h}(d,u)$, where dependence of outcomes on $\mathbf{U}$ through $U_i$ is without loss of generality. In particular, the resulting potential outcomes are of the form $Y_i(\mathbf{D})\equiv Y(D_i;U_i)=\tilde{h}(D_i,U_i)$.
\end{ex}

The following example considers exogenous spillovers with a known exposure mapping and a fixed radius of interaction, and was analyzed in depth e.g. by \cite{LWa22}:

\begin{ex}\textbf{(Exogenous Exposure Models)} \label{ex:exog_exposure} Suppose that the outcome for unit $i$ depends on their own assignment $D_i$ as well as the average assignment among its network neighbors, $T_i(\mathbf{D}):=\frac1{|\mathcal{N}(i)|}\sum_{j\neq i}L_{ij}D_j$, where $|\mathcal{N}(i)|=\sum_{j\neq i}L_{ij}$. This corresponds to the model (\ref{equilibrium_condition}) with
\[h(\mathbf{D},\mathbf{L},\mathbf{Y},\mathbf{U};i)=\tilde{h}\left(D_i,T_i(\mathbf{D}),U_i\right)\]
Hence the resulting potential outcomes $Y_i(\mathbf{D})\equiv Y(D_i,T_i(\mathbf{D}),U_i)=\tilde{h}\left(D_i,T_i(\mathbf{D}),U_i\right)$ depend on $\mathbf{D}_{-i}$ only through the known exposure mapping $T_i(\mathbf{D})$.
\end{ex}

Exposure models may be useful e.g. to measure aggregate effects of immunizing, or giving a prophylaxis to an individual against an infectious disease with the potential for human-to-human transmission. A prominent example of this problem is the experiment by \cite{MKr03} on the effect of deworming drugs on health and educational outcomes in schools in Kenya, which estimated the total effect of assigning an entire school to the health intervention.

If instead only randomly selected individual students or students in randomly selected classrooms had been assigned to treatment in the experimental intervention, estimated direct and peer effects would have been contingent on the identity and share of treated units or classrooms in the presence of externalities across classrooms. The immunization decision then affects not only the measured outcome for the treated unit directly, but also outcomes along any potential transmission chain that includes that unit. Hence, there is not only a potential benefit of immunization to other subjects, but the individual benefit to the immunized individual also varies in turn with how likely that unit is going to be exposed. In extreme cases, that aggregate disease environment may in itself be very sensitive to small changes in the environment, and therefore the specific assignment of treatment, if e.g. a major infectious wave could have been triggered by a single transmission event.

\subsection{Running Examples}

The three examples in this subsection recur throughout the paper and are developed further in later sections. In the more general case allowing for endogenous spillover effects from network neighbors' outcomes $Y_j$ on $Y_i$, one canonical framework is the linear-in-means model:

\begin{ex}\textbf{(Linear-in-Means Model)}\label{ex:linear_in_means_ex} Suppose the model (\ref{equilibrium_condition}) holds with a mapping
\[h(\mathbf{D},\mathbf{L},\mathbf{Y},\mathbf{U};i)=\beta_0 + \beta_1 D_i + \gamma_1\sum_{j\neq i}L_{ij}D_j + \gamma_2\sum_{j\neq i}L_{ij}Y_j + U_i\]
and $\mathbb{E}[U_i|\mathbf{D},\mathbf{L}]=0$. The potential values for $\mathbf{Y}$ can then be found by solving the linear system of $n$ equations and are given by
\[\mathbf{Y}(\mathbf{D}) = \left(\mathbf{I}_n - \gamma_2\mathbf{L}\right)^{-1}\left(\beta_0 + (\beta_1\mathbf{I}_n + \gamma_1\mathbf{L})\mathbf{D}+\mathbf{U}\right)\]
provided that the inverse exists.
\end{ex}

Identification of the model parameters $\boldsymbol{\beta},\boldsymbol{\gamma}$ was analyzed in \cite{BDF09} who propose an instrumental variable strategy based on assignments $D_k$ among nodes $k$ in a network neighborhood of radius 2 around node $i$. This approach does not generally require sparsity of $\mathbf{L}$ or approximate neighborhood interference; however, it relies on the assumption of a homogeneous, parametric structure of interaction effects with additively separable heterogeneity.

To better understand the consequences of heterogeneous interaction effects, we also analyze the following problem in greater detail below:

\begin{ex}\textbf{(Infinitesimal Shift)}\label{ex:infinitesimal_ex} Suppose (\ref{equilibrium_condition}) derives from an equilibrium model with a continuous policy variable $\mathbf{X}=(X_1,\dots,X_n)$, and outcomes are determined according to
\begin{equation}\label{equilibrium_infinitesimal}Y_i = h(\mathbf{X},\mathbf{L},\mathbf{Y},\mathbf{U};i) \end{equation}
for $i=1,\dots,n$. We assume that $h(\cdot)$ is twice continuously differentiable with respect to $\mathbf{D},\mathbf{Y}$ with probability 1. We can then consider the effect of an incremental intervention that changes an initial continuous assignment $\mathbf{X}_0$ to $\mathbf{X}_1:=\mathbf{X}_0+\delta\mathbf{D}$ for a small value of $\delta>0$ and $\mathbf{D}:=(D_i)_{i=1}^n$.

Denoting the equilibrium given the experimental assignment $\mathbf{D}$ with $\mathbf{Y}^*\equiv\mathbf{Y}^*(\mathbf{D})$, under regularity conditions to be discussed below, for any other assignment $\tilde{\mathbf{D}}$, there exists a new equilibrium $\tilde{\mathbf{Y}}$ satisfying
\[\tilde{\mathbf{Y}}\equiv \tilde{\mathbf{Y}}(\tilde{\mathbf{D}})  = \mathbf{Y}^* + \delta\left(\mathbf{I}_n-\mathbf{H_Y}\right)^{-1}\mathbf{H_X}(\tilde{\mathbf{D}}-\mathbf{D}) + O(\delta^2)\]
where $\mathbf{H_Y}:=\nabla_{\mathbf{Y}}h(\mathbf{X}_0,\mathbf{L},\mathbf{Y}^*,\mathbf{U})$ and $\mathbf{H_X}:=\nabla_{\mathbf{X}}h(\mathbf{X}_0,\mathbf{L},\mathbf{Y}^*,\mathbf{U})$, and the $i,j$th elements of $\mathbf{H_Y},\mathbf{H_X}$ are zero whenever $L_{ij}=0$. Under regularity conditions, there exists a unique equilibrium $\tilde{\mathbf{Y}}$ for the perturbed problem in a neighborhood of the experimental outcome $\mathbf{Y}^*$. If we restrict attention to counterfactuals where the economy is assumed to settle in that ``local" equilibrium, we can therefore describe causal effects in terms of the linearized equilibrium mapping. We analyze this problem in more detail in Section \ref{sec:infinitesimal_model}.
\end{ex}

This framework can be useful e.g. in analyzing equilibrium effects in models of trade where shocks may affect either the level of economic activity in specific regions, or the cost of trading between region pairs. Such equilibrium effects have been estimated by \cite{AKM20}'s analysis of the impact of China's accession to the World Trade Organization on economic variables at the commuting zone level. \cite{DHo16} used the 19th century expansion of the national railroad network in the US to estimate the effect of lower transportation costs to the national market on the value of agricultural land across counties.

Estimation in that literature relies on parametric assumptions that represent the equilibrium in terms of closed-form indices which are no longer sufficient for equilibrium outcomes if those assumptions were relaxed, so that it is unclear ex ante how these estimates should be interpreted if the model is not taken at face value. We look at simpler versions of this problem in a design-based approach that does not make assumptions on the nature of interference but instead on the mechanism for assigning the exogenous variables creating identifying variation, where causal parameters are specific to the realized experimental population.

Furthermore, these studies are also retrospective assessments of a change in previous equilibria in response to a change in the environment, where effects are understood to be contingent on time and place.  Issues arising from applying a design-based approach to observational rather than experimental data have been analyzed among others by \cite{LHB16} and \cite{AAIW17}.

To describe scenarios in which global outcomes may be very sensitive to unit-specific assignments, we also consider the following conceptual example:

\begin{ex}\textbf{(``Patient-Zero"-Scenario)}\label{ex:patient_zero} Suppose the aim is to protect a population against an emerging infectious disease, where the outcome of interest $Y_{i}\in\{0,1\}$ is an indicator for unit $i$ contracting the disease. For illustrative purposes, we assume that any individual with an infected neighbor also gets infected unless they receive a preventive treatment (prophylaxis, immunization, isolation), denoted by an indicator variable $D_i\in\{0,1\}$. In terms of the model (\ref{equilibrium_condition}),
\[h(\mathbf{D},\mathbf{L},\mathbf{Y},\mathbf{U};i)=(1-W_iD_i)\max_{j\neq i}\left\{L_{ij}Y_{j}\right\}\]
where the unit-specific effectiveness $W_i\in\{0,1\}$ is not observed. Unbeknownst to the policy maker, there is furthermore a single injection point for the disease, so that $Y_{i_0} = 1-W_{i_0}D_{i_0}$ and any other solutions to the structural equilibrium conditions are assumed not to occur. Potential values are then given by $Y_i(\mathbf{D})\equiv (1-W_{i_0}D_{i_0})\max_S\max_{i_0=j_0,\dots,j_S=i}\prod_{s=0}^{S-1}L_{j_sj_{s+1}}(1-W_{j_{s+1}}D_{j_{s+1}})$, i.e. $i$ gets infected if there exists at least one transmission chain $j_0,\dots,j_S$ of untreated units that connect $i_0$ to $i$ under the network $\mathbf{L}$.
\end{ex}

This example will illustrate how common causal estimands, e.g. of the individual benefit of receiving the treatment, may generally depend on epidemiological events, which are in general uncertain ex ante. This admittedly somewhat extreme version of the problem also exemplifies how the design-based approach differs from predictive models in terms of where it locates the source of that uncertainty, where our approach treats the disease status of a unit as deterministic for any given treatment allocation, whereas the latter models the dynamic of infections as stochastic given the units' immunization status.

\section{Average Causal Effects}

\label{subsec:general_approach}

In this section we propose a class of average causal effects that help understand the mechanism of interference by comparing counterfactuals that represent local changes to the experimental assignment. This approach generalizes the average direct effect of the causal variable $D_i$ on the outcome $Y_i$ of that same unit to average spillovers from changing the exposure of that reference unit $i$.

\subsection{Contingent Average Exposure Effects}

We define these estimands in terms of hypothetical local experiments that vary the treatment allocation for the network neighborhood of the reference unit. We parametrize these local experiments with summary exposure measures in the sense of \cite{Man11} and \cite{ASa17}; however, following \cite{Sav24}, we do not assume that these exposures accurately represent the effect of the social allocation $\mathbf{D}$ on potential outcomes under the true causal mechanism. Specifically, let
\begin{equation}
\label{exposure_dfn} T:\left\{\begin{array}{lcl}\mathcal{D}^n\times\mathcal{L}\times\mathcal{N}&\rightarrow&\mathcal{T}\\(\mathbf{D},\mathbf{L};i)&\mapsto&T_i(\mathbf{D})\end{array}\right.
\end{equation}
be a mapping of unit-specific assignments to unit-specific exposures in a set $\mathcal{T}$, where we generally suppress the dependence of $T_i$ on the network $\mathbf{L}$. We let $\mathcal{N}_T(i)$ denote the subset of $\mathcal{N}$ on which the value of $T_i(\mathbf{D})$ is determined.  Formally, $T_i(\mathbf{D})=T_i(\tilde{\mathbf{D}})$ for any $\tilde{\mathbf{D}}=(\tilde{D}_j)_j$ such that $D_j=\tilde{D}_j$ for any $j\in\mathcal{N}_T(i)$.

\begin{ex}\textbf{(Exposure Mappings)} The direct treatment effect on unit $i$ is the causal effect of assigning unit $i$ to treatment, corresponding to the exposure $T_i(\mathbf{D}) = D_i$. We may also be interested in average causal effects of varying the number of treated neighbors on $Y_i$, corresponding to an exposure $T_i(\mathbf{D})=\sum_{j\neq i}L_{ij}D_j$, the effect of the proportion of neighbors receiving the treatment $T_i(\mathbf{D})=\frac{\sum_{j\neq i}L_{ij}D_j}{\sum_{j\neq i}L_{ij}}$, or an indicator whether at least one network neighbor receives the treatment, $T_i(\mathbf{D})=\max_{j\neq i}\left\{L_{ij}D_j\right\}$.
\end{ex}

An \emph{exposure model} is a model that assumes that a known exposure mapping correctly parametrizes the effect of the social assignment on individual outcomes. In the terminology of \cite{Sav24}, the exposure model is correctly specified if $Y_i(\mathbf{D})\equiv \tilde{Y}_i(T_i(\mathbf{D}))$, and otherwise we refer to it as misspecified. If the exposure model is misspecified, there is therefore neglected treatment heterogeneity conditional on a given exposure $T_i(\mathbf{D})=t_0\in\mathcal{T}$, and realized outcomes therefore vary nontrivially  according to the exact assignment $\mathbf{D}|_{T_i=t_0}$.

We propose to estimate average exposure effects on the outcome of a given unit that correspond to counterfactual experiments that vary the exposure $T_i$ of unit $i$ while leaving other aspects of the social treatment unchanged.\footnote{One complication arises from the fact that the exact value $t$ of the exposure may not be achieved for some units, e.g. due to integer constraints. For example, if $T_i = T_i(\mathbf{D})$ is the fraction of units in the network neighborhood $\mathcal{N}_T(i)$ receiving the treatment, then $T_i$ can only take values that are integer multiples of the reciprocal of $|\mathcal{N}_T(i)|$. Such a challenge can be sidestepped by discretizing exposure levels, e.g. by modifying the exposure measure according to $\tilde{T}_i:=\frac1{N_0}\lceil N_0T_i\rceil$ for some integer $N_0$.} Using $\mathcal{D}_i(t)\subset\mathcal{D}^n$ to denote the set of assignments producing nominal exposure level $t$ to unit $i$,
\[\mathbf{D}\in \mathcal{D}_i(t) \Leftrightarrow T_i(\mathbf{D}) = t,\]
we define an average exposure effect via the following type of counterfactual assignment:

\begin{dfn} A counterfactual assignment of unit $i$ to exposure $T_i=t$ given the experimental assignment $\mathbf{D}$ is a random assignment $\mathbf{D}^*(t,i)=(D_j^*(t,i))_j$ with p.d.f. $\pi_T(t,i)\equiv\pi_T(\mathbf{D}^*;t,i)$ such that almost surely $\mathbf{D}^*(t,i)\in\mathcal{D}_i(t)$, and $D_j^*(t,i)=D_j$ for all $j\notin\mathcal{N}_T(i)$.
\end{dfn}

Such a mechanism corresponds to a thought experiment in which we randomly select a single reference unit $i$ from a given network and apply the counterfactual assignment only to the neighborhood of that unit defining $T_i$, leaving the allocation $\mathbf{D}$ unchanged at the realized experimental assignment for all nodes outside $\mathcal{N}_T(i)$. We define the corresponding average counterfactual as
\[V^*(t)\equiv V^*(t|\mathbf{Y}(\cdot),\mathbf{D}):= \frac1n\sum_{i=1}^n\mathbb{E}_{\pi_T(t,i)}[Y_i(\mathbf{D}^*)|\mathbf{Y}(\cdot),\mathbf{D}] \]
Note that this parameter depends on the distribution of assignments $\mathbf{D}|_{T_i=t}$ induced by the chosen mechanisms $\pi_T(t,i)$.

We can then define the \textbf{contingent average exposure effect} between exposure levels $t_1$ and $t_0$ as the contrast
\begin{eqnarray}
\nonumber\tau_{CAE}(t_1,t_0)&:=&V^*(t_1)-V^*(t_0)\\
\label{coa_exposure_dfn}&=&
 \frac1n\sum_{i=1}^n\mathbb{E}_{\pi_T(t_1,i)}[Y_i(\mathbf{D}^*)|\mathbf{Y}(\cdot),\mathbf{D}]- \frac1n\sum_{i=1}^n\mathbb{E}_{\pi_T(t_0,i)}[Y_i(\mathbf{D}^*)|\mathbf{Y}(\cdot),\mathbf{D}]
\end{eqnarray}
This average exposure effect corresponds to the ceteris paribus effect of an intervention that changes one unit's exposure on the outcome of only that particular unit. Such a contrast is a well-defined causal parameter even though the average is over treatment assignments to nodes $j\notin\mathcal{N}_T(i)$ that differ across units $i\in\mathcal{N}$ and cannot be realized simultaneously in the cross-section of nodes.

\begin{ex}\textbf{(Direct Effect)} If $T_i(\mathbf{D})=D_i$ and $D_i$ is binary, then
\[\tau_{CAE}(1,0)=V^*(1)-V^*(0)= \frac1n\sum_{i=1}^n\left(Y_i(1;\mathbf{D}_{-i})-Y_i(0;\mathbf{D}_{-i})\right)=\tau_{dir}\]
where $\mathbf{D}_{-i}:=(D_j)_{j\neq i}$ given the realized experimental assignment $\mathbf{D}$. We can interpret this effect under various assumptions regarding the spillover model $\mathbf{Y}(\cdot)$. Under the conventional SUTVA assumption, $Y_i(\mathbf{D})=Y_i(D_i)$, so that $\tau_{CAE}(1,0)$ is equal to the sample average treatment effect (see e.g. \cite{AAIW14}), independent of the experimental assignment $\mathbf{D}$. Under the exposure model $Y_i(\mathbf{D})=Y_i(D_i,T_i(\mathbf{D}))$ where $T_i(\mathbf{D}) = \frac{\sum_{j\neq i}L_{ij}D_j}{\sum_{j\neq i}L_{ij}}$ is the fraction of network neighbors receiving the treatment, $\tau_{CAE}(1,0)$ is the average direct treatment effect of $D_i$ on $Y_i$ given the realized exposures $T_i(\mathbf{D})$.
\end{ex}

In a general endogenous interactions model (\ref{equilibrium_condition}), the ``direct effect" $\tau_{CAE}(1,0)$ in the previous example is the total effect from changing $D_i$ between zero and one on the same unit's outcome, which includes general equilibrium ``feedback" effects from the interactions with other units' outcomes. Hence, in the absence of SUTVA, the direct effect of $D_i$ on $Y_i$ generally varies with the assignment $\mathbf{D}_{-i}$ to other units.

\begin{ex}\textbf{(Spillover Effect)} Suppose that $T_i(\mathbf{D}) = \frac{\sum_{j\neq i}L_{ij}D_j}{\sum_{j\neq i}L_{ij}}$. Under SUTVA, $\tau_{CAE}(t_1,t_0):=V^*(t_1)-V^*(t_0)=0$.  Under the exposure model $Y_i(\mathbf{D})=Y_i(D_i,T_i(\mathbf{D}))$ where $T_i(\mathbf{D}) = \frac{\sum_{j\neq i}L_{ij}D_j}{\sum_{j\neq i}L_{ij}}$, $\tau_{CAE}(t_1,t_0)$ is the indirect (spillover) treatment effect from increasing the treated fraction of a unit's neighbors from $t_0$ to $t_1$. In the equilibrium model (\ref{equilibrium_condition}), that indirect effect also includes feedback effects from neighbors' outcomes responding to changes in the reference unit's outcome and vice versa.
\end{ex}

In particular, if the exposures $T_1,\dots,T_n$ are correctly specified, the contingent average exposure effects do not vary across experimental assignments $\mathbf{D}$ and coincide with the (unconditional) average exposure effect in \cite{ASa17}. However, if exposures $T_i$ are not sufficient for $Y_i$, average exposure effects may vary across different social treatments even if they result in the same exposures.

\begin{ex}\textbf{(Patient-Zero Scenario)}\label{ex:patient_zero_cae} (a) In the scenario introduced in Example \ref{ex:patient_zero}, the contingent exposure effect with respect to $T_i(\mathbf{D})=D_i$, i.e. the direct effect $\tau_{dir}$, given the assignment $\mathbf{D}$, is equal to the share of units $i$ for which there exists a chain $i_0=j_0,j_1,\dots,j_S=i$ such that $(1-W_{j_s}D_{j_s})L_{j_{s-1}j_s}=1$ for all $s$, and $W_{i_0}D_{i_0}=0$. If under the experimental assignment $W_{i_0}D_{i_0}=1$, then the direct effect $\tau_{dir}=0$. (b) If we consider the exposure $T_i(\mathbf{D})=\frac{\sum_{j\neq i}L_{ij}D_j}{\sum_{j\neq i}L_{ij}}$ and $t_1>t_0\in[0,1]$, then if $D_{i_0}=0$, then $\tau_{CAE}(t_1,t_0)$ is equal to the share of units $i$ for which $D_i=0$ and whose only active transmission chains given $\mathbf{D}$ run through neighbors $j$ who are treated under $T_i=t_1$ but not under $T_i=t_0$. If $D_{i_0}=1$, then $\tau_{CAE}(t_1,t_0)=0$ regardless of the other components of the assignment $\mathbf{D}$.
\end{ex}

In this example, the measured effectiveness of the vaccine is therefore greater in the presence of an infectious wave, which in this setting is endogenous to the treatment. As this example illustrates, the ``local" effects vary with ``global" states that may be affected by the experiment and/or the eventual policy after scale-up. The way the example is constructed, there remains ``irreducible" aggregate uncertainty regarding average exposure effects even under a purely design-based perspective due to the outsize influence of the unit $i_0$ over the disease environment for the rest of the population.

\subsection{Estimation}

We next discuss estimation of average exposure effects. For the purposes of this paper, we focus on Hurvitz-Thompson-type, inverse probability weighting (IPW) estimators. For the problem at hand, inverse probability weighting effectively consists in reweighting realized outcomes from the experimental assignment using importance weights that represent the relative likelihood of a particular social treatment $\mathbf{D}$ under the target policy relative to the experimental assignment, and is therefore intuitively appealing. Large-sample properties for these estimators will be given in Section \ref{sec:asy_prop} below.

The definition in (\ref{coa_exposure_dfn}) immediately suggests a modification of the inverse probability weighting approach in (\ref{hurvitz_thompson_direct}). Since the counterfactual assignment $\pi_T(\mathbf{D};t,i)$ leaves the assignment $\mathbf{D}_{\mathcal{N}_T(i)}$ unchanged, the marginal distribution of these components under $\pi_T$ is the same as under the experimental mechanism $\pi_0$,  $\pi_T(\mathbf{D}_{-\mathcal{N}_T(i)};t,i)=\pi_0(\mathbf{D}_{-\mathcal{N}_T(i)};i)$, so that we can factor $\pi_T$ according to
\[\pi_T(\mathbf{D}_{\mathcal{N}_T(i)},\mathbf{D}_{-\mathcal{N}_T(i)};t,i)
= \pi_{T}(\mathbf{D}_{\mathcal{N}_T(i)}|\mathbf{D}_{-\mathcal{N}_T(i)};t,i)
\pi_{T}(\mathbf{D}_{-\mathcal{N}_T(i)};t,i)\]
Hence, the unconditional likelihood ratio is equal to its conditional analog,
\[\frac{\pi_{T}(\mathbf{D}_{\mathcal{N}_T(i)},\mathbf{D}_{-\mathcal{N}_T(i)};t,i)}{\pi_0(\mathbf{D})}
=\frac{\pi_{T}(\mathbf{D}_{\mathcal{N}_T(i)}|\mathbf{D}_{-\mathcal{N}_T(i)};t,i)}
{\pi_0(\mathbf{D}_{\mathcal{N}_T(i)}|\mathbf{D}_{-\mathcal{N}_T(i)})}\]

The generalized IPW estimator therefore takes the form
\begin{equation}\label{coa_ipw_estimator}\hat{V}_n(t)\equiv \hat{V}(t|\mathbf{Y}(\cdot),\mathbf{D}):= \frac1n\sum_{i=1}^nY_i\frac{\pi_{T}(\mathbf{D}_{\mathcal{N}_T(i)}|\mathbf{D}_{-\mathcal{N}_T(i)};t,i)}
{\pi_0(\mathbf{D}_{\mathcal{N}_T(i)}|\mathbf{D}_{-\mathcal{N}_T(i)})} \end{equation}
The likelihood ratios
\[r_{it}(\mathbf{d}_{\mathcal{N}_T(i)}):= \frac{\pi_{T}(\mathbf{d}_{\mathcal{N}_T(i)}|\mathbf{D}_{-\mathcal{N}_T(i)};t,i)}
{\pi_0(\mathbf{d}_{\mathcal{N}_T(i)}|\mathbf{D}_{-\mathcal{N}_T(i)})}\] serve as importance weights and can be interpreted as a statistical distance between the experimental and the target distribution.

We can immediately verify that
\[\mathbb{E}_{\pi_0}\left[\left.Y_i\frac{\pi_{T}(\mathbf{D}_{\mathcal{N}_T(i)}|\mathbf{D}_{-\mathcal{N}_T(i)};t,i)}
{\pi_0(\mathbf{D}_{\mathcal{N}_T(i)}|\mathbf{D}_{-\mathcal{N}_T(i)})}\right|\mathbf{D}_{-\mathcal{N}_T(i)}=\mathbf{d}_{-\mathcal{N}_T(i)}\right] =
\mathbb{E}_{\pi_{Ti}}\left[Y_i\right]\]
where we use the shorthand $\pi_{Ti}$ for $\pi_T(\cdot|\mathbf{D}_{-\mathcal{N}_T(i)}=\mathbf{d}_{-\mathcal{N}_T(i)};t,i)$. In order for the estimator (\ref{coa_ipw_estimator}) to have useful large-sample properties, it is generally necessary that the variance of the importance weights $r_{it}(\mathbf{d}_{\mathcal{N}_T(i)})$ does not increase too fast in $n$. This imposes a statistical constraint on ``how far" it is feasible to extrapolate from the experimental assignment. For formal conditions for consistency and asymptotic normality, we refer to Section \ref{sec:asy_prop} below.

The resulting contingent exposure effects still differ in terms of the conditional distributions of $\mathbf{D}_{\mathcal{N}_T(i)}|T_i=t$. A natural choice is $\pi_{T,0}(\mathbf{D}_{\mathcal{N}_T(i)},\mathbf{D}_{-\mathcal{N}_T(i)};t,i):=\pi_0(\mathbf{D}|T_i(\mathbf{D})=t,\mathcal{D}_{-\mathcal{N}_T(i)})$, in which case the likelihood ratio simplifies to
\[\frac{\pi_{T,0}(\mathbf{D}_{\mathcal{N}_T(i)},\mathbf{D}_{-\mathcal{N}_T(i)};t,i)}{\pi_0(\mathbf{D})}
=\frac{\dum\left\{T_i(\mathbf{D}_{\mathcal{N}_T(i)})=t\right\}}{\mathbb{P}_{\pi_0}(T_i(\mathbf{D})=t|\mathcal{D}_{-\mathcal{N}_T(i)})}
\]
where $\mathbb{P}_{\pi_0}(T_i(\mathbf{D})=t|\mathcal{D}_{-\mathcal{N}_T(i)})$ is the conditional probability of $T_i(\mathbf{D})=t$ given $\mathcal{D}_{-\mathcal{N}_T(i)}$ under the experimental distribution $\pi_0$.

\subsection{Total Effect of Intervention and Global States}

One important quantity when considering scaling up an intervention from an experimental trial is the total effect of assigning the treatment to all units in the networked population. For simplicity, consider the case of a binary treatment, $D_i\in\{0,1\}$, so that
\[\tau_{tot}:=V^*(1)-V^*(0)=\frac1n\sum_{i=1}^n(Y_i(1,\dots,1) - Y_i(0,\dots,0))\]

In some settings, such a global effect may be expressed as an aggregate of partial responses. However, in this and the next section we give stylized examples to identify scenarios in which there are ``irreducible" global features of the assignment $\mathbf{D}$ whose causal effect is not identified without additional assumptions, posing a challenge to extrapolating from partial responses to global effects. As these examples illustrate, this challenge is not restricted to a particular estimation approach but reflects a fundamental failure of the network experiment to generate identifying variation on certain aspects of the social treatment.

\begin{ex}\textbf{(Known Exposure Mapping)} Consider again the scenario in Example \ref{ex:exog_exposure} with potential outcomes $Y_i(\mathbf{D}) = Y_i(D_i,T_i)$ where $T_i=\frac{\sum_{j\neq i}L_{ij}D_j}{\sum_{j\neq i}L_{ij}}$. The experimental assignment $D_1,\dots,D_n$ is i.i.d. Bernoulli with probability $\mathbb{P}_{\pi_0}(D_i=1)=\pi_0$, and the policy concerns an i.i.d. assignment with probability $\mathbb{P}_{\pi_1}(D_i=1)=\pi_1$. Since the exposures $D_i,T_i$ are sufficient for potential outcomes $Y_i(\cdot)$, the contingent average exposure effects are equal to their unconditional expectations, \begin{eqnarray}\nonumber\tau_{dir}&:=&(\pi_1-\pi_0)\left(\mathbb{E}_{\pi_0}\left[Y_i(1,T_i)|\mathbf{Y}(\cdot),\mathbf{D}\right]-\mathbb{E}_{\pi_0}\left[Y_i(0,T_i)|\mathbf{Y}(\cdot),\mathbf{D}\right]\right)\\
\nonumber&=&(\pi_1-\pi_0)\left(\mathbb{E}_{\pi_0}\left[Y_i(1,T_i)|\mathbf{Y}(\cdot)\right]-\mathbb{E}_{\pi_0}\left[Y_i(0,T_i)|\mathbf{Y}(\cdot)\right]\right)\end{eqnarray} and \[\tau_{ind}:=\sum_{d=0,1}(1-\pi_1)^{1-d}\pi_1^d\left(\mathbb{E}_{\pi_1}\left[Y_i(d,T_i)|\mathbf{Y}(\cdot)\right]-\mathbb{E}_{\pi_0}\left[Y_i(d,T_i)|\mathbf{Y}(\cdot)\right]\right)\]
Since $D_i$ and $T_i$ are independent under both $\pi_0$ and $\pi_1$, the total effect from changing the policy from $\pi_0$ to $\pi_1$ is \[\tau_{tot}:=\frac1n\sum_{i=1}^n\left\{\mathbb{E}_{\pi_1}\left[Y_i(D_i,T_i)|\mathbf{Y}(\cdot)\right]-\mathbb{E}_{\pi_0}\left[Y_i(D_i,T_i)|\mathbf{Y}(\cdot)\right]\right\}
=\tau_{dir} + \tau_{ind}\]
If the network is sparse enough, the experiment induces variation in $T_i$ independent of $D_i$, so that the importance weights for the IPW estimator for $\mathbb{E}_{\pi_1}\left[Y_i(D_i,T_i)|\mathbf{Y}(\cdot)\right]$ remain bounded. If the network is dense, the experimental distribution of $T_i$ under independent assignment concentrates around $\pi_0$, whereas the counterfactual distribution of $T_i$ concentrates around $\pi_1$, so that the positivity condition for identification of the counterfactual fails asymptotically.
\end{ex}

In this example, the exposure model is assumed to be correctly specified in the sense that potential outcomes depend on the social treatment only through the measured exposure. Hence, neither of the contingent average exposure effects $\tau_{dir}$ and $\tau_{ind}$ depends on the experimental assignment, and therefore the total effect is generally identified from the experiment. \cite{LWa22} show that the rate at which $\tau_{ind}$ can be estimated depends on the sparsity sequence for the network.

\addtocounter{ex}{-1}
\begin{ex}\textbf{(Patient-Zero Scenario, continued)} In the scenario introduced in Example \ref{ex:patient_zero}, if $W_{i_0}=1$ the potential outcome with all units receiving the treatment, $Y_i(1,\dots,1)=0$ for all units, whereas the average potential outcome with no unit receiving the treatment equals the proportion of units pertaining to the same connected component of $\mathbf{L}$ as $i_0$. If on the other hand $W_{i_0}=0$, then the total effect is the difference in the population share between the connected component under $L_{ij}$ ($(1-W_i)(1-W_j)L_{ij}$, respectively) including the unit $i_0$. Hence a single realization of $\mathbf{D}$ reveals the value of $W_{i_0}$ only if $P(D_{i_0}=1)=1$; however, not knowing the identity of $i_0$, this can only be achieved by the experimental assignment $D_1=\dots=D_n=1$. Given that initial allocation, the counterfactual $Y_i(0,\dots,0)$ is not known unless all units of the population are connected under $\mathbf{L}$, because otherwise we do not know which connected component of $\mathbf{L}$ includes $i_0$.
\end{ex}

This is a stylized example for a scenario in which realized outcomes for all units vary with a ``global" state $D_{i_0}$ that is determined by the initial assignment. Contingent average exposure effects do represent proper causal effects of local changes to the assignment and are therefore informative about the mechanism through which treatments affect outcomes. However, these quantities are contingent on the shared global state $D_{i_0}$, and experimental estimates given $D_{i_0}=0$ are therefore generally not sufficient to predict counterfactuals with $D_{i_0}=1$, and vice versa.

A similar issue arises in models with multiple equilibria, where even if the selection rule favors equilibria close to the same reference point, the number of equilibria may vary across assignments, potentially introducing a discontinuity in the set of possible outcomes:

\begin{ex}\textbf{(Multiple Equilibria)} Consider a model of peer effects for youth smoking, where students' preference for smoking depends on the fraction of their peers who smoke. Specifically, suppose that there are $n$ students at the school, where the peer network $\mathbf{L}$ is the complete graph, i.e. $L_{ij}=1$ for all $j\neq i$. We denote student $i$'s decision to smoke with a binary indicator $Y_i\in\{0,1\}$, and the fraction of other students smoking with $T_i:=\frac1{n-1}\sum_{j=1}^nL_{ij}Y_j=\frac1{n-1}\sum_{j\neq i}Y_j$. There is an intervention $D_i\in\{0,1\}$ affecting the student's attitude towards smoking. Student $i$'s chosen action is then to satisfy \[Y_i=\left\{\begin{array}{lcl}1&\hspace{0.5cm}&\textnormal{if }\alpha_i + \beta D_i + \gamma T_i>0\\
0&&\textnormal{otherwise}\end{array}\right.\] There are two types of students: $20\%$ are of type A (``always smokers") and have $\alpha_i=1.2$, the remaining $80\%$ are of type F (``followers") and have $\alpha_i=-0.5$. We also assume that $\beta=-1$, $\gamma=1$, and $n\geq20$. We can see immediately that a student $i$ of type F who is assigned $D_i=0$ smokes if and only if $T_i>0.5$, but doesn't smoke if given $D_i=1$ regardless of $T_i$. Hence, if less than half of the type-F students receive the treatment, then there are three possible equilibria, one in which only type-A students smoke, one in which type-A and all untreated type-F students smoke, and (subject to integer constraints) one in which all type-A and a fraction of untreated type-F students smoke. If the fraction of type-F students receiving the treatment is greater than one half, then there is a unique equilibrium in which all type-A students smoke, and all type-F students do not, regardless of their unit-specific treatment status.

If the population of students always selects the largest equilibrium (with respect to the partial order on $\{0,1\}^n$), the global effect of a change of the proportion of treated type-F students from $0.4$ to $0.6$ (say) amounts to reducing the number of students who smoke from $60\%$ to $20\%$. However, for an experimental assignment in which $D_i=1$ for more than half of the type-F students, the direct effect of the treatment is zero, and therefore is uninformative with respect to the substantial global effect that would result from a moderate scaling down (or different assignment between students of either type) of the experimental assignment.
\end{ex}

These examples are very stylized but meant to illustrate the conceptual point that with endogenous interference there are plausible scenarios under which nontrivial components of the implied reduced-form exposure vary globally in a way that precludes estimation of a response from a single realization of the networked population. Certain average exposure responses can still be identified, but represent partial responses that are contingent on the experimental assignment, and may therefore fail to anticipate the full general equilibrium response to a counterfactual allocation $\mathbf{D}$.

\section{Local Changes in Smooth Equilibrium Models}

\label{sec:infinitesimal_model}

In order to illustrate how to interpret contingent average exposure effects in the presence of additional structural assumptions, we now return to the setting in Example \ref{ex:infinitesimal_ex}. In this section we will re-express average exposure effects, where we investigate whether reduced-form causal parameters can be used to recover structural features or predict policy effects. Our main qualitative findings are that (a) policy effects can typically be represented as sums of exposure effects over network neighborhoods of increasing radius, which are difficult to estimate at a longer radius, and (b) ``reduced-form" average exposure effects can be combined to identify certain weighted average structural derivatives. However, as in the classical case of a single endogenous variable, those weights cannot, in general, be directly recovered from the data and may also vary with the realization of the experimental allocation $\mathbf{D}$. Furthermore, (c) average exposure effects only capture local changes around the equilibrium selected in the realization of the experiment. If equilibrium selection depends on $\mathbf{D}$, a design-based inferential theory therefore has to treat the estimand of interest as stochastic in general.

In this section, we analyze the model (\ref{equilibrium_infinitesimal}) where $\mathbf{X}$ is continuous. Throughout we consider a marginal intervention that changes an initial continuous assignment $\mathbf{X}_0$ to $\mathbf{X}:=\mathbf{X}_0+\delta\mathbf{D}$ for a small value of $\delta>0$ and $\mathbf{D}:=(D_i)_{i=1}^n$. We assume that the mapping $\mathbf{h}(\cdot)$ is differentiable with respect to $\mathbf{X},\mathbf{Y}$ where we use $\nabla_Y\mathbf{h}(\cdot)$ and $\nabla_X\mathbf{h}(\cdot)$ to denote the Jacobian matrices of partial derivatives with respect to the components of $\mathbf{Y}$ and $\mathbf{X}$ respectively.

\subsection{Equilibria and Potential Values}

Following \cite{AFr90}, we say that a fixed point $\mathbf{Y}^* = \mathbf{h}(\mathbf{X},\mathbf{L},\mathbf{Y}^*,\mathbf{U})$ is a \emph{regular point} of the mapping if for the Jacobian $\mathbf{H}_{\mathbf{Y}}:=\nabla_{\mathbf{Y}}\mathbf{h}(\mathbf{X}_0,\mathbf{L},\mathbf{Y}^*,\mathbf{U})$, the $n\times n$ matrix $(\mathbf{I}_n-\mathbf{H}_Y)$ is nonsingular.

\begin{ass}\label{ass:infinitesimal_ass1}\textbf{(Regular Equilibrium)} The shared support of $Y_1,\dots,Y_n$ is a compact set $\mathcal{Y}$ and the mapping $h(\cdot)$ is twice continuously differentiable with respect to $\mathbf{D},\mathbf{Y}$ with probability 1. Furthermore, with probability 1, the realized equilibrium $\mathbf{Y}^*:=\mathbf{Y}^*(\mathbf{D})$ is a regular point of the fixed-point mapping $\mathbf{Y}^*=\mathbf{h}(\mathbf{D},\mathbf{L},\mathbf{Y}^*,\mathbf{U})$.
\end{ass}

By Proposition 5.4.8 in \cite{AFr90}, regular equilibria are locally unique. Also, by Assumption \ref{ass:infinitesimal_ass1}, $\mathcal{Y}^n$ is furthermore compact for any fixed value of $n$, so that there exist only finitely many regular equilibria for any experimental allocation $\mathbf{D}$. We next show that under this assumption we can conveniently parametrize equilibrium selection:

We assume throughout that $\delta>0$ is sufficiently small so that for each value pair of $\mathbf{d},\mathbf{\tilde{d}}\in\mathcal{D}^n$ and regular equilibrium $\mathbf{Y}^*(\mathbf{d})$ given the treatment allocation $\mathbf{X}=\mathbf{X}_0 + \delta\mathbf{d}$, there also exists a regular equilibrium $\mathbf{Y}^*(\mathbf{\tilde{d}})$ given treatment allocation $\tilde{\mathbf{X}}=\mathbf{X}_0 + \delta\mathbf{\tilde{d}}$ in the neighborhood of $\mathbf{Y}^*(\mathbf{d})$. Under Assumption \ref{ass:infinitesimal_ass1}, such a value of $\delta$ is guaranteed to exist. The regular equilibria given $\mathbf{\tilde{d}}=\mathbf{0}$, therefore constitute a finite set $\mathcal{Y}_0^*=\{\mathbf{Y_1}^*(0),\dots,\mathbf{Y_K}^*(0)\}\subset\mathcal{Y}^n$. Any regular equilibrium $\mathbf{Y}^*(\mathbf{\tilde{d}})$ for any alternative assignment $\mathbf{\tilde{d}}$ is therefore in a neighborhood of exactly one of the $K$ elements of $\mathcal{Y}_0^*$ and furthermore locally unique.

We can therefore define an equilibrium selection mechanism as a mapping $\nu:\mathcal{D}^n\rightarrow\{1,\dots,K\}$, which associates each treatment allocation $\mathbf{d}$ with an index $k=\nu(\mathbf{d})\in\{1,\dots,K\}$ identifying the point $\mathbf{Y}_k^*(0)\in\mathcal{Y}_0^*$, and then selects the (locally unique) regular equilibrium given that allocation that is closest to $\mathbf{Y}_k^*(0)$. We can therefore parametrize potential outcomes as $\mathbf{Y}(\mathbf{d})\equiv \mathbf{Y}(\mathbf{d},\nu(\mathbf{d}))$. Given the realized assignment $\mathbf{D}$ of the experiment, we observe outcomes satisfying \[Y_i = h(\mathbf{D},\mathbf{L},
\mathbf{Y}(\mathbf{D},\nu(\mathbf{D})),\mathbf{U};i),\] which are therefore responses to the equilibrium selected in the experimental realization according to $\nu(\mathbf{D})$.

The average exposure effects can therefore be expressed in terms of the ``local" counterfactuals $\tilde{\mathbf{Y}}^*(\tilde{\mathbf{D}}):=\mathbf{Y}(\tilde{\mathbf{D}},\nu(\mathbf{D}))$ relative to the equilibrium selected in the experiment. Using the implicit function theorem, we can then linearize the equilibrium mapping around $\mathbf{Y}^*$ to obtain
\[\tilde{\mathbf{Y}}^*(\tilde{\mathbf{D}})= \mathbf{Y}^* + \delta\left(\mathbf{I}_n-\mathbf{H_Y}\right)^{-1}\mathbf{H_X}(\tilde{\mathbf{D}}-\mathbf{D}) + O(\delta^2)\]
where $\mathbf{H_Y}:=\nabla_{\mathbf{Y}}h(\mathbf{X}_0,\mathbf{L},\mathbf{Y}^*,\mathbf{U})$ and $\mathbf{H_X}:=\nabla_{\mathbf{X}}h(\mathbf{X}_0,\mathbf{L},\mathbf{Y}^*,\mathbf{U})$, and the $i,j$th elements of $\mathbf{H_Y},\mathbf{H_X}$ are zero whenever $L_{ij}=0$. In what follows, we will ignore the $O(\delta^2)$ remainder term in this expansion and instead take local counterfactuals to be
\begin{equation}\label{inf_eq_RF}\tilde{\mathbf{Y}}(\tilde{\mathbf{D}})\equiv \tilde{\mathbf{Y}}(\tilde{\mathbf{D}};\mathbf{D})= \mathbf{Y}^* + \delta\left(\mathbf{I}_n-\mathbf{H_Y}\right)^{-1}\mathbf{H_X}(\tilde{\mathbf{D}}-\mathbf{D})\end{equation}

We can use the representation (\ref{inf_eq_RF}) to represent the average exposure effects proposed in Definition \ref{coa_exposure_dfn} in terms of this structural model. For example, the contingent average direct effect on $Y_i$ of changing $X_i$ to $X_i+\delta$ is given by
\begin{eqnarray}\nonumber\tau_{dir} &=&\frac1n\sum_{i=1}^n(\tilde{Y}_i(\mathbf{D}_{1i};\mathbf{D})-\tilde{Y}_i(\mathbf{D}_{0i};\mathbf{D}))\\ \nonumber&=&\frac{\delta}n\sum_{i=1}^n\mathbf{e}_i'\left(\mathbf{I}_n-\mathbf{H_Y}\right)^{-1}\mathbf{H_X}\mathbf{e}_i = \frac{\delta}n\tr\left(\left(\mathbf{I}_n-\mathbf{H_Y}\right)^{-1}\mathbf{H_X}\right)
\end{eqnarray}
where we define $\mathbf{D}_{1i}$ and $\mathbf{D}_{0i}$ as the assignment vectors obtained from $\mathbf{D}$ after changing $D_i$ to one (or zero, respectively), and $\mathbf{e}_i$ denotes the $i$th unit vector. Similarly, the contingent average exposure effect on $Y_i$ of changing $X_j$ to $X_j+\delta$ for all $j$ with $L_{ij}=1$ is
\[\tau_1 = \frac{\delta}n\sum_{i=1}^n\mathbf{e}_i'\left(\mathbf{I}_n-\mathbf{H_Y}\right)^{-1}\mathbf{H_X}\mathbf{L}\mathbf{e}_i = \frac{\delta}n\tr\left(\left(\mathbf{I}_n-\mathbf{H_Y}\right)^{-1}\mathbf{H_X}\mathbf{L}\right)\]
As discussed in the previous section, these parameters can then be estimated using inverse probability weighting as in (\ref{coa_ipw_estimator}).

If equilibria are not unique, the local counterfactuals $\tilde{\mathbf{Y}}(\tilde{\mathbf{D}};\mathbf{D})$ need in general not coincide with the ``ex ante" potential values $\mathbf{Y}(\mathbf{D})$ defined in (\ref{potential_outcomes_generic}). Rather, they correspond to the closest equilibrium to the point initially selected under the experimental allocation $\mathbf{D}$. This distinction is important for the interpretation of the average exposure effects defined in this paper: the estimators (\ref{coa_ipw_estimator}) are obtained by re-weighting outcomes observed under the experimental allocation. Therefore they represent average counterfactuals corresponding to local equilibria in the neighborhood of the equilibrium selected under the realization of the experiment and do not contain information about distant alternative equilibria that were not realized in the experiment.

This illustrates point (c) from the introduction to this section: in the present setting, the average exposure effects for the general case therefore correspond to ``local" counterfactuals constructed in the neighborhood of the experimental draw. Sensitivity of a causal estimand of the form (\ref{coa_exposure_dfn}) to the experimental draw is (at least to the leading term in the approximation) driven by selection among the finitely many regular equilibria given the experimental allocation.

\subsection{Policy Effects}

We are now interested in analyzing these estimands more closely to understand how ``agnostic" reduced-form estimators relate to structural features of the equilibrium model, and what additional structure may be necessary for experimental estimates to speak to policy counterfactuals. Considering a change from $X_i$ to $X_i + \delta \tilde{D}_i$ for a small value of $\delta\neq0$, one question is whether the total effect of that intervention can be represented in terms of partial effects that can be estimated nonparametrically using this strategy. Here we focus on the total effect of such an intervention,
\[\tau_{tot} = \frac1n\sum_{i=1}^n(\tilde{Y}_i(\mathbf{1},\mathbf{1})-\tilde{Y}_i(\mathbf{0},\mathbf{0}))\]
representing the difference in average outcomes from treating all units at $\tilde{D}_i=1$ for all $i$, versus not treating any of the units at $\tilde{D}_i=0$.

To represent the total effect of an intervention, we can extend the definition of $\tau_1$ to higher-order indirect effects at any given path distance from $i$. Here, let the matrix $\tilde{\mathbf{L}}_s:=(\tilde{L}_{ij,s})_{ij}$ indicate the units $j$ at a path distance equal to $s$ from $i$, where $\tilde{L}_{ij,s}=1$ if the shortest path from $i$ to $j$ through $\mathbf{L}$ is of length $s$. In particular, $\tilde{\mathbf{L}}_0 = \mathbf{I}_n$, the identity matrix. We can then define the estimands
\begin{equation}\label{tau_s_defn}\tau_s:= \frac{\delta}n\sum_{i=1}^n\mathbf{e}_i'\left(\mathbf{I}_n-\mathbf{H_Y}\right)^{-1}
\mathbf{H_X}\tilde{L}_{\mathnormal{s}}\mathbf{e}_i\end{equation} corresponding to the contingent average exposure effect on $Y_i$ of changing $X_j$ to $X_j+\delta$ for all units $j$ at a path distance $s$ from $i$. In particular, $\tau_{dir}\equiv \tau_0$. If we let $S$ be the diameter of the largest connected component of the network, we can write $\tilde{\mathbf{L}}_0 + \tilde{\mathbf{L}}_1 + \dots + \tilde{\mathbf{L}}_S=\mathbf{L}_{\infty}$, a matrix of indicators $L_{ij,\infty}$ of whether $i$ and $j$ belong to the same connected component of the network. Note that both $\mathbf{I}_n$ and $\mathbf{H_Y}$ are block-diagonal with blocks corresponding to connected components of $\mathbf{L}$, so that the inverse is block-diagonal as well. Since $\mathbf{H_X}$ is also block-diagonal with no cross partials across distinct connected components, we have \[\left(\mathbf{I}_n-\mathbf{H_Y}\right)^{-1}\mathbf{H_X}\boldsymbol{\iota}_n = \left(\mathbf{I}_n-\mathbf{H_Y}\right)^{-1}\mathbf{H_X}\mathbf{L}_{\infty}\mathbf{e}_i.\]
Using (\ref{tau_s_defn}), the global effect of the network can therefore be written
\begin{eqnarray}\label{global_decomp}\nonumber\tau_{tot} &=& \frac{\delta}{n}\sum_{i=1}^n\mathbf{e}_i'\left(\mathbf{I}_n-\mathbf{H_Y}\right)^{-1}\mathbf{H_X}\boldsymbol{\iota}_n\\
\nonumber&=&\frac{\delta}{n}\sum_{i=1}^n\mathbf{e}_i'\left(\mathbf{I}_n-\mathbf{H_Y}\right)^{-1}\mathbf{H_X}(
\tilde{\mathbf{L}}_0 + \tilde{\mathbf{L}}_1 +\dots +\tilde{\mathbf{L}}_S)\mathbf{e}_i\\
&=&\tau_{dir} + \tau_1+\dots+\tau_S\end{eqnarray}
This establishes point (a) from the introduction to this section: the theory in this paper provides conditions for estimability of $\tau_s$ for $s=0,1,\dots$, where we find that in typical cases average indirect effects can not be estimated consistently for large values of $s$. We derive rates for estimation of average spillover effects for two stylized network models in Appendix \ref{sec:design_rate_app}, where we find that it is generally challenging to estimate average effects of exposures at longer network distances.

We may ask whether there exists an alternative consistent estimator for the total effect $\tau_{tot}$ that could sidestep those challenges, but some basic considerations suggest that the answer to that question is likely negative: Since by Assumption \ref{ass:infinitesimal_ass1} the eigenvalues of $\mathbf{H}_{\mathbf{Y}}$ are bounded in absolute value by a constant less than one, we can replace the inverse $\left(\mathbf{I}_n-\mathbf{H_Y}\right)^{-1}$ with its Neumann expansion and obtain
\begin{equation}\label{tau_tot}\tau_{tot} = \frac{\delta}n\sum_{i=1}^n \mathbf{e}_i'\left(\mathbf{I}_n-\mathbf{H_Y}\right)^{-1}\mathbf{H_X}\cdot{\boldsymbol{\iota_n}} = \frac{\delta}n\sum_{i=1}^n\mathbf{e}_i'\left(\sum_{s=0}^{\infty}\mathbf{H_Y}^s\right)
\mathbf{H_X}{\boldsymbol{\iota_n}}\end{equation}
where $\mathbf{e}_i$ is again the $i$th unit vector and $\boldsymbol{\iota}_n$ is an $n$-dimensional vector of ones. The interesting case is that in which the spillover network is fully connected. It turns out that under that scenario, interference at longer network distances has ergodic properties in the sense that the higher-order effect of an assignment $\mathbf{D}$ can, to an approximation, be summarized by a global exposure measure $\mathbf{T}_{\infty,i}(\tilde{\mathbf{D}})\equiv\mathbf{T}_{\infty}(\tilde{\mathbf{D}})$ that does not vary across units.

\begin{prp}\label{prp:inf_ergodic} Suppose that there exists a positive integer $s_1$ such that all elements of $\mathbf{H_Y}^{s_1}$ are strictly positive, and furthermore that all eigenvalues of $\mathbf{H}_{\mathbf{Y}}$ are less than one in absolute value. Then there exists a global exposure measure $\mathbf{T}_{\infty}(\tilde{\mathbf{D}}):=\sum_{i=1}^nv_i^*\tilde{X}_i$ such that for any $s_0$,
\begin{eqnarray}\tilde{\tau}&:=&\frac1n\sum_{i=1}^n(\tilde{Y}_i(\tilde{\mathbf{D}};\mathbf{D})-Y_i^*)\\
\nonumber& =& \frac{\delta}n\sum_{i=1}^n \mathbf{e}_i'\left(\sum_{s=0}^{s_0}\mathbf{H_Y}^s\right)\mathbf{H_X}(\tilde{\mathbf{D}}-\mathbf{D}) + \lambda^{s_0}\left(\frac{\delta}n\sum_{i=1}^nw_i^*\mathbf{T}_{\infty}(\tilde{\mathbf{D}})
+ O\left(\delta\varrho^{s_0}\right)\right)   \end{eqnarray}
with constants $0\leq|\varrho|,|\lambda|<1$, $(v_i^*)_i$ and $(w_i^*)_i$ only depending on $\mathbf{H}_{\mathbf{Y}}$.
\end{prp}

This result is a consequence of the Frobenius-Perron theory for the matrix $\mathbf{H}_{\mathbf{Y}}$ and a proof is given in the appendix. We can interpret the bounds on the eigenvalues of $\mathbf{H}_Y$ assumed by this proposition as a local stability condition for the equilibrium $\mathbf{Y}^*$. If all spillover effects are nonnegative, i.e. $\mathbf{H}_{\mathbf{Y}}\geq0$, we can interpret the condition that $\mathbf{H_Y}^{s_1}>0$ as a requirement that the interference network is fully connected in the sense that for every $s\geq s_1$ and node pair $i_0,i_s$ there exists a path of length $s$ of nodes $i_0,\dots,i_s$ such that the $(i_{b+1},i_b)$ element of $\mathbf{H}_{\mathbf{Y}}$ is nonzero for every $b=1,\dots,s$.

An important consequence of this result is that any experimental assignment $\tilde{\mathbf{D}}$ will fail to generate any cross-sectional variation in the exposure  $\mathbf{T}_{\infty,i}(\tilde{\mathbf{D}})\equiv \mathbf{T}_{\infty}(\tilde{\mathbf{D}})$ that captures the leading component of the total treatment effect. In the absence of any additional assumptions, we cannot therefore identify any weighted average of the unit-specific responses $w_i$ from a single experiment on that population. This will generally pose a challenge for estimating the total effect for a policy change that does not leave $\mathbf{T}_{\infty}(\tilde{\mathbf{D}})$ unchanged.

\subsection{Structural Exposure Model} We now turn to the interpretation of contingent average exposure effects when the researcher is willing to make some assumptions regarding the structure of the equilibrium mappings in (\ref{equilibrium_condition}). Specifically, we consider versions of the problem in which interference is mediated by the model exposure
$T_i:=T_i(\mathbf{X},\mathbf{Y},\mathbf{L})$ for a known, scalar-valued function $T_i(\cdot)$, so that
\begin{equation}\label{struct_exp_model}h(\mathbf{X},\mathbf{Y},\mathbf{L};i) = \tilde{h}(X_i,T_i(\mathbf{Y},\mathbf{L});i).\end{equation}
Notice that this exposure mapping enters the \emph{structural} representation, and does not imply knowledge of the exposure mapping for the \emph{reduced form} in the sense that is assumed in \cite{ASa17}. Moreover, even if the exposure $T_i$ only depends on outcomes and treatments of, say, immediate neighbors of $i$ in the network $\mathbf{L}$, the reduced-form (equilibrium) outcome $Y_i$ can generally vary with unit-specific assignments at an arbitrarily large network distance from $i$.

We show that under certain broad assumptions, we can interpret certain functions of contingent average exposure effects as LATE-type complier weighted average derivatives with respect to exposures, in the spirit of \cite{AIR96}. We consider the following framework:

\begin{ass}\label{ass:endog_exposure}\textbf{(Structural Exposure Model)} (a) The equilibrium mapping in (\ref{equilibrium_condition}) is of the form (\ref{struct_exp_model}) for a known exposure mapping $T_i:=T_i(\mathbf{X},\mathbf{Y},\mathbf{L})$. (b) The mapping $\mathbf{h}(\mathbf{X},\mathbf{Y},\mathbf{L}) = (h(\mathbf{X},\mathbf{Y},\mathbf{L};i))_i$ is differentiable with respect to $\mathbf{Y},\mathbf{X}$, where all entries in $\mathbf{H}_{\mathbf{Y}}$ and $\mathbf{H}_{\mathbf{X}}$ are nonnegative.
\end{ass}

Part (b) implies that (at least locally to $\mathbf{Y}^*$) the model exhibits strategic complementarities. The main purpose of this condition is to constrain the sign of equilibrium responses, in analogy with the Monotonicity condition in the classical LATE framework, so that unit-specific effects are guaranteed to enter structural estimands with nonnegative weights. While a more comprehensive theory  for monotone comparative statics (\cite{Top78},\cite{MRo90}) is available for this problem, we can determine the direction of equilibrium responses directly due to the linear local structure of the problem.

\subsection{Linear-in-Means Model}

To frame thoughts, consider first the special case of the linear-in-means model in Example \ref{ex:linear_in_means_ex},
\[h(\mathbf{X},\mathbf{L},\mathbf{Y},\mathbf{U};i)=\beta_0 + \beta_1 X_i + \gamma_1\sum_{j\neq i}L_{ij}X_j + \gamma_2\sum_{j\neq i}L_{ij}Y_j + U_i\]
with potential values given by
\[\mathbf{Y}(\tilde{\mathbf{D}}) = \delta\left(\mathbf{I}_n - \gamma_2\mathbf{L}\right)^{-1}\left(\beta_0 + (\beta_1\mathbf{I}_n + \gamma_1\mathbf{L})(\tilde{\mathbf{D}}-\mathbf{D})+\mathbf{U}\right)\]

This is a special case of the setting in Assumption \ref{ass:endog_exposure}, where we do not need to assume strategic complementarities. Identification of this parametric model is well understood and has been analyzed by \cite{BDF09} who propose a linear instrumental variables strategy for the general specification that also includes covariates and exogenous spillovers. We now consider a special case of their setup without covariates to illustrate how the reduced-form estimands (\ref{coa_exposure_dfn}) can also serve as a basis for estimation of structural model parameters. For simplicity we also assume in addition that $\gamma_1=0$.

As in the general case, we can calculate the contingent average exposure effect on $(\mathbf{LY})_i$ of changing $D_j$ from zero to one for all units $j$ with $L_{ij}=1$ as
\begin{eqnarray}\nonumber \tau_{1,LY} &=&\frac{\delta}n\sum_{i=1}^n\mathbf{e}_i'\left(\mathbf{L}(\mathbf{I}-\gamma_2\mathbf{L})^{-1}\beta_1\mathbf{L}\right)\mathbf{e}_i\\
\nonumber&=&\frac{\delta}n\tr\left(\mathbf{L}(\mathbf{I}-\gamma_2\mathbf{L})^{-1}\beta_1\mathbf{L}\right)
\end{eqnarray}
Similarly, the effect of that change on $Y_i$ is given by
\[\tau_{1,Y} =  \frac{\delta}n\tr\left((\mathbf{I}-\gamma_2\mathbf{L})^{-1}\beta_1\mathbf{L}\right)\]
Note that $(\mathbf{I}-\gamma_2\mathbf{L})^{-1} = \mathbf{I} + \gamma_2\mathbf{L}(\mathbf{I}-\gamma_2\mathbf{L})^{-1}$, and furthermore, in the absence of self-links $L_{ii}=0$, $\tr(\mathbf{L})=0$. We can therefore recover the structural parameter
\[\frac{\tau_{1,Y}}{\tau_{1,LY}} = \frac{\beta_1\tr(\mathbf{L})+\gamma_2\tr\left(\mathbf{L}(\mathbf{I}-\gamma_2\mathbf{L})^{-1}\beta_1\mathbf{L}\right)}
{\tr\left(\mathbf{L}(\mathbf{I}-\gamma_2\mathbf{L})^{-1}\beta_1\mathbf{L}\right)}
=\gamma_2\]
Since the model is exactly linear, this argument also does not require $\delta$ to be small.

We can similarly recover $\beta_1$ from the average direct effect $\tau_{dir}\equiv\tau_{0,Y}$ of $D_i$ on $Y_i$ and the average direct effect $\tau_{0,LY}$ of $D_i$ on $(\mathbf{LY})_i$ using the same expansion:
\begin{eqnarray}\nonumber\beta_1&=&\frac1n\beta_1\tr(\mathbf{I}_n)\\
\nonumber&=& \frac1n\tr\left((\mathbf{I}-\gamma_2\mathbf{L})^{-1}\beta_1\mathbf{I}\right)
 - \frac1n\gamma_2\tr\left(\mathbf{L}(\mathbf{I}-\gamma_2\mathbf{L})^{-1}\beta_1\mathbf{I}\right)\\
\nonumber&=& \delta^{-1}(\tau_{0,Y} -\gamma_2\tau_{0,LY})
\end{eqnarray}
where $\gamma_2$ is identified from the previous step.

Hence for this simple version of the linear-in-means equilibrium model, we can interpret certain functions of contingent average exposure effects structurally. Conversely, the contingent average exposure effects can form the basis for identification of all structural model parameters, so that given this specification, the total effect of an intervention can be computed directly from the formula (\ref{tau_tot}) given our knowledge of the parameters $\beta_1,\gamma_2$. So even though for the purposes of this paper, the contingent exposure effects can be interpreted causally without imposing structure on the interference model, within this parametric framework, we can also replicate a special case of the identification argument in \cite{BDF09} without covariates.

\subsection{General Case}

In order to understand the change to the identification analysis when responses may be heterogeneous, we first turn to the generalization of the familiar linear-in-means model to a nonlinear response but for the special case in which unit $i$'s response depends on the sum of outcomes among their immediate network neighbors, i.e.
\[T_i(\mathbf{X},\mathbf{Y},\mathbf{L}) = (\mathbf{LY})_i\]
where $(\mathbf{LY})_i$ denotes the $i$th entry of the matrix product $\mathbf{LY}$.

We can then evaluate the Jacobians using the chain rule, $\mathbf{H_Y}=\diag(h_{Ti})\mathbf{L}$ and $\mathbf{H_X} = \diag(h_{Xi})\mathbf{I_n}$, where $h_{Ti}:=\left.\frac{\partial}{\partial t}\tilde{h}(x,t;i)\right|_{x=X_i,t=(\mathbf{LY})_i}$ and $h_{Xi}:=\left.\frac{\partial}{\partial x}\tilde{h}(x,t;i)\right|_{x=X_i,t=(\mathbf{LY})_i}$.

Similar to the previous example, the contingent average exposure effect on $(\mathbf{LY})_i$ of changing $D_j$ from zero to one for all units $j$ with $L_{ij}=1$ can then be written as
\[\tau_{1,LY} =\frac{\delta}n\tr\left(\mathbf{L}(I-\mathbf{H_Y})^{-1}\diag(h_{Xi})\mathbf{L}\right)\]
Similarly, the effect of that change on $Y_i$ is given by
\[\tau_{1,Y} =  \frac{\delta}n\tr\left((\mathbf{I}-\mathbf{H_Y})^{-1}\diag(h_{Xi})\mathbf{L}\right)\]

In analogy to our calculation for the linear-in-means model, we can again combine these two estimators to obtain the following:

\begin{prp}\label{end_exp_prpLY} Suppose Assumptions \ref{ass:infinitesimal_ass1} and \ref{ass:endog_exposure} hold with $T_i(\mathbf{X},\mathbf{Y},\mathbf{L}) = (\mathbf{LY})_i$. Then the ratio
\[\frac{\tau_{1,Y}}{\tau_{1,LY}} = \sum_{i=1}^n \frac{a_i}{\sum_{j=1}^na_j}h_{Ti}\]
is a weighted average of the derivatives $h_{Ti}$, where $a_i = \left(\mathbf{L}(\mathbf{I}-\mathbf{H_Y})^{-1}\diag(h_{Xi})\mathbf{L}\right)_i\geq0$ is the marginal change in $(\mathbf{LY})_i$ from increasing $X_j$ for each unit $j$ with $L_{ij}=1$.
\end{prp}

\flushleft\textsc{Proof:} To analyze the expression for $\tau_{1,Y}$ we can write
\[(\mathbf{I}-\mathbf{H_Y})^{-1} = \mathbf{H_Y}(\mathbf{I}-\mathbf{H_Y})^{-1} + \mathbf{I}\]
Furthermore, in the absence of self-links, $L_{ii}=0$, we have for any diagonal matrix $\diag(a_i)$ that $\tr(\diag(a_i)\mathbf{L})=0$. Since $\mathbf{H_Y}=\diag(h_{Ti})\mathbf{L}$, it therefore follows that
\begin{eqnarray}\nonumber\tau_{1,Y}
&=& \frac1n\left\{\tr\left(\mathbf{H_Y}(\mathbf{I}-\mathbf{H_Y})^{-1}\diag(h_{Xi})\mathbf{L}\right) + \tr(\diag(h_{Xi})\mathbf{L})\right\}\\
\nonumber&=& \frac1n\tr\left(\diag(h_{Ti})\mathbf{L}(\mathbf{I}-\mathbf{H_Y})^{-1}\diag(h_{Xi})\mathbf{L}\right)
\end{eqnarray}
Hence, the ratio
\[\frac{\tau_{1,Y}}{\tau_{1,LY}} = \sum_{i=1}^n \frac{a_i}{\sum_{j=1}^na_j}h_{Ti}\]
where $a_i = \left(\mathbf{L}(\mathbf{I}-\mathbf{H_Y})^{-1}\diag(h_{Xi})\mathbf{L}\right)_i$ is the marginal change in $(\mathbf{LY})_i$ from increasing $D_j$ for each unit $j$ with $L_{ij}=1$ from zero to one. Using the Neumann representation of the inverse, and noting that by Assumption \ref{ass:endog_exposure} (b) all entries of the matrices $\mathbf{H}_{\mathbf{Y}}$ and $\mathbf{H}_{\mathbf{X}}$ are nonnegative, $a_i$ is the product of the $i$th unit vector and an infinite sum of products of nonnegative matrices, and therefore nonnegative for each $i=1,\dots,n$\qed\\[4pt]

This result shows that if interference is exclusively channeled through a known exposure mapping in the structural equilibrium conditions, we can combine different contingent average exposure effects to identify complier weighted average derivatives with respect to exposures. Since selection among multiple equilibria can depend on the assignment $\mathbf{D}$, the weights $\frac{a_i}{\sum_{j=1}^na_j}$ also generally vary with the experimental allocation. If the equilibrium responses are not monotone in unit-specific treatment assignments, then in general the weights on unit-specific marginal effects are not guaranteed to be nonnegative, but in settings with strategic complements, monotone comparative statics can deliver the analog of the monotonicity condition in the classical LATE framework in \cite{AIR96}. In general, we may use contingent average exposure effects with respect to different changes in treatment assignments to network neighbors of the reference unit to identify different weighted averages of $h_{Ti}$.

We can directly extend this analysis to the case in which the spillover is mediated by the model exposure
$T_i:=T_i(\mathbf{Y},\mathbf{L})$ so that  $h(\mathbf{X},\mathbf{Y},\mathbf{L};i) =
\tilde{h}(X_i,T_i(\mathbf{Y},\mathbf{L});i)$. We assume that $\tilde{h}(x,t;i)$ is differentiable with respect to $x,t$ for each $i$ with $\tilde{h}_{Ti}:=\frac{\partial}{\partial t}\left.\tilde{h}(x,t;i)\right|_{x=X_i,t=T_i}$ and $\tilde{h}_{Xi}:=\frac{\partial}{\partial x}\left.\tilde{h}(x,t;i)\right|_{x=X_i,t=T_i}$. We furthermore assume that the mapping $T(\cdot)$ is differentiable with respect to $\mathbf{Y}$ with Jacobian $\mathbf{T_Y} = (T_{Yij})_{i,j}:=\nabla_{\mathbf{Y}}T_i(\mathbf{Y},\mathbf{L})$ where we assume strategic complements, $(T_{Yij})_{i,j}\geq0$ for all $i,j$. We do not necessarily assume that the correct exposure mapping is known to the researcher, but that the measured network represents the correct structure of spillovers, $T_{Yij}\neq0$ only if $L_{ij}\neq0$. Proposition \ref{end_exp_prpLY} then generalizes as follows:

\begin{prp}\label{end_exp_prpT} Suppose Assumptions \ref{ass:infinitesimal_ass1} and \ref{ass:endog_exposure} hold and let $a_{ij} = \left(\mathbf{L}(\mathbf{I}-\mathbf{H_Y})^{-1}\diag(\tilde{h}_{Xk})\mathbf{L}\right)_{ij}\geq0$
denote the marginal change in $(\mathbf{LY})_i$ from increasing $X_k$ for each unit $k$ with $L_{jk}=1$. Then the ratio
\[\frac{\tau_{1,Y}}{\tau_{1,LY}} = \sum_{i=1}^n \frac{b_i}{\sum_{j=1}^nb_j}\left(\tilde{h}_{Ti}\bar{T}_{Yi}\right)\]
is a weighted average of the derivatives $\tilde{h}_{Ti}\bar{T}_{Yi}$, where
\[\bar{T}_{Yi}:=\frac{\sum_{j=1}^nT_{Yij}L_{ij}a_{ji}}{\sum_{j=1}^nL_{ij}a_{ji}}\]
is a weighted average of $T_{Yij}$ across units $j$ that are connected to $i$, and $b_i:=\sum_{j=1}^n L_{ij}a_{ji}$.
\end{prp}

See the appendix for a proof. If the structure of responses is relaxed to also allow for exogenous interaction effects of the form
\[h(\mathbf{X},\mathbf{Y},\mathbf{L};i) = \tilde{h}(X_i,T_i(\mathbf{X},\mathbf{Y},\mathbf{L});i)\]
we can instead consider the respective contingent average exposure effects regarding $Y_i$ and $(\mathbf{LY})_i$ of changing $D_j$ from zero to one for all units $j$ that are at a network distance of 2 or greater from $i$.

\subsection{Higher-Order Effects} We can extend this approach to higher-order effects at network distances greater than 1. Specifically, consider the contingent average exposure effect on $Y_i$ of changing the treatment status of one or several units at a network distance $s>0$, and let $\tilde{\mathbf{L}}_s:=(\tilde{L}_{ij,s})_{ij}$ where $\tilde{L}_{ij,s}=1$ if the shortest path from $i$ to $j$ through $\mathbf{L}$ is of length $s$ and the treatment status of unit $j$ is changed under the counterfactual experiment with respect to the outcome of unit $i$. We can verify recursively that for any $s\geq0$, we can expand the inverse
\[(\mathbf{I}-\mathbf{H_Y})^{-1} = \sum_{t=0}^{s-1}\mathbf{H_Y}^t + \mathbf{H_Y}^s(\mathbf{I}-\mathbf{H_Y})^{-1}\]
Since by assumption $(\mathbf{H_Y})_{ij}=0$ if $L_{ij}=0$, it follows that $\left(\mathbf{H_Y}^t\right)_{ij}=0$ whenever $\left(\mathbf{L}^t\right)_{ij}=0$. In particular, $\tr\left(\mathbf{H_Y}^t\tilde{\mathbf{L}}_s\right)=0$ for every $t<s$.

Hence, we can evaluate the contingent average exposure effect on $Y_i$ of changing $D_j$ from zero to one for all units $j$ with $\tilde{L}_{ij,s}=1$ according to
\[\tau_{s,Y} =  \frac1n\tr\left((\mathbf{I}-\mathbf{H_Y})^{-1}\diag(h_{Xi})\tilde{\mathbf{L}}_s\right)\]
We also define the contingent average exposure effect on $(\mathbf{L}^s\mathbf{Y})_i$ of changing $D_j$ from zero to one for all units $j$ with $\tilde{\mathbf{L}}_{s,ij}=1$ as
\[\tau_{s,L^sY} =\frac1n\tr\left(\mathbf{L}^s(I-\mathbf{H_Y})^{-1}\diag(h_{Xi})\tilde{\mathbf{L}}_s\right)\]
We can then generalize Proposition \ref{end_exp_prpLY} to higher-order effects as follows:

\begin{prp}\label{end_exp_prp_higher} Suppose Assumptions \ref{ass:infinitesimal_ass1} and \ref{ass:endog_exposure} hold. Then the ratio
\[\frac{\tau_{s,Y}}{\tau_{s,L^sY}} = \frac{\sum_{j_0,\dots,j_s}(h_{Yj_0}L_{j_0j_1}\cdots h_{Yj_{s-1}}L_{j_{s-1}j_s})a_{j_s}}{\sum_{j_0,\dots,j_s}(L_{j_0j_1}\cdots L_{j_{s-1}j_s})a_{j_s}}
\] where $a_{j_s}\geq0$.
\end{prp}

See the appendix for a proof. Hence, that ratio is an average over $s$-fold products of the derivative $h_{Ti}$ over all paths of length $s$ (including potential cycles) connecting node pairs $j_0$ to $j_s$, weighted by initial responses $a_{j_s}$. If the derivatives $h_{Ti}$ are independent of $\mathbf{L}$, then first- and higher-order average responses would be connected mechanically; however, since $h_{Ti}$ is the derivative of $\tilde{h}(D_i,T_i;i)$ evaluated at $T_i$, such an assumption is generally implausible when $\tilde{h}(\cdot)$ is a nonlinear function in its second argument, and this approach therefore only identifies LATE-type weighted average responses instead. In particular, the magnitude of these average exposure effects generally also depends on the experimental assignments of the policy variable to units that are left unchanged under the counterfactual assignments.

\subsection{Discussion}

This establishes point (b) from the introduction to this section: a comparison of Proposition \ref{end_exp_prp_higher} to the analysis of the linear-in-means model shows how adding parametric structure greatly strengthens the conclusions the researcher may draw from these ``agnostic" contingent average exposure effects. In this setting the parametric model imposes homogeneity on unit-responses, which then suffices to extrapolate responses estimated from ``local" response to ``global" counterfactuals in general equilibrium. For extrapolation to policy counterfactuals with equilibrium effects, auxiliary structural assumptions of this kind may often be indispensable. However our results show that the estimated ``local" average responses are also causal parameters in their own right whose interpretation is robust to those added restrictions.

\section{Statistical Properties}

\label{sec:asy_prop}

In this section we develop an asymptotic theory for the estimator of the contingent average exposure effect in (\ref{coa_ipw_estimator}). Our results are design-based and concern the randomization distribution of the estimator. Statistical properties are therefore evaluated unconditionally over realizations of $\mathbf{D}$ from the experimental distribution $\pi_0$.

\subsection{Preliminaries}

Specifically, we denote the likelihood ratio weights for the generalized IPW estimator with
\[r_{it}(\mathbf{d}):=\frac{\pi_{T}(\mathbf{d}_{\mathcal{N}_T(i)}|\mathbf{d}_{-\mathcal{N}_T(i)};t,i)}
{\pi_0(\mathbf{d}_{\mathcal{N}_T(i)}|\mathbf{d}_{-\mathcal{N}_T(i)};i)}\]
and we also write $r_{it}:= r_{it}\left(\mathbf{D}_{\mathcal{N}_T(i)}\right)$. Under our design-based approach, the statistical properties of the estimator of the contingent average exposure effect are largely determined by the joint distribution of $r_{it}$ that results from a particular experimental design $\pi_0$. Using that notation, we can rewrite
\begin{eqnarray}\nonumber\hat{V}_n(t)-V^*(t|\mathbf{Y}(\cdot),\mathbf{D})&=& \frac1n\sum_{i=1}^n\left\{\sum_{\tilde{\mathbf{d}}\in\mathcal{D}^{|\mathcal{N}_T(i)|}}
\left(r_{it}\left(\tilde{\mathbf{d}}\right)
-\pi_{T}(\tilde{\mathbf{d}}|\mathbf{D}_{-\mathcal{N}_T(i)};t,i)\right)
Y_i(\tilde{\mathbf{d}};\mathbf{D}_{-\mathcal{N}_T(i)})\right\}\\
\label{coa_est_error}& =:&\frac1n\sum_{i=1}^nu_i(t)\end{eqnarray}

To understand the nature of the estimation error with respect to the contingent average exposure effect given $\mathbf{D}$, consider the problem of estimating the direct effect, i.e. $T_i(\mathbf{D})=D_i$, where for the purposes of this example $D_i\in\{0,1\}$. In that case $\pi_T(d|\mathbf{D}_{-\mathcal{N}_T(i)};d,i)$ corresponds to a point mass at $D_i=d\in\{0,1\}$ with $\tilde{\mathbf{D}}_{-\mathcal{N}_T(i)}$ held fixed at $\mathbf{D}_{-\mathcal{N}_T(i)}$, so that $u_i(d) = (r_{it}-1)Y_i(d;\mathbf{D}_{-i})$. Hence, while conditional on $\mathbf{D}_{-i}$ and $\mathbf{Y}(\cdot)$, the observed outcome for a given unit $Y_i(D_i;\mathbf{D}_{-i})$ is fixed, the difference in average responses $V^*(1)-V^*(0)$ also depends on the unobserved counterfactual $Y_i(1-D_i;\mathbf{D}_{-i})$. The estimation error therefore reflects the ex-ante uncertainty as to which of the two relevant potential outcomes is observed.

Under a design-based interpretation, the statistical properties of the estimator are therefore determined by the distribution of $u_1(t),\dots, u_n(t)$ over possible assignments induced by the experimental protocol $\pi_0$. We establish conditions for unbiasedness, consistency, and asymptotic normality, as well as identification of an upper bound of the asymptotic variance of the estimator.

\subsection{Assumptions}

While our analysis is conditional on potential outcomes as a mapping from unit-specific assignments to outcomes, $\mathbf{Y}\equiv\mathbf{Y}_n:\mathcal{D}^n\rightarrow \mathbf{R}^n$, we need to constrain their variability along the asymptotic sequence $n,n+1,\dots$ in order to justify large-population approximations. This could in principle be done by modeling the set of units $\mathcal{N}$ representing a random sample
from a suitably defined super-population. However, given the potential complexity in how unit-specific heterogeneity may interact with the social treatment, we formulate an alternative requirement constraining the realized mapping $\mathbf{Y}_n$ along that sequence:

\begin{ass}\textbf{(Potential Outcomes)}\label{ass:bd_outcome_ass}
For each $n$, potential outcomes satisfy $|Y_i(\mathbf{D})|\leq B_Y<\infty$ for all $\mathbf{D}\in\mathcal{D}^n$ and $i=1,\dots,n$.
\end{ass}
Imposing bounded support for potential outcomes greatly simplifies the notation and calculations for our proof of asymptotic normality. An extension of our results to the case of potential outcomes with unbounded support but bounded moments will be left for future research.

An important condition for identification of treatment effects is that each relevant treatment arm is assigned with nonzero probability under the experimental mechanism---typically referred to as ``probabilistic assignment" (e.g. \cite{IRu15}) or ``positivity" (e.g. \cite{ASa17}). In network experiments, assignment probabilities for exposure values, summarizing particular aspects of the social treatment, are generally determined jointly by the network structure of the mapping of unit-specific assignments to exposures, as well as the dependence of $D_1,\dots,D_n$ under the experimental assignment. As evident from the examples discussed below, this poses a particular challenge for asymptotics since probabilities for exposure values may degenerate at a rate that also depends on the asymptotic sequence (e.g. sparse or dense) assumed for the network $\mathbf{L}$. We therefore state the positivity condition as a high-level assumption in the main text and relegate a discussion of primitive conditions for some leading cases to Appendix \ref{sec:asy_ex_app}.

\begin{ass}\textbf{(Positivity)}\label{ass:positivity_ass} For each $i$ and $\mathbf{d}_{\mathcal{N}_T(i)}\in\mathcal{D}^{|\mathcal{N}_T(i)|}$, the weights $r_{it}\left(\mathbf{d}_{\mathcal{N}_T(i)}\right)^s$ have variance bounded by a sequence $B_{Tn}^s$ for some $s>0$, where $\mathbf{D}$ is distributed according to the experimental assignment mechanism $\pi_0$.
\end{ass}

The sequence $B_{Tn}^s$ effectively controls how fast the experimental assignment probability of exposure levels may converge to zero along the asymptotic sequence relative to their target distribution, where $s$ specifies for which moment the bound needs to hold. Consistency will rely on a bound for $s=1$, whereas asymptotic normality requires a stronger version of this requirement with $s=2$. By formulating this requirement in terms of likelihood ratios $r_{it}$ given the target counterfactual $\pi_T$, exposure values are implicitly weighted by their relative importance under the policy counterfactual under consideration. This reflects the fact that it is generally easier to assess counterfactuals that concentrate probability on exposure values that are also realized under the experimental assignment with sufficiently high probability.

By construction of the estimator (\ref{coa_ipw_estimator}), the conditional likelihood ratio $r_{it}$ is a function of a subvector $\mathbf{D}_{\mathcal{N}_T(i)}$ of the unit-specific assignments $\mathbf{D}$. We refer to the set $\mathcal{N}_T(i)\subset\mathcal{N}$ as the domain of $r_{it}$. The importance weights $r_{1t},\dots,r_{nt}$ are therefore generally dependent because the domains $\mathcal{N}_T(i)$ and $\mathcal{N}_T(j)$ may overlap for a node pair $i,j$, and furthermore the experimenter may often want to choose randomization designs under which components of $\mathbf{D}$ are not independent.

We formulate high-level sufficient conditions for consistency and asymptotic normality in terms of dependency neighborhoods (see \cite{CSh04}), allowing for dependence among discrete subsets of units: Specifically, for each node $i$, we let the dependency neighborhood be the smallest set $\mathcal{A}_T(i)$ such that \[\mathbf{D}_{(\mathcal{A}_T(i))^c}\ind\mathbf{D}_{\mathcal{N}_T(i)}.\]
Our results require the size of network neighborhoods to grow at a sufficiently slow rate:

\begin{ass}\textbf{(Design Dependence)}\label{ass:design_dep_ass} Let $\mathcal{A}_{T}(i)$ denote the dependency neighborhood of $\mathcal{N}_T(i)$ with respect to the unit-specific assignments $D_1,\dots,D_n$.  Then there exists a sequence $A_{Tn}$ of finite constants such that $|\mathcal{A}_T(i)|\leq A_{Tn}$ for all $n$, $i=1,\dots,n$.
\end{ass}

We give different rate conditions on $A_{Tn}$ relative to the rate $B_{Tn}^s$ from Assumption \ref{ass:positivity_ass} that are sufficient for consistency and asymptotic normality, respectively, of the estimator (\ref{coa_ipw_estimator}). A characterization of design dependence in terms of dependency neighborhoods is best suited for the case of unweighted graphs $\mathbf{L}$ and exposure mappings that are defined on strict subsets of $\mathcal{N}$. We derive rates $A_{Tn}$ and $B_{Tn}^s$ for two classes of network models in Appendix \ref{sec:design_rate_app} below. A statement of alternative conditions for the case of weighted graphs and exposure mappings with large domains will be left for future research.

Even though the design-based weights $\tilde{r}_{it}:=(r_{it}-\pi_{T}(\cdot|\cdot;i))$ are mean-independent by construction, conditioning on $\mathbf{D}_{-i}$ in the definition of the estimand introduces other dependencies that are relevant for statistical properties of the estimator under the (ex-ante) randomization distribution. To illustrate the challenge, suppose that the object of interest is the direct effect $T(\mathbf{D};i) = D_i$, and that for simplicity, $D_1,\dots,D_n$ are independent under $\pi_0$. We then have that
\begin{eqnarray}\nonumber\cov_{\pi_0}(u_i(1),u_j(1))&=&\cov_{\pi_0}(\tilde{r}_{i1}Y_i(1;\mathbf{D}_{-i}),\tilde{r}_{j1}Y_j(1;\mathbf{D}_{-j}))\\
\nonumber&=&\cov_{\pi_0}(\tilde{r}_{i1}Y_i(1,D_j;\mathbf{D}_{-i,j}),\tilde{r}_{j1}Y_j(1,D_i;\mathbf{D}_{-i,j}))\\
\nonumber&=&\mathbb{E}_{\pi_0}\left[\tilde{r}_{i1}Y_j(1,D_i;\mathbf{D}_{-i,j})\right]\mathbb{E}_{\pi_0}\left[\tilde{r}_{j1}Y_i(1,D_j;\mathbf{D}_{-i,j})\right]
\end{eqnarray}
which is generally not equal to zero despite the fact that assignments were independent.

To establish consistency and asymptotic normality, we therefore assume the following restriction on the average magnitude of the spillover effect of $\mathbf{D}_{\mathcal{N}_T(i)}$ across all units $j=1,\dots,n$.

\begin{ass}\textbf{(Bounded Influence)}\label{ass:bd_influence_ass} Let $\mathcal{D}_T(i):=\mathcal{D}^{|\mathcal{A}_T(i)|}$. Then for the random variable
\[\varphi_{ij}(t):=\sup_{\mathbf{d}_i\in\mathcal{D}_T(i)}Y_j(\mathbf{d}_j,\mathbf{d}_i;\mathbf{D}_{-\mathcal{A}_T(i)\cup\mathcal{N}_T(j)})-
\inf_{\mathbf{d}_i\in\mathcal{D}_T(i)}Y_j(\mathbf{d}_j,\mathbf{d}_i;\mathbf{D}_{-\mathcal{A}_T(i)\cup\mathcal{N}_T(j)})\]
there exists a sequence $C_n$ such that \[\frac1{n}\sum_{i\neq j}\mathbb{E}_{\pi_0}|\varphi_{ij}(t)| \leq C_n\]
for all $j$ and $\mathbf{d}_j\in\mathcal{D}^{|\mathcal{N}_T(j)|}$.
\end{ass}

The derivation of asymptotic properties for causal estimates will require different conditions on the rate of the bound $C_n$, where for the case of an exposure measure with bounded support, our consistency result imposes $C_n=o(1)$. This condition restricts the average impact on the outcome of a unit $j$ from the assignment to other units. Since that average is taken across source nodes, this requirement of a vanishing upper bound $C_n$ does not preclude the existence of influential units with a non-vanishing impact on a large share of the population, potentially all of $\mathcal{N}$, or other nontrivial long-range spillover effects which would otherwise pose challenges to estimation of the conventional unconditional causal parameters. To illustrate this, we return to some of our leading examples in Appendix \ref{sec:asy_ex_app}.

\subsection{Asymptotic Results}

Given these assumptions, we can now state our main results characterizing the statistical properties of estimators of the form (\ref{coa_ipw_estimator}). Our first result concerns the bias of the estimator (\ref{coa_ipw_estimator}):

\begin{thm}\textbf{(Unbiasedness)}\label{unbiased_thm} Suppose that Assumptions \ref{ass:experimental_assg} and \ref{ass:bd_outcome_ass}-\ref{ass:positivity_ass} hold. Then the estimator in (\ref{coa_ipw_estimator}) is \emph{unbiased} conditional on $\mathbf{Y}(\cdot)$,
\[\mathbb{E}_{\pi_0}\left[\left.\hat{V}_n(t)-V^*(t|\mathbf{Y}(\cdot),\mathbf{D})\right|\mathbf{Y}(\cdot)\right]=0\]
\end{thm}

See the appendix for a proof. It is understood that in general $V^*(t|\mathbf{Y}(\cdot),\mathbf{D})$ varies with $\mathbf{D}$ and the estimator is generally not unbiased conditional on $\mathbf{D}$. This is analogous to \cite{HSW22} with the important difference that in our analysis the estimand itself varies with the assignment $\mathbf{D}$ and is therefore random ex ante.

We furthermore find that the estimator (\ref{coa_ipw_estimator}) is consistent:

\begin{thm}\textbf{(Consistency)}\label{consistency_thm} Suppose Assumptions \ref{ass:experimental_assg} and \ref{ass:bd_outcome_ass}-\ref{ass:design_dep_ass} hold with $s=1$, $B_{Tn}A_{Tn}/n\rightarrow0$, and $B_{Tn}C_n^2\rightarrow0$. Then the estimator in (\ref{coa_ipw_estimator}) is consistent,
\[|\hat{V}_n(t)-V^*(t|\mathbf{Y}(\cdot),\mathbf{D})|\stackrel{p}{\rightarrow}0\]
\end{thm}

See the appendix for a proof. It is important to note that this result relies entirely on properties of the assignment mechanism and does not make any assumptions on the structure of interference in $\mathbf{Y}(\cdot)$. In particular, there is no presumption that the exposure mapping $T:\mathcal{D}^n\times\mathcal{N}\rightarrow\mathcal{T}^n$ accurately represents the ``structural" mechanism that generates interference in outcomes. We analyze the rates of $A_{Tn}$ and $B_{Tn}$ for some scenarios of interest in Proposition \ref{app_est_rate_prp} in the appendix.

We next establish asymptotic normality of a suitably studentized version of the estimator in (\ref{coa_ipw_estimator}). The present setup is nonstandard in that the estimand as well as the scale parameter for the estimation error are potentially stochastic regardless of population size. Our main result concerns the unconditional distribution of the estimation error induced by the randomization device $\pi_0$, which is relevant for the analysis of ex-ante statistical guarantees for inference.

In our setting, assignments are generally dependent, due to dependence of individualized assignments under the mechanism $\pi_0(\mathbf{D})$, as well as the choice of exposure measure $T(\cdot)$. The asymptotic rate of the estimation error is therefore not necessarily standard but is sensitive to design-dependence in $r_{1t},\dots,r_{nt}$ where the marginal variance of $r_{it}$ may also diverge according to the rate in Assumption \ref{ass:positivity_ass}.

To characterize the scale of the estimation error we denote
\[\omega_{ij}(t_1,t_2)\equiv\omega_{ij}(t_1,t_2;\mathbf{Y}(\cdot),\mathbf{D}):=\cov(u_i(t_1),u_j(t_2)|\mathbf{Y}(\cdot),\mathbf{D}_{-\mathcal{A}_T(i)})\]
for any $i,j\in \mathcal{N}$. For exposure values $t_1,t_2$, we then define
\begin{equation}\label{Omega_n_def}
\omega_n(t_1,t_2)\equiv\omega_n(t_1,t_2;\mathbf{Y}(\cdot),\mathbf{D}):=\frac{n^{2\varrho}}{n^2}\sum_{i=1}^n\sum_{j\in \mathcal{A}_T(i)}\omega_{ij}(t_1,t_2)\end{equation}
where $\varrho\leq\frac12$ is chosen such that $\omega_n(t_1,t_1)$ converges to a finite and strictly positive limit. Since potential outcomes $\mathbf{Y}(\cdot)$ are assumed to be bounded, and given the sequences specified in Assumptions \ref{ass:positivity_ass} and \ref{ass:design_dep_ass}, $\varrho$ has to be chosen such that $n^{2\varrho-1}B_{Tn}A_{Tn}=O(1)$, so that
\[n^{\varrho} = O\left(\sqrt{\frac{n}{B_{Tn}A_{Tn}}}\right).\] The rate exponent $\varrho$ therefore depends only on properties of the assignment mechanism $\pi_0$ and the choice of exposure measure, which are presumed to be known to the researcher.

For the joint distribution of the estimator across a finite set $\mathcal{T}=\{t_1,\dots,t_S\}$ of exposure levels, we also let
\[\Omega_n\equiv\Omega_n(\mathbf{Y}(\cdot),\mathbf{D}):=\left(\omega_n(t_1,t_2)\right)_{t_1,t_2\in\mathcal{T}}.\]
We then assume the following:

\begin{ass}\label{ass:asy_rate_ass}\textnormal{(Asymptotic Rate)} There exist constants $\varrho\in\left(0,\frac12\right]$ and $\kappa\in(0,1)$ such that all eigenvalues of the asymptotic variance matrix $\Omega_n\equiv \Omega_n(\mathbf{Y}(\cdot),\mathbf{D})$ are almost surely bounded between $\kappa$ and $1/\kappa$ for all $n$ sufficiently large.
\end{ass}

We derive expressions for the exponent $\varrho$ of the rate of consistency for several examples in Proposition \ref{app_est_rate_prp} in the appendix. It is important to note here that the conditioning variables $\mathbf{D}_{-\mathcal{A}_T(i)}$ in the definition of $\omega_{ij}(\cdot)$ are not nested. $\Omega_n$ is therefore not a proper conditional variance matrix and may generally also vary with $\mathbf{D}$. Rather, $\Omega_n$ serves as a normalizing sequence that is constructed in a way that ensures that the re-scaled estimation error is asymptotically normal even when the non-studentized estimation error is not.

We can now state our second main asymptotic result:

\begin{thm}\textbf{(Asymptotic Normality)}\label{normality_thm} Suppose Assumptions \ref{ass:experimental_assg} and \ref{ass:bd_outcome_ass}-\ref{ass:design_dep_ass} hold with $s=2$, $\frac{n}{B_{Tn}A_{Tn}}=O(n^{2\varrho})$ for some $\varrho>0$, and $\frac{nC_n^2}{A_{Tn}}\rightarrow0$. Furthermore suppose that the target distribution $\pi_T(\mathbf{d}_{\mathcal{N}_T(i)}|\mathbf{D}_{-\mathcal{N}_T(i)})\equiv\pi_T(\mathbf{d}_{\mathcal{N}_T(i)})$ does not depend on its conditioning argument. Then the estimator in (\ref{coa_ipw_estimator}) scaled by $\Omega_n$ is asymptotically normal,
\[n^{\varrho}\Omega_n(\mathbf{Y}(\cdot),\mathbf{D})^{-1/2}(\hat{V}_n(\mathbf{t})-V^*(\mathbf{t}|\mathbf{Y}(\cdot),\mathbf{D}))\stackrel{d}{\rightarrow}N(0,I)\]
\end{thm}

See the appendix for a proof. We rely on Stein's method, which requires that we control cross-unit dependencies between assignments and outcomes in terms of moments up to the fourth order. Since a change of the assigned treatment status of influential units may also affect the scale of potential outcomes for all other units, the standardization by the assignment-dependent sequence $\Omega_n$ is generally necessary for a Gaussian asymptotic distribution.

\subsection{Variance Bound}

In accordance with the design-based approach taken in this paper, Theorem \ref{normality_thm} concerns the unconditional distribution of the estimator $\hat{V}_n(\mathbf{t})$. It can therefore serve as a basis to analyze ex-ante statistical properties of inference procedures over the randomization distribution of the estimator or a test statistic. The Central Limit Theorem \ref{normality_thm} suggests Gaussian asymptotic inference using a consistent, or appropriately conservative, estimator of $\Omega_n$. As in more standard situations (see e.g. \cite{AAIW14}, \cite{AGL14}, and \cite{ASa17}), the asymptotic distribution of the estimation error in (\ref{coa_est_error}) is typically not identified from experimental data.

Specifically, consider the problem of inference for the treatment contrast $V^*(t_1)-V^*(t_2)$ for a pair of exposure levels $t_1,t_2\in\mathcal{T}$. Multiplying out the numerator in the expression for $\omega_{ii}(t_1,t_2)$, we can write $\var_{\pi_0}(u_i(t_1)-u_i(t_2))=\var_{\pi_0}(u_i(t_1))+\var_{\pi_0}(u_i(t_2))
-2\cov_{\pi_0}(u_i(t_2),u_i(t_1))$. As in the leading case of SUTVA, the covariance term is not point-identified since it depends on the joint distribution of potential values for $Y_i$ under two different exposures $t_1,t_2$ whereas the outcome is observed for at most one of the two exposure values for any given unit. While the distributions of the relevant potential outcomes given different exposures, $Y_i(\mathbf{D})|T(\mathbf{D};i)=t$, are point-identified conditional on the realized assignment $\mathbf{D}$, conditional inference in a setting that allows for multiple equilibria poses some additional challenges that have to be addressed separately and are beyond the current scope of this paper.

The definition of $\Omega_n$ in (\ref{Omega_n_def}) suggests estimation using sample analogs to averages of pairwise covariances. To this end, denote \[\pi_{0,ij}(\mathbf{d}_{t_1},\mathbf{d}_{t_2}|\mathbf{D},i,j):=
\mathbb{P}_{\pi_0}(\mathbf{D}_{\mathcal{N}_T(i)}=\mathbf{d}_{t_1},\mathbf{D}_{\mathcal{N}_T(j)}=\mathbf{d}_{t_2}|\mathbf{D}_{-\mathcal{N}_T(i)\cup\mathcal{N}_T(j)}).\]
We then propose the following inverse probability weighting (IPW) variance estimator for $\omega_n(t_1,t_1)$,
\[\hat{\omega}_n(t_1,t_1):=\frac1n\sum_{i=1}^n\sum_{j\in\mathcal{A}_T(i)}\sum_{\mathbf{d}_{t_1}\in\mathcal{D}_i(t_1)}
\frac{\dum\{\mathbf{D}_{\mathcal{N}_T(i)}=\mathbf{d}_{t_1},\mathbf{D}_{\mathcal{N}_T(j)}=\mathbf{d}_{t_1}\}}
{\pi_{0,ij}(\mathbf{d}_{t_1},\mathbf{d}_{t_1}|\mathbf{D},i,j)}Y_iY_j
\cov(r_{i}(\mathbf{d}_{t_1}),r_j(\mathbf{d}_{t_1}))\]
where $\mathcal{D}_i(t_1):=\{\mathbf{d}_{t_1}\in \{0,1\}^{|\mathcal{N}_T(i)|}:T(\mathbf{d}_{t_1};\mathbf{D}_{-\mathcal{N}_T(i)})=t_1\}$.  Under regularity conditions, this estimator can be shown to be consistent for $\omega_n(t_1,t_1)$.

In contrast, the covariance $\omega_n(t_1,t_2)$ with $t_1\neq t_2$ is generally not identified: from (\ref{Omega_n_def}),
\begin{equation}\label{omega_12_formula}\omega_n(t_1,t_2) = \frac{n^{2\varrho}}{n^2}\left\{\sum_{i=1}^n\sum_{j\in \mathcal{A}_T(i)\backslash\{i\}}\omega_{ij}(t_1,t_2)
 + \sum_{i=1}^n\omega_{ii}(t_1,t_2)\right\}\end{equation}
The first term can be estimated using a similar inverse probability weighting approach as before. The second term
\begin{eqnarray}\nonumber W_n(t_1,t_2)&:=& \frac{n^{2\varrho}}{n^2}\sum_{i=1}^n\omega_{ii}(t_1,t_2)\\
\nonumber&=& \frac{n^{2\varrho}}{n^2}\sum_{i=1}^n\sum_{\tilde{\mathbf{d}}_{t_1}\in\mathcal{D}_i(t_1)}
\sum_{\tilde{\mathbf{d}}_{t_2}\in\mathcal{D}_i(t_2)}\cov(r_i(\tilde{\mathbf{d}}_{t_1}),r_i(\tilde{\mathbf{d}}_{t_2}))
Y_i(\tilde{\mathbf{d}}_{t_1};\mathbf{D}_{-\mathcal{N}_T(i)})
Y_i(\tilde{\mathbf{d}}_{t_2};\mathbf{D}_{-\mathcal{N}_T(i)})\end{eqnarray}
depends on products of different potential values for the same unit, $Y_i(\mathbf{d}_{t_1};\mathbf{D}_{-\mathcal{N}_T(i)})Y_i(\mathbf{d}_{t_2};\mathbf{D}_{-\mathcal{N}_T(i)})$ with $\mathbf{d}_{t_1}\neq\mathbf{d}_{t_2}$ which are not simultaneously observed. For the IPW strategy of constructing an estimator, this problem manifests itself in that $\pi_{0,ii}(\mathbf{d}_{t_1},\mathbf{d}_{t_2}|\mathbf{D}_{-\mathcal{N}_T(i)\cup\mathcal{N}_T(j)};i,j)\equiv 0$ whenever $t_1\neq t_2$.

\subsection{Variance Estimation}

We propose conservative estimation for the problem of inference for a treatment contrast between exposure levels $t_1$ and $t_2$ via
\[\hat{\tau}_n(t_1,t_2):= \hat{V}_n(t_1)-\hat{V}_n(t_2)\]
The asymptotic variance of $\hat{\tau}_n(t_1,t_2)$ is given by
\[\sigma_n(t_1,t_2):= \omega_n(t_1,t_1)+\omega_n(t_2,t_2)-2\omega_n^L(t_1,t_2)\]
Noting that the first two terms are point-identified, it is evident from (\ref{omega_12_formula}) that an upper bound for $\sigma_n(t_1,t_2)$ can be obtained by replacing the term $W_n(t_1,t_2)$ with a lower bound $W_n^L(t_1,t_2)$.

For a given value $t\in\mathcal{T}$ of the exposure measure, define conditional c.d.f.s
\[F_n(y;t):=\frac1n\sum_{i=1}^n\mathbb{P}_{\pi_0} \left(\left.Y_i(\mathbf{D})\leq y \right|T_i(\mathbf{D})=t,\mathbf{D}_{-\mathcal{N}_T(i)}\right)\]
where
\[\mathbb{P}_{\pi_0} \left(\left.Y_i(\mathbf{D})\leq y \right|T_i(\mathbf{D})=t,\mathbf{D}_{-\mathcal{N}_T(i)}\right): =\sum_{\mathbf{d}_{t}\in\mathcal{D}_i(t)}\dum\left\{Y_i(\mathbf{d}_{t};\mathbf{D}_{-\mathcal{N}_T(i)})\leq y\right\}\pi_0(\mathbf{d}_{t}|t;i)\]
and $\pi_0(\mathbf{d}_{t}|t;i) :=\pi_0(\mathbf{d}_{t}|T_i(\mathbf{d}_{t},\mathbf{D}_{-\mathcal{N}_T(i)})=t,\mathbf{D}_{-\mathcal{N}_T(i)};i)$. From the previous argument, these distributions are point-identified from the data, and a natural estimator is given by
\[\hat{F}_n(y;t):=\frac1n\sum_{i=1}^n\frac{\dum\{Y_i\leq y,T_i = t\}}{\pi_0(t|\mathbf{D}_{-\mathcal{N}_T(i)};i)}\]
where $\pi_0(t|\mathbf{D}_{-\mathcal{N}_T(i)};i):=\mathbb{P}_{\pi_0} \left(\left.T_i(\mathbf{D})=t\right|\mathbf{D}_{-\mathcal{N}_T(i)}\right)$.

We now state a lower variance bound under the additional assumption that the assignment mechanism is symmetric conditional on $\mathbf{D}$ in the sense that $\pi_0(t|\mathbf{D}_{-\mathcal{N}_T(i)};i)$ does not depend on $i$ (see the footnote below for an adjustment for the general case). Following an argument similar to \cite{IMe18}, a lower bound to $W_n(t_1,t_2)$ subject to the marginal distributions $F_n(y;t_1),F_n(y;t_2)$ can be stated in terms of the isotone matching between those marginal distributions, which we define as
\[\tilde{Y}^{iso}(y;t_1,t_2):=F_{n}^{-1}(F_{n}(y;t_1);t_2)\]
We then let
\[W_n^L(t_1,t_2):= \frac{n^{2\varrho}}{n^2}\sum_{i=1}^n\sum_{\tilde{\mathbf{d}}_{t_1}}
\sum_{\tilde{\mathbf{d}}_{t_2}}\cov(r_i(\tilde{\mathbf{d}}_{t_1}),r_i(\tilde{\mathbf{d}}_{t_2}))
Y_i(\tilde{\mathbf{d}}_{t_1};\mathbf{D}_{-\mathcal{N}_T(i)})Y_i^{iso}(Y_i(\tilde{\mathbf{d}}_{t_1};\mathbf{D}_{-\mathcal{N}_T(i)});t_1,t_2)
\]
denote the analog of $W_n(t_1,t_2)$ under that isotone assignment.\footnote{For the general case in which $\pi_0(t|\mathbf{D}_{-\mathcal{N}_T(i)};i)$ varies with $i$, we construct the least favorable coupling as follows: We define $W_i:=Y_i\cov(r_{it_1},r_{it_2})$ and denote
\[G_n(w;t_1):=\frac1n\sum_{i=1}^n\sum_{\mathbf{d}_{t_1}\in\mathcal{D}_i(t_1)}\dum\left\{\cov(r_{it_1},r_{it_2})Y_i(\mathbf{d}_{t_1};\mathbf{D}_{-\mathcal{N}_T(i)})\leq y\right\}\pi_0(\mathbf{d}_{t_1}|T_i(\mathbf{d}_{t_1},\mathbf{D}_{-\mathcal{N}_T(i)})=t_1,\mathbf{D}_{-\mathcal{N}_T(i)};i)\]
We then let
\[\tilde{Y}^{iso}(y;t_1,t_2):=F_{n}^{-1}(G_{n}(w;t_1);t_2)\] and apply the analogous formula for $W_n^L$ given that coupling.} We can then show that $W_n^L(t_1,t_2)$ is indeed a lower bound for $W_n(t_1,t_2)$:

\begin{prp}\textbf{(Lower Bound)}\label{variance_bd_prp} Suppose that $\pi_0(t|\mathbf{D}_{-\mathcal{N}_T(i)};i)=\pi_0(t|\mathbf{D}_{-\mathcal{N}_T(i)})$ almost surely. Then $W_n^L(t_1,t_2)\leq W_n(t_1,t_2)$ $\pi_0$-almost surely.
\end{prp}

See the appendix for a proof. It is important to note that the resulting bound is not necessarily sharp: for one, we minimize the term $W_n$ without consideration of the remaining two terms in the expression (\ref{omega_12_formula}), i.e. ignoring the additional constraint that the potential values for each unit under the least favorable assignment have to be the same in all three terms. Furthermore, we minimize the lower bound without distinguishing different ``local" assignments $\tilde{\mathbf{d}}_{t}$ that result in the same exposure value $t=t_1,t_2$. As shown in the proof, this constitutes a relaxation of constraints relative to the original problem, and therefore results in a lower value for the resulting constrained minimization problem.

We propose estimating this bound using its sample analog, where we let
\[\hat{Y}^{iso}(y;t_1,t_2):=\hat{F}_{n}^{-1}(\hat{F}_{n}(y;t_1);t_2)\]
denote the isotone assignment given the estimated c.d.f.s $\hat{F}_n(y;t_1),\hat{F}_n(y;t_2)$. We then let
\[\hat{W}_n^L(t_1,t_2):= \frac{n^{2\varrho}}{n^2}\sum_{i=1}^n\sum_{\tilde{\mathbf{d}}_{t_1}\in\mathcal{D}_i(t_1)}
\sum_{\tilde{\mathbf{d}}_{t_2}\in\mathcal{D}_i(t_2)}\cov(r_i(\tilde{\mathbf{d}}_{t_1}),r_i(\tilde{\mathbf{d}}_{t_2}))\frac{Y_i\hat{Y}_i^{iso}(Y_i;t_1,t_2)\dum\{T_i=t_1\}}
{\pi_0(t_1|\mathbf{D}_{-\mathcal{N}_T(i)};i)}\]
and obtain the resulting lower bound estimate for that covariance,
\begin{eqnarray}\nonumber\hat{\omega}_n^L(t_1,t_2)&:=&\frac1n\sum_{i=1}^n\sum_{j\in\mathcal{A}_T(i)\backslash\{i\}}
\sum_{\mathbf{d}_{t_1}\in\mathcal{D}_i(t_1)}\sum_{\mathbf{d}_{t_2}\in\mathcal{D}_j(t_2)}
\frac{\dum\{\mathbf{D}_{\mathcal{N}_T(i)}=\mathbf{d}_{t_1},\mathbf{D}_{\mathcal{N}_T(j)}=\mathbf{d}_{t_2}\}}
{\pi_{0,ij}(\mathbf{d}_{t_1},\mathbf{d}_{t_2}|\mathbf{D};i,j)}Y_iY_j
\cov(r_{i}(\mathbf{d}_{t_1}),r_j(\mathbf{d}_{t_2}))\\
\nonumber&& + \hat{W}_n^L(t_1,t_2)\end{eqnarray}
In the present version of this paper we do not provide regularity conditions for consistent estimation of the components of $\hat{\Omega}_n$, which require strengthening of Assumptions \ref{ass:positivity_ass} and \ref{ass:design_dep_ass} to ensure consistent estimation of covariances across unit pairs.

\subsection{Inference}

Combining the previous results, we propose the following procedure for constructing a confidence interval for the treatment contrast $V^*(t_1|\mathbf{D})-V^*(t_2|\mathbf{D})$ for the exposure pair $t_1,t_2$.
\begin{itemize}
\item Obtain the point estimate $\hat{\tau}_n(t_1,t_2):= \hat{V}_n(t_1)-\hat{V}_n(t_2)$ for the contingent average exposure effect.
\item Using the previous formulae for estimation of $\Omega_n$, compute the asymptotic standard error
\[\hat{\sigma}_n(t_1,t_2):= \hat{\omega}_n(t_1,t_1)+\hat{\omega}_n(t_2,t_2)-2\hat{\omega}_n^L(t_1,t_2)\]
where $\hat{\omega}_n^L(t_1,t_2)$ is the lower bound estimator for the covariance $\omega_n(t_1,t_2)$.
\item Form the confidence interval
\[CI = \left[\hat{\tau}_n(t_1,t_2)-n^{\frac12-\varrho}\hat{\sigma}_n(t_1,t_2)^{1/2}z_{1-\frac{\alpha}2},
\hat{\tau}_n(t_1,t_2)+n^{\frac12-\varrho}\hat{\sigma}_n(t_1,t_2)^{1/2}z_{1-\frac{\alpha}2}\right]\]
for the $1-\frac{\alpha}2$ quantile of the standard normal distribution $z_{1-\frac{\alpha}2}$.
\end{itemize}
The variance estimator $\hat{\sigma}_n(t_1,t_2)$ is by construction asymptotically conservative in the sense that the negative part of $\hat{\sigma}_n(t_1,t_2)-\var_{\pi_0}(\hat{\tau}_n(t_1,t_2))$ converges in probability to zero. Hence, by Theorem \ref{normality_thm} and Slutsky's Lemma, the interval $CI$ has asymptotic confidence size greater than or equal to $1-\frac{\alpha}2$ with respect to the randomization distribution $\pi_0$. Hypothesis tests for contingent average exposure effects based on Gaussian critical values can be found using the same approach. Note that under this construction, size and coverage are controlled only \textbf{unconditionally}, i.e. ex-ante with respect to the realization of assignments $D_1,\dots,D_n$, while the potential values $\mathbf{Y}(\cdot)$ are regarded as fixed.

\section{Conclusion}

In this paper, we define causal estimands that are identified and estimable from a single experiment on the network under minimal assumptions on the structure of interference. Since we cannot rule out in particular that the outcome for any reference unit may depend on unit-specific assignments to any other unit in the network, the contingent average exposure effects proposed here are partial average responses that may vary with other global features of the realized assignment.

To understand what these ``agnostic" reduced-form estimands can reveal about the underlying structural mechanism, we analyzed in detail the case of an equilibrium model with smooth response functions. Under additional structural assumptions---most transparently in the linear-in-means model---contingent average exposure effects can be combined to recover Local Average Treatment Effect (LATE)-type weighted averages of structural derivatives, and in that parametric case are in fact sufficient to extrapolate from ``local" responses to the ``global" equilibrium response to a counterfactual intervention. Absent such structure, the reduced-form estimands retain their less restrictive causal interpretation but need not identify the corresponding global effect: our stylized examples, including the Patient-Zero and Multiple Equilibria scenarios, illustrate that some policy-relevant global features of the assignment can remain irreducibly unidentified from a single experiment, and Proposition \ref{prp:inf_ergodic} shows formally that in a fully connected network, the leading-order contribution of long-range interference collapses onto a single global exposure measure that cannot be identified without additional assumptions.

Under our design-based approach, this type of estimand variability requires a non-standard statistical theory for estimation and inference. We therefore developed a fixed-population, design-based asymptotic theory for the proposed class of IPW estimators, treating the experimental assignment as the principal source of estimation uncertainty, and established that these estimators are unbiased, consistent and asymptotically normal. As in the classical case of randomization inference without interference, the asymptotic variance of these estimators is generally not point-identified; we therefore also proposed a conservative variance estimator and corresponding asymptotically valid confidence intervals.

For practical work, the researcher may often choose to impose the structure necessary to arrive at a definitive policy conclusion, and use that additional model structure to define exposure measures to operationalize an estimation approach. Our theory can potentially be used to give contrasts with respect to such an exposure measure a more robust causal interpretation that does not rely on correct specification of the underlying model. Throughout, we have restricted attention to unweighted graphs and to potential outcomes with bounded support; extending the design-based theory developed here to weighted interference networks and to outcomes with unbounded support, and deriving primitive conditions for a broader class of experimental designs, are natural directions for future work.

\appendix

\section{Proofs for Results in Section \ref{sec:infinitesimal_model}}

\flushleft\textsc{Proof of Proposition \ref{prp:inf_ergodic}} By assumption, there is $s_1$ such that $\mathbf{H_Y}^{s_1}>0$, and the matrix $\mathbf{H}_{\mathbf{Y}}$ is therefore primitive according to the definition of the term in \cite{Sen81}. Theorem 1.2 in \cite{Sen81} therefore implies that there exists a spectral representation of $\mathbf{H_Y}$ where the largest singular value (in absolute value) $\lambda_1$ is associated with a unique eigenvector pair $\tilde{\mathbf{w}}_1=(\tilde{w}_i)_i,\mathbf{v}_1=(v_i)_i$, and is well separated from the second largest singular value $\lambda_2$ so that
\[\sum_{s=0}^{\infty}\mathbf{H_Y}^s = \sum_{s=0}^{\tilde{s}}\mathbf{H_Y}^s + \frac{\lambda_1^{\tilde{s}}}{1-\lambda_1}\tilde{\mathbf{w}}\mathbf{v}' + O(\lambda_2^{\tilde{s}})\]
Since the equilibrium was assumed to be a regular point of the mapping $\mathbf{h}$, $|\lambda_1|<1$.

Hence the leading contribution of the assignment change $t\boldsymbol{\mathbf{D}}=\delta(D_i)_i$ among units at a network distance greater than $\tilde{s}$ to the outcome of unit $i$ is given by $\frac{\delta\lambda_1^{\tilde{s}}}{1-\lambda_1}\tilde{w}_i\left(\mathbf{v}'\mathbf{H}_X(\tilde{\mathbf{D}}- \mathbf{D})\right)=:\frac{\delta\lambda_1^{\tilde{s}}}{1-\lambda_1} \tilde{w}_i T_i^*(\tilde{\mathbf{D}})$. Hence, the claim follows by setting $\lambda\equiv\lambda_1$, $\varrho\equiv\lambda_2$, $w_i^* = \tilde{w}_i/(1-\lambda_1)$ and $\mathbf{v}^*:=\mathbf{v}'\mathbf{H}_X$\qed

\flushleft\textsc{Proof of Proposition \ref{end_exp_prpT}:} By the chain rule the Jacobian of $h(\mathbf{D},\mathbf{Y},\mathbf{L};i)$ then evaluates to $\mathbf{H_Y}=\diag\left(\tilde{h}_{Ti}\right)\mathbf{T_Y}$.

As in the proof of Proposition \ref{end_exp_prpLY}, the contingent average exposure effect on $Y_i$ of changing $D_j$ from zero to one for all units $j$ with $L_{ij}=1$ is then given by
\[\tau_{1,Y} =  \frac1n\tr\left((\mathbf{I}-\mathbf{H_Y})^{-1}\diag(\tilde{h}_{Xi})\mathbf{L}\right)
= \frac1n\tr\left(\mathbf{H_Y}(\mathbf{I}-\mathbf{H_Y})^{-1}\diag(\tilde{h}_{Xi})\mathbf{L}\right)=\frac1n\tr(\mathbf{H_Y}A)\]
where we write $\mathbf{A}:=(\mathbf{I}-\mathbf{H_Y})^{-1}\diag(\tilde{h}_{Xi})\mathbf{L}\equiv (a_{ij})_{ij}$. Similarly, $\tau_{1,LY}$ equals
\[\tau_{1,LY} =  \frac1n\tr\left(\mathbf{L}(\mathbf{I}-\mathbf{H_Y})^{-1}\diag(\tilde{h}_{Xi})\mathbf{L}\right)=\frac1n\tr(\mathbf{L}\mathbf{A})\]
We can then write
\[\tr(\mathbf{L}\mathbf{A})=\sum_{i=1}^n\left(\sum_{j=1}^nL_{ij}a_{ij}\right)=:\sum_{i=1}^n b_i\]
so that we also have
\begin{eqnarray}
\nonumber\tr(\mathbf{H_Y}\mathbf{A})&=&\sum_{i=1}^n\sum_{j=1}^n\left((\mathbf{H_Y})_{ij}a_{ji}\right)=\sum_{i=1}^n \tilde{h}_{Ti}\sum_{j=1}^nT_{Yij}a_{ji}\\
\nonumber&=&\sum_{i=1}^n \tilde{h}_{Ti}b_i\left(\frac{\sum_{j=1}^nT_{Yij}L_{ij}a_{ji}}{\sum_{j=1}^nL_{ij}a_{ji}}\right)
=\sum_{i=1}^n \tilde{h}_{Ti}\bar{T}_{Yi}b_i
\end{eqnarray}
where the third equality uses that $L_{ij}=1$ whenever $T_{Yij}\neq 0$.

Hence, following the same steps as before the ratio of $\tau_{1,Y}$ over the contingent average exposure effect on $(\mathbf{LY})_i$ is given by
\[\frac{\tau_{1,Y}}{\tau_{1,LY}} = \sum_{i=1}^n \frac{b_i}{\sum_{j=1}^nb_j}\left(\tilde{h}_{Ti}\bar{T}_{Yi}\right)\]
Since $b_j$ and $T_{Yij}$ are nonnegative for all $i,j$ as in the proof of Proposition \ref{end_exp_prpLY}, this ratio of contingent average exposure effects gives a weighted average of the derivative of $\tilde{h}(x,t;i)$ with respect to $t$ times the average derivative of $T_i$ with respect to $Y_j$ among those units $j$ that are connected to $i$\qed

\subsection{Proof of Proposition \ref{end_exp_prp_higher}} As in the proof of Proposition \ref{end_exp_prpLY}, we can write
\[\tau_{s,Y} =  \frac1n\tr\left((\mathbf{I}-\mathbf{H_Y})^{-1}\diag(h_{Xi})\tilde{\mathbf{L}}_s\right)
= \frac1n\tr\left(\mathbf{H_Y}^s(\mathbf{I}-\mathbf{H_Y})^{-1}\diag(h_{Xi})\tilde{\mathbf{L}}_s\right)\]
noting that $\tr(\tilde{\mathbf{L}}_s)=0$ as well. Since
\[\tau_{s,L^sY} =\frac1n\tr\left(\mathbf{L}^s(I-\mathbf{H_Y})^{-1}\diag(h_{Xi})\tilde{\mathbf{L}}_s\right)\]
the ratio of the two effects then equals
\[\frac{\tau_{s,Y}}{\tau_{s,L^sY}} = \frac{\sum_{j_0,\dots,j_s}(h_{Tj_0}L_{j_0j_1}\cdots h_{Tj_{s-1}}L_{j_{s-1}j_s})a_{j_s}}{\sum_{j_0,\dots,j_s}(L_{j_0j_1}\cdots L_{j_{s-1}j_s})a_{j_s}}
\]
establishing the claim\qed

\section{Rate Calculations}

\label{sec:design_rate_app}

The asymptotic behavior of the estimator (\ref{coa_ipw_estimator}) crucially depends on the rate sequences $A_{Tn}$ and $B_{Tn}^s$ specified in Assumptions \ref{ass:positivity_ass} and \ref{ass:design_dep_ass} which need to be checked on a case-by-case basis. The relationship between network topology and randomization design and rates is subtle, so we derive results for some leading cases.

\subsection{Network Structure}
In order to illustrate how rates may vary with the network topology, we conduct our analysis for two alternative network structures: a spatial grid model and Erd\H{o}s-R\'enyi random graphs.

\begin{ass}\textbf{(Erd\H{o}s-R\'enyi Random Graph Model, ERRGM)}\label{errgm_ass}
Links $L_{ij}^{ERRGM}$ are formed independently at random with probability $p:=\eta_n$ for any node pair $i,j$.
\end{ass}

The link density for this model is $h_n^{ERRGM}:=\frac{\sum_{i,j}\mathbb{E}\left[L_{ij}^{ERRGM}\right]}{n(n-1)}=\eta_n$.

For a setting in which the network has a natural dimension, we consider a simple latent space network formation model in which connections are fully determined by homophilous node attributes.

\begin{ass}\textbf{(Geometric Random Graph Model, GRGM)}\label{grgm_ass}
We assume that nodes are associated with positions $Z_i$ in a $d$-dimensional Euclidean space, where link formation is determined according to
\[L_{ij}^{GRGM} = \dum\{ \|Z_i-Z_j\|_{\infty}<\delta_n\}\]
using the max norm, $\|z_1-z_2\|_{\infty}:=\max_{s=1,\dots,d}|z_{1s}-z_{2s}|$.
\end{ass}

We choose the threshold sequence $\delta_n:=\frac12\eta_n^{1/d}$ for some sequence $\eta_n$ to control sparsity of the network sequence. Since the distance of nodes is defined as the max-norm, the radius-$r$ network neighborhood of node $i$ corresponds to the set of all nodes with positions in a $d$-dimensional hypercube of side length $2r\delta_n$.

We assume that node positions are deterministic and evenly spaced on a grid in $\mathcal{Z}:=[-1,1]^d$. For $\delta_n$ small enough, we can ignore boundary effects, and rates are determined by the properties of reference nodes in the $\delta_n$-interior of $\mathcal{Z}$, which we define as $\mathcal{N}^{\delta_n}:=\{i\in\mathcal{N}:B_{\delta_n}(Z_i)\subset\mathcal{Z}\}$. The link density for this model is
$h_n^{GRGM}:=\frac{\sum_{i,j\in\mathcal{N}^{\delta_n}}\mathbb{E}\left[L_{ij}^{GRGM}\right]}{n(n-1)}
\propto \eta_n$.

\subsection{Hierarchical Randomization Design}

\label{subsec:hierarch_design}

We investigate experimental designs $\pi_0$ that allow for dependence in individual assignments among neighboring units on the graph. For illustrative purposes, we focus on a simple scheme where we first partition $\mathcal{N}$ into components (\emph{clusters}) $\mathcal{N}_1,\dots,\mathcal{N}_K$ where the number of such clusters increases according to $K=K_n$ as $n$ grows.

We consider star partitions of network radius $r_H$, where the $k$th cluster contains a reference node (\emph{centroid}) $i(k)$ and the network neighborhood of radius $r_H$ of that node, i.e. $\mathcal{N}_{r_H}(i(k))\subset \mathcal{N}_k$. All nodes that are at a network distance greater than $r_H$ from any of the centroids $i(1),\dots,i(K)$ are assigned to an arbitrarily chosen cluster among the partition elements closest to that node.

We choose that set of centroids as a \emph{maximal independent set} at radius $r_H$, that is, a set of nodes $\mathcal{I}_{r_H}\subset\mathcal{N}$ that is (a) \emph{independent} in the sense that neighborhoods at radius $r_H$ don't overlap, $\mathcal{N}_{r_H}(i(k))\cap \mathcal{N}_{r_H}(i(l))=\emptyset$ whenever $k\neq l$, and (b) that is \emph{maximal} in the sense that there exists no other independent set $\tilde{\mathcal{I}}_{r_H}$ such that $|\tilde{\mathcal{I}}_{r_H}|>|\mathcal{I}_{r_H}|$. By construction, the resulting collection of clusters is a partition of $\mathcal{N}$.

Note furthermore that there is no index $j\in\mathcal{N}_k$ such that $d(j,i(k))>2r_H$ because otherwise $j$ would either not have been assigned to the closest cluster, or $\mathcal{I}_{r_H}\cup\{j\}$ would also have been an independent set, contradicting the premise that $\mathcal{I}_{r_H}$ was maximal. For the spatial grid network with edges $\left(L_{ij}^{GRGM}\right)_{ij}$, up to integer constraints such a partition can be constructed by choosing centroids on a $d$-dimensional evenly spaced grid at distance $2r_H\delta_n$.

The experimental assignment $\pi_0$ is then determined as follows:

\begin{ass}\label{hierarch_des_ass}\textbf{(Hierarchical Randomization Design)}
In the first stage we draw assignment probabilities $\tilde{p}_1,\dots,\tilde{p}_K$ independently and uniformly at random from a two-point distribution with $\mathbb{P}_{\pi_0}(\tilde{p}_k = \bar{p}_0)=\mathbb{P}_{\pi_0}(\tilde{p}_k = \bar{p}_1)=\frac12$ given values $\bar{p}_0,\bar{p}_1\in[0,1]$.  In the second stage, treatment indicators $D_i$ are drawn conditionally independently at random according to
\[D_i|\tilde{p}_1,\dots,\tilde{p}_K \sim\textnormal{Bernoulli}(\tilde{p}_{k(i)})\]
where $k(i)$ is the index of the cluster containing $i$, i.e. $i\in\mathcal{N}_{k(i)}$.
\end{ass}

\subsection{Exposure Measures}

To show how those rates depend on the support of the exposure measures, we focus on counterfactuals that vary the proportion of treated units in a sphere of radius $r_T=0,1,\dots$ around a reference unit. That is, we define the sphere at radius $r_T$ around node $i$ as $\mathcal{N}_{r_T}^{\circ}(i):=\{j\in\mathcal{N}:d(i,j)=r_T\}$ and then let
\[T_{i,r_T}:=\frac{1}{|\mathcal{N}_{r_T}^{\circ}(i)|} \sum_{j\in\mathcal{N}_{r_T}^{\circ}(i)}D_j\]

\subsection{Construction of Counterfactual}

We define the contingent average exposure effect with respect to $T_{i,r_T}$ in terms of a counterfactual assignment where all units $j\in\mathcal{N}_{r_T}^{\circ}(i)$ receive treatment according to $D_j^*|\mathbf{D}\stackrel{iid}{\sim}\textnormal{Bernoulli}(p^*)$, so that
\[\pi_T^*(\mathbf{d}_{\mathcal{N}_{r_T}^{\circ}(i)}^*|\mathbf{D}_{-\mathcal{N}_{r_T}^{\circ}(i)};p^*):=
(p^*)^{\sum_{j\in\mathcal{N}_{r_T}^{\circ}(i)}d_j^*}
(1-p^*)^{\sum_{j\in\mathcal{N}_{r_T}^{\circ}(i)}(1-d_j)}\]
This corresponds to a binomial design for the exposure measure $T_{i,r_T}$ with $|\mathcal{N}_{r_T}^{\circ}(i)|$ trials and success probability $p^*$.

On the other hand, the distribution of $T_{i,r_T}$ resulting from the experimental design is a mixture over Poisson binomial sequences,
\[\pi_0(\mathbf{d}_{\mathcal{N}_{r_T}^{\circ}(i)}|\mathbf{D}_{-\mathcal{N}_{r_T}^{\circ}(i)}) = \mathbb{E}\left[\prod_k\prod_{j\in\mathcal{N}_{r_T}^{\circ}(i)\cap\mathcal{N}_k}p_{k}^{d_j}(1-p_{k})^{1-d_j}\right] \]
Hence, the likelihood ratio for allocation $\mathbf{D}_{\mathcal{N}_{r_T}^{\circ}(i)}^*$ between the two mechanisms is
\begin{eqnarray}
\nonumber r_{ip^*}(\mathbf{D}_{\mathcal{N}_{r_T}^{\circ}(i)})&=&\frac{\pi_T^*(\mathbf{D}_{\mathcal{N}_{r_T}^{\circ}(i)}^*|\mathbf{D}_{-\mathcal{N}_{r_T}^{\circ}(i)};p^*)}
{\pi_0(\mathbf{D}_{\mathcal{N}_{r_T}^{\circ}(i)}|\mathbf{D}_{-\mathcal{N}_{r_T}^{\circ}(i)})}\\
\nonumber&=&\mathbb{E}\left[\prod_k\prod_{j\in\mathcal{N}_{r_T}^{\circ}(i)\cap\mathcal{N}_k}\left(\frac{p_{k}}{p^*}\right)^{d_j}
\left(\frac{1-p_{k}}{1-p^*}\right)^{1-d_j}\right]^{-1}
\end{eqnarray}
The behavior of $A_{Tn}$ therefore depends not only on the size of $\mathcal{N}_{r_T}^{\circ}(i)$ but also on the distribution of sizes $|\mathcal{N}_{r_T}^{\circ}(i)\cap\mathcal{N}_k|$ of elements of the randomization partition of $\mathcal{N}_{r_T}^{\circ}(i)$.

\subsection{Distribution of the Exposure $T_{i,r_T}$}

To describe the asymptotic behavior of the likelihood ratio $r_{ip^*}(\mathbf{D}_{\mathcal{N}_{r_T}^{\circ}(i)})$, we first do the calculations for fixed $i$ and $r_T$ so that we omit those subscripts. Denote $N_{i,r_T}:=|\mathcal{N}_{r_T}^{\circ}(i)|$, where we also take $N$ to be $N_{i,r_T}$ as long as this does not cause any ambiguities. In general $N\equiv N_n$ may grow along the asymptotic sequence $n\rightarrow\infty$ at a rate depending on the exact structure of the spillover network.

If the network expands according to a sparse sequence so that network neighborhoods at any fixed radius remain bounded in size, the sequences $A_{Tn},B_{Tn}^s$ are of order $O(1)$. The cases of dense and semi-sparse network asymptotics with $|\mathcal{N}_{r_T}^{\circ}(i)|\rightarrow\infty$ are more complex and will therefore be the focus of the remainder of this analysis. For the remainder of this appendix, we therefore focus on dense designs with $|\mathcal{N}_r^{\circ}|\rightarrow\infty$.

Conditional on the probabilities $\tilde{p}_1,\dots,\tilde{p}_n$, the experimental assignment generates the exposure measure $T_{i,r_T}$ according to the conditional Poisson binomial experiment with p.d.f.
\[f_{T_r}(t|\tilde{p}_1,\dots,\tilde{p}_n;i):=\sum_{\mathbf{d}:\bar{d}_{i,r_T} = t}\prod_{j\in\mathcal{N}_{r_T}^{\circ}(i)}\tilde{p}_j^{d_j}(1-\tilde{p}_j)^{1-d_j} \]
Since $\tilde{p}_j$ takes only two values, $\bar{p}_0,\bar{p}_1$, this expression simplifies to
\[f_{T_{r_T}}(t|\tilde{p}_1,\dots,\tilde{p}_n;i):=\sum_{k_0=0}^{Nt}\binom{N_0}{k_0}\binom{N_1}{k-k_0} \bar{p}_0^{k_0}(1-\bar{p}_0)^{N_0-k_0}\bar{p}_1^{k-k_0}(1-\bar{p}_1)^{N_1-k+k_0}\]
assuming that $Nt$ is integer, where $N_0\equiv N_{0i}$ denotes the number of units $j\in\mathcal{N}_{r_T}^{\circ}(i)$ with $\tilde{p}_j=\bar{p}_0$ and $N_1$ denotes the number of units $j\in\mathcal{N}_{r_T}^{\circ}(i)$ with $\tilde{p}_j=\bar{p}_1$, so that in particular $N=N_0+N_1$.

Conversely, under the counterfactual assignment $\pi_T^*$, the unconditional distribution of $T_{i,r}$ is given by
\[f_{T_{r_T}}(t|p^*,\dots,p^*;i):=\binom{N}{Nt}\left((p^*)^t(1-p^*)^{1-t}\right)^{Nt} \]

If $N_{i,r_T}:=|\mathcal{N}_{r_T}^{\circ}(i)|\rightarrow\infty$, we can approximate the distribution of $T_{i,r_T}$ with the Gaussian distribution according to
\[\left.\sqrt{N_{i,r_T}}\frac{T_{i,r_T}-\mu_{ir}^*}{\sigma_{i,r_T}^*}\right|p_1,\dots,p_n\stackrel{d}{\rightarrow}N\left(0,1\right)\]
where
\begin{eqnarray}
\nonumber \mu_{ir_T}^*&:=&\sum_{j\in\mathcal{N}_{r_T}^{\circ}(i)}p_j\\
\nonumber \sigma_{i,r_T}^*&:=&\frac1N\sum_{j\in\mathcal{N}_{r_T}^{\circ}(i)}p_j(1-p_j)
\end{eqnarray}
Similarly, under the counterfactual assignment $\pi_T^*$,
\[\sqrt{N_{i,r_T}}(T_{i,r_T}^*-p^*)\stackrel{d}{\rightarrow}N\left(0,p^*(1-p^*)\right)\]

\subsection{Rate for $B_{Tn}^s$}

From the previous calculations, for large $N$ the counterfactual distribution of $\hat{p}_n$ is asymptotically normal and concentrated in a neighborhood of $\pi^*$, and the experimental distribution $\pi_0$ behaves like a mixture of normals governed by the cross-sectional distribution of $\alpha_{ni}:=\frac{N_{0i}}{N_{i,r_T}}$. Note again that for a given choice of exposure measures, the researcher controls the distribution of $\alpha_{ni}$ via the choice of the randomization design and is therefore understood to be known throughout.

Denoting $a^*:=\frac{p^*-p_1}{p_2-p_1}$, for a fixed mixture probability $\alpha= a^*$, the conditional likelihood ratio given $\alpha_{ni}=a^*$ is bounded from above and away from zero almost surely, and with bounds independent of $n$, so that its variance remains bounded. If on the other hand $\alpha\neq a^*$, we can similarly verify that the likelihood ratio between the corresponding sums of binomials diverges exponentially in $N$. We therefore need to distinguish three cases depending on whether the limit of the cross-sectional distribution of $\alpha_{ni}$ (a) has a point mass at $a^*$, or (b) is continuous and has strictly positive density at $a^*$, or whether (c) $a^*$ is outside of the limit of the support of $\alpha_{ni}$.

For case (b), if the asymptotic distribution of $\alpha_{ni}$ is continuous with strictly positive density $f_A(a^*)$ at $a^*$, then the experimental distribution of $\hat{p}_n$ also has strictly positive density at $p^*$, whereas the target distribution is asymptotically Gaussian and concentrated in a neighborhood of $\pi^*$.

We will therefore first consider the likelihood ratio between a target distribution $N(0,\sigma^2)$ for small values of $\sigma$, and an experimental distribution $Z$ with continuous density $f_Z(z)$ at $z=0$. Defining the random variable
\[Y:=\frac1{\sqrt{2\pi}\sigma f_Z(Z)}\exp\left\{-\frac{Z^2}{2\sigma^2}\right\}\]
Taking expectations with respect to $Z$, we then have that for $\sigma$ sufficiently small, we have the approximations
\[\mathbb{E}[Y]\approx 1\]
and
\[\var(Y) \approx \frac{1}{2f_Z(0)\sqrt{\pi}\sigma}-1\]
For any fixed $s>1$, we can modify these calculations to conclude that under the same conditions, $\var(Y^s)\approx O(\sigma^{2s-1})$. Since the target distribution behaves according to $\sqrt{N}(\hat{p}_n-p^*)\approx N(0,p^*(1-p^*))$, so that we evaluate these moments along the sequence $\sigma_n = \sqrt{\frac{p^*(1-p^*)}{N}}$. Hence, for case (b), Assumption \ref{ass:positivity_ass} holds with
\[B_{Tn}^s=\var(Y^s) = O\left(N^{s-\frac12}\right)\]
On the other hand, under (a), $B_{Tn}^s = O(1)$, whereas under (c) $B_{Tn}^s = O(\exp\{N\})$ for any $s$.

We therefore need to determine which of these three regimes describes the large-sample behavior of the distribution of $\alpha_{ni}$ for the hierarchical design described in Assumption \ref{hierarch_des_ass}. In particular, we need to determine whether the distribution of $\alpha_{ni}$ from a given design includes $a^*$ in its support.

\subsubsection{Geometric Random Graph Model}

We can write $W_k = \dum\{\tilde{p}_k=p_1\}$ $n_{ki}:=|\mathcal{N}_{r_T}^{\circ}(i)\cap\mathcal{N}_k|$, and
\begin{equation}\label{alpha_n_formula}\alpha_{ni}= \frac{\sum_{k=1}^Kn_{ki}W_k}{\sum_{k=1}^Kn_{ki}}.\end{equation}
for any network model.

For the Geometric Random Graph Model in Assumption \ref{grgm_ass}, $N$ generally grows at the rate of the expected degree, $n\eta_n$. Consider first the case in which the randomization groups are at least as large as the support of the exposure measure, $r_H\geq r_T$. In that case, $\mathcal{N}_{r_T}^{\circ}(i)$ intersects with at most $2^d$ randomization clusters $\mathcal{N}_k$, corresponding to nodes with positions in the adjacent hypercubes of length $2r_H\delta_n$ in $[-1,1]^d$. On the other hand, the nodes in $\mathcal{N}_{r_T}^{\circ}(i)$ correspond to the set $\mathcal{B}_{r_T}^{\circ}(i)\subset[-1,1]$ given by the complement of the hypercube of length $2(r_T-1)\delta_n$ centered at $Z_i$ with respect to the hypercube of side length $2r_T\delta_n $ centered at the same point. Depending on the position of node $i$ relative to that of the centroid $k(i)$ of the hypercube containing $i$, the share of $\mathcal{B}_{r_T}^{\circ}(i)$ that overlaps with any one of those adjacent hypercubes can take any value on the unit interval. Since those fractions do not change with $n$ and the node positions are evenly spaced on $[-1,1]^d$, the distribution of $\frac{n_{ki}}{\sum_{k=1}^Kn_{ki}}$ therefore converges to a continuous limit with full support on $[0,1]$.

For the case $r_T>r_H$, the previous argument changes only insofar as that $\mathcal{N}_{r_T}^{\circ}(i)$ may now intersect with up to $\left(\lceil\frac{r_T}{r_H}\rceil + 1\right)^d$ randomization clusters, and that the volume in attribute space corresponding to the overlap $\mathcal{N}_{r_T}^{\circ}(i)\cap\mathcal{N}_k$ may correspond to a fraction of at most $\left(\frac{r_H}{r_T}\right)^d$ of the volume corresponding to $\mathcal{N}_{r_T}^{\circ}(i)\cap\mathcal{N}_k$. In this case, the distribution of $\frac{n_{ki}}{\sum_{k=1}^Kn_{ki}}$ therefore converges to a continuous limit whose support includes the interval $\left[0,\left(\frac{r_H}{r_T}\right)^d\right]$.

Since for any fixed $r_T,r_H$, the number $\left(\lceil\frac{r_T}{r_H}\rceil + 1\right)^d$ of randomization clusters with $n_{ki}\geq 1$ is finite, $B_{Tn}^s$ for the GRGM therefore always behaves according to case (a) if $p^*\in\{\bar{p}_0,\bar{p}_1\}$ and according to case (b) if $\bar{p}_0<p^*<\bar{p}_1$.

\subsubsection{Erd\H{o}s-R\'enyi Random Graph Model}
For the Erd\H{o}s-R\'enyi Random Graph Model in Assumption \ref{errgm_ass} with $N$ growing at the same rate as the expected degree $n\eta_n$, network neighborhoods of large radii typically saturate the entire node set $\mathcal{N}$. We therefore restrict ourselves to the case in which $r_H + r_T<\log(n\eta_n)$, and we also maintain that $n\eta_n\rightarrow\infty$.

If the network distance between $i$ and the closest centroid $k(i)$ of a randomization cluster is given by $d_{ik(i)}$, then by construction $\mathcal{N}_{k(i)}$ includes the network neighborhood $\mathcal{N}_{r_H-d_{ik(i)}}(i)$ of node $i$. Hence, if $r_H\geq r_T + d_{ik(i)}$, the number of nodes in the intersection $\mathcal{N}_{r_T}^{\circ}(i)\cap\mathcal{N}_{k(i)}$ is equal to $|\mathcal{N}_{r_T}^{\circ}(i)|$. If on the other hand $r_T>r_H-d_{ik(i)}$, then with probability approaching 1, no node other than $k(i)$ is included in the intersection $\mathcal{N}_{r_T}^{\circ}(i)\cap\mathcal{N}_{k(i)}$. Hence as $n$ grows large, $n_{ki}\in\{0,1,|\mathcal{N}_{r_T}^{\circ}(i)|\}$ with probability approaching one.

Furthermore, if $r_H\geq r_T$, $\mathcal{N}_{r_T}^{\circ}(i)\subset\mathcal{N}_{k(i)}$ for the closest centroid $k(i)$ if and only if $d_{ik(i)}\leq r_H-r_T$, which will be the case for a fraction $|\mathcal{N}_{r_H-r_T}(k(i))|/|\mathcal{N}_{r_H}(k(i))|$ of nodes $j\in\mathcal{N}_{r_H}(k(i))$. For the ERRGM with average degree $n\eta_n$ and $r_H,r_T$ small enough relative to $n$, that fraction will be of the order $(n\eta_n)^{-r_T}$, which converges to zero under our assumptions for any $r_T\geq1$. If on the other hand $r_T>r_H$, $n_{ki}\in\{0,1\}$ for all $i,k$ with probability approaching 1.

Since in the formula (\ref{alpha_n_formula}), we therefore have two scenarios regarding the weights $\frac{n_{ki}}{\sum_{j}n_{ji}}$: we could either have one cluster such that $n_{ki} =|\mathcal{N}_{r_T}^{\circ}(i)|$ so that one of the weights equals one and all other are zero, or $n_{ki}=0$ for most clusters but $n_{ki}=1$ for a number of clusters growing at the rate $n\eta_n$. Since $W_k$ are i.i.d. Bernoulli with success probability $\frac12$, the support of the cross sectional distribution of $\alpha_{ni}$ therefore converges to a discrete limit with support $\{0,\frac12,1\}$. Furthermore, by our previous argument, if $r_H\geq r_T$ the relative frequency of $\alpha_{ni}$ in the neighborhood of $\frac12$ converges to zero except in the trivial case $r_T=0$. On the other hand if $r_T>r_H$ and therefore $n_{ki}\in\{0,1\}$ for all $i,k$ with probability approaching 1, we have that by the law of large numbers the distribution of $\alpha_{ni}$ concentrates in the neighborhood of $\frac12$. For the ERRGM with increasing average degree the behavior of $B_{Tn}^s$ therefore corresponds to cases (a) or (c), but never (b).

We can then summarize these derivations as follows:

\begin{prp}\label{positivity_rate_prp}
Suppose that $r_T>0$ and $r_H$ are fixed, and that $\pi_0$ follows the hierarchical design in Assumption \ref{hierarch_des_ass} with $\bar{p}_0\leq p^*\leq\bar{p}_1$. Then under the Geometric Random Graph Model (GRGM) in Assumption \ref{grgm_ass}, we have for any $s>0$ that
\[B_{Tn}^s = O\left((n\eta_n)^{2s-1}\right)\]
if $\bar{p}_0<p^*<\bar{p}_1$ and $B_{Tn}^s = O(1)$ if $p^*\in\{\bar{p}_0,\bar{p}_1\}$. Under the Erd\H{o}s-R\'enyi Random Graph Model (ERRGM) in Assumption \ref{errgm_ass}, we have for any $s$ that $B_{Tn}^s = O(1)$ if $p^*\in\{\bar{p}_0,\bar{p}_1\}$ and $r_H\geq r_T$; $B_{Tn}^s = O(1)$ if $p^*=\frac12(\bar{p}_0+\bar{p}_1)$ if $r_H<r_T$, and $B_{Tn}^s = O(\exp\{n\eta_n\})$ otherwise.
\end{prp}
While the stated rate for the GRGM does not depend on the relative magnitudes of $r_T,r_H$, it can be seen from the previous calculations that the constants in that approximation deteriorate significantly for increasing values of $\frac{r_T}{r_H}$ or $d$, so that estimation of spillover effects at network distances far exceeding the size of randomization clusters will also be unreliable.

\subsection{Rate for $A_{Tn}$}

The rate $A_{Tn}$ gives an upper bound on the number of nodes in randomization neighborhoods $\mathcal{N}_k$ that intersect with the support of the exposure measure, $\mathcal{N}_{r_T}^{\circ}(i)$. In the geometric random graph model (GRGM), two nodes are connected if and only if they are less than a distance $\delta_n$ apart in the latent attribute space. Therefore the network topology directly translates into the geometry of the underlying attribute space.

The set $\mathcal{N}_{r_T}^{\circ}(i)$ consists of the nodes that are exactly at network distance $r_T$ from the reference unit $i$. For the GRGM, there always exists a network path of length less than or equal to $r_T$ between nodes $i$ and $j$ if and only if $\|Z_i-Z_j\|<r_T\delta_n$. Similarly, there exists a path of length strictly less than $r_T$ between the two nodes if and only if $\|Z_i-Z_j\|<(r_T-1)\delta_n$, so that
\[\mathcal{N}_{r_T}^{\circ}(i):=\left\{j\in\mathcal{N}: (r_T-1)\delta_n\leq \|Z_i-Z_j\|<r_T\delta_n\right\}\]
The dependency neighborhood of $\mathcal{N}_{r_T}^{\circ}(i)$ is the union $\mathcal{A}_{T}(i):=\bigcup_{k:\mathcal{N}_{r_T}^{\circ}(i)\cap\mathcal{N}_k\neq\emptyset}\mathcal{N}_k$. Under the GRGM, the corresponding set in latent attribute space is therefore contained in the shell of thickness $\delta_n(1+2r_H):=r_W\delta_n$ of a hypercube of side length $2(r_H+r_T+1)\delta_n\equiv r_L\delta_n$.

The volume of that hypercube shell corresponds to a fraction less than or equal to $dr_L^{d-1}r_W2^{-d}$ of the hypercube $[-1,1]^d$ and therefore behaves like $O(\delta_n^d)$ with constants depending on $r_T,r_H$ as $n$ increases. Therefore the upper bound on the rate $A_{Tn}$ in the GRGM behaves according to
\[A_{Tn} = O(n\eta_n)\]
Hence the rate is determined by the sparsity sequence $\eta_n$ and the dimension $d$ of the latent attribute space. The respective radii $r_T,r_H$ of the exposure measure and the randomization clusters only affect $A_{Tn}$ through a multiplicative constant that is not explicitly given here.

For the Erd\H{o}s-R\'enyi Random Graph model with expected degree $n\eta_n$, network neighborhoods of large radii typically saturate the entire node set $\mathcal{N}$, so we restrict ourselves to the case in which $r_H + r_T<\log(n\eta_n)$. In that case we have that $|\mathcal{N}_{r_T}^{\circ}(i)|$ grows at a rate $(n\eta_n)^{r_T}$, so that the dependency neighborhood grows at a rate no faster than \[|\mathcal{A}_{T}(i)|\leq A_{Tn}= O\left((n\eta_n)^{r_T+r_H}\right)\]
Hence the growth rate depends critically on the sparsity sequence $\eta_n$ and the respective radii of the exposure measure and the randomization clusters.

We can summarize the results of these calculations as follows:

\begin{prp}\label{dependency_rate_prp}
Suppose that $\pi_0$ follows the hierarchical design in Assumption \ref{hierarch_des_ass}. Then under the Geometric Random Graph Model (GRGM) in Assumption \ref{grgm_ass},
\[A_{Tn} = O(n\eta_n)\]
Under the Erd\H{o}s-R\'enyi Random Graph Model (ERRGM) in Assumption \ref{errgm_ass},
\[A_{Tn}= O\left((n\eta_n)^{r_T+r_H}\right)\]
\end{prp}

\subsection{Asymptotic Rates for Estimation}

We can now combine the results from these calculations to give rates of consistency, where the convergence rate in Theorem \ref{normality_thm} is defined by $n^{2\varrho} = \frac{A_{Tn}B_{Tn}^1}{n}$. The estimator is consistent under Theorem \ref{consistency_thm} if this rate condition holds for some $\varrho>0$. If furthermore $\frac{n}{B_{Tn}^2A_{Tn}}=O(n^{2\varrho})$, then the conclusions of Theorem \ref{normality_thm} hold.

\begin{prp}\label{app_est_rate_prp} Suppose that the assumptions of Theorem \ref{consistency_thm} hold. Furthermore, $r_T>0$ and $r_H$ are fixed, and that $\pi_0$ follows the hierarchical design in Assumption \ref{hierarch_des_ass} with $\bar{p}_0\leq p^*\leq\bar{p}_1$. (a) Then for the GRGM in Assumption \ref{grgm_ass}, the estimator (\ref{coa_ipw_estimator}) for the average exposure effect is consistent if $n^{\frac12}\eta_n\rightarrow0$. If the stronger condition $n^{\frac34}\eta_n\rightarrow0$ holds, then the estimator is asymptotically normal at the rate $n^{\varrho}:=n^{\frac32}\eta_n^2$. (b) For the ERRGM in Assumption \ref{errgm_ass}, if $r_H\geq r_T$ and $p^*\in\{\bar{p}_0,\bar{p}_1\}$,  the estimator (\ref{coa_ipw_estimator}) for the average exposure effect is consistent and asymptotically normal if $n^{\frac{r_T+r_H-1}{r_T+r_H}}\eta_n\rightarrow0$.
\end{prp}

These claims follow immediately from Theorems \ref{consistency_thm} and \ref{normality_thm} together with the rates in Propositions \ref{positivity_rate_prp} and \ref{dependency_rate_prp}. All cases require that the link density $\eta_n$ decreases at some rate as $n$ grows.

For dense sequences of Erd\H{o}s-R\'enyi random graphs, we can therefore only consistently estimate counterfactuals corresponding to exposures equal to values in the support of the cluster-specific assignment probabilities $p_0^*,p_1^*$, so a hierarchical design would have to account for the target values of that measure. In particular, if we are interested in the treatment contrast between two Bernoulli designs with marginal treatment probabilities $p_0^*,p_1^*$, the rate in (b) can only be achieved if $\{\bar{p}_0,\bar{p}_1\}=\{p_0,p_1\}$, whereas the rate for the GRGM in (a) applies for any $p_0^*,p_1^*$ in the convex hull of $\{\bar{p}_0,\bar{p}_1\}$. If the researcher is interested in designing an experiment that simultaneously estimates exposure effects at various spillover radii $r_{T1},\dots,r_{TS}$, then the above rates apply simultaneously subject to the constraint $r_H\geq r_{T1},\dots,r_{TS}$. That is, the randomization partition has to be coarse enough to include network neighborhoods of radius greater or larger than the longest spillover distance under consideration.

\section{Primitive Conditions for Assumptions \ref{ass:positivity_ass}-\ref{ass:bd_influence_ass}}
\label{sec:asy_ex_app}

In this appendix, we discuss the high-level conditions in Assumptions \ref{ass:positivity_ass}-\ref{ass:bd_influence_ass}.

\subsection{Conditions for Assumption \ref{ass:positivity_ass}}

We first consider Assumption \ref{ass:positivity_ass}, where we consider two different exposure mappings, (a) the unit-specific assignment, $T_1(\mathbf{D};i) = D_i$, and (b) the fraction among the direct neighbors of $i$ receiving treatment, $T_2(\mathbf{D};i)=\frac{\sum_{j\neq i}L_{ij}D_j}{\sum_{j\neq i} L_{ij}}$.

\subsubsection{Bernoulli Design}\label{sec:bernoulli_des} Consider an experiment in which a binary unit-specific treatment $D_i\in\{0,1\}$ was assigned independently at random with $\mathbb{P}(D_i=1) = p_0$, i.e. $D_i\stackrel{iid}{\sim}\textnormal{Bernoulli}(p_0)$, so that $\pi_0(\mathbf{D}) = p_0^{\sum_i D_i}(1-p_0)^{n-\sum_iD_i}$. Suppose also that we are interested in a counterfactual policy under which $\tilde{D}_i\stackrel{iid}{\sim}\textnormal{Bernoulli}(p_1)$, with $p_1\geq p_0$ i.e. the random assignment $\pi_1(\mathbf{D}) = p_1^{\sum_i D_i}(1-p_1)^{n-\sum_iD_i}$.

For the direct effect based on exposure $T_1(\mathbf{D};i) = D_i$, we can see that $\var(r_{it})\leq p_0^{-2}$ and Assumption \ref{ass:positivity_ass} holds with $B_{Tn}$ bounded by a finite constant. So in particular, the weights $r_{it}$ in the IPW estimator for the counterfactual $\mathbb{E}_{\pi_1}[V(T_1(\mathbf{D};i))]$ satisfy Assumption \ref{ass:positivity_ass} with a constant upper bound.

If we are instead interested in the average effect of shifting the distribution of $T_2(\mathbf{D};i)$ from $\pi_0$ to that generated by $\pi_1$, the likelihood ratio weights for the estimator for $\mathbb{E}_{\pi_1}[V(T_2(\mathbf{D};i))]$ is $r_{it}=\left(\frac{p_1}{p_0}\right)^{\sum_{j\neq i}L_{ji}D_j}\left(\frac{1-p_1}{1-p_0}\right)^{\sum_{j\neq i}L_{ji}(1-D_j)}$. If $0<p_0<1$ and the network degree $N_i:=\sum_{j\neq i}L_{ji}$ is bounded by $\bar{N}$ along the asymptotic sequence, then $\var(r_{it})\leq\left(p_1/p_0\right)^{2N_i}\leq \left(p_1/p_0\right)^{2\bar{N}}<\infty$ is bounded uniformly across $i$ and $n$.

If we assume instead that $N_i<\bar{N}_n$ for a sequence $\bar{N}_n\rightarrow\infty$, then $\var(r_{it})\leq\left(p_1/p_0\right)^{2N_i}\leq \left(p_1/p_0\right)^{2\bar{N}_n}=:B_{Tn}$ with a bound that grows to infinity as long as $p_1> p_0$. The rate $\bar{N}_n$ depends on the asymptotic sequence of networks $\mathbf{L}=\mathbf{L}_n$, where we refer to the network sequence as dense if $\bar{N}_n/n$ does not vanish as $n$ grows large, and sparse otherwise.

Intuitively, the cross-sectional distribution of $T_2(\mathbf{D};i)$ under the experimental assignment $\pi_0$ concentrates near $p_0$ for diverging degree sequences by a law of large numbers, whereas the counterfactual distribution under $\pi_1$ concentrates near $p_1\neq p_0$, so that the experimental assignment asymptotically fails to generate exposures at the values most relevant to the counterfactual assignment.

\subsubsection{Two Stage Randomization Design}\label{sec:2stage_des} As the calculations for the previous example show,  there is in general insufficient variation in the social treatment under Bernoulli experimental designs in the case of dense network asymptotics. The IPW estimator for the derivative $\left.\frac{\partial}{\partial p_1}\mathbb{E}_{\pi_1}[V(T_1(\mathbf{D};i))]\right|_{p_1=p_0}$ under Bernoulli assignments was analyzed by \cite{LWa22} who also showed that an adjustment for network principal components could achieve consistency at slower sparse sequences, but not for dense network sequences. A slower rate $B_{Tn}$ in Assumption \ref{ass:positivity_ass} can be potentially achieved by dependent designs $\pi_0$, e.g. \cite{HHa08} suggested two-stage randomization designs to achieve greater experimental variation in exposure to identify indirect effects.

Consider an experiment using a two-stage randomization design where we vary the likelihood of treatment in the neighborhood around reference units but evaluate the effect on all units, including but not restricted to reference units. Specifically, we may consider an experiment that implements a two-stage randomization procedure that first selects $K$ reference units $i_1,\dots,i_K$ from $\mathcal{N}$, and a treatment probability $p_{i_k}\in[0,1]$ for the neighborhood $\mathcal{N}_T(i_k)$ at random according to a distribution $F(p)$, independently across $k=1,\dots,K$. Finally, unit-specific treatment $D_j=1$ is assigned at random with probability $\tilde{p}_j:=\frac{\sum_{k=1}^K\dum\{j\in\mathcal{N}_T(i_k)\}p_{i_k}}{\sum_{k=1}^K\dum\{j\in\mathcal{N}_T(i_k)\}}$, and conditionally independent given $\tilde{p}_1,\dots,\tilde{p}_n$.

Suppose again that the estimand is the average counterfactual corresponding to assigning exposures $T_2(\mathbf{D};i)=\frac{\sum_{j\neq i}L_{ij}D_j}{\sum_{j\neq i} L_{ij}}$ generated according to a counterfactual assignment $\pi_1$ where we assign $D_j\stackrel{iid}{\sim}\textnormal{Bernoulli}(p_1)$ to all units $j$ at network distance 1 from a randomly selected unit $i$. The cross-sectional distribution for the exposures $T_2(\mathbf{D};i)$ under the experimental assignment generally depends on the number of reference units $K$ and the fraction of units pertaining to intersections between neighborhoods $\mathcal{N}_T(i_k)$ for multiple reference units. However, for a given choice of the distribution $F(p)$, the variance of likelihood ratios $r_{it}$ need in general not diverge to infinity even if $\mathbb{E}_{\pi_0}[p_i]\neq p_1$ and $\bar{N}_n\rightarrow\infty$ at some rate.

\subsection{Examples for Assumption \ref{ass:design_dep_ass}}

We next revisit the previous examples to discuss Assumption \ref{ass:design_dep_ass}.

\subsubsection{Bernoulli Design} Consider again the Bernoulli design from Example \ref{sec:bernoulli_des}. If $D_1,\dots,D_n$ are assigned independently at random, then $\mathcal{A}_T(i)=\mathcal{N}_T(i)$, so that $A_{Tn}=\max_{i}|\mathcal{N}_T(i)|$.

\subsubsection{Two-Stage Randomization Design} Consider again the two-stage design from Example \ref{sec:2stage_des}. By construction, unit-specific treatment $D_j=1$ are conditionally independent given $\tilde{\pi}_1,\dots,\tilde{\pi}_n$, so that $\mathcal{A}_T(i)=\bigcup_{k:i\in\mathcal{N}_T(i_k)}\mathcal{N}_T(i_k)$. Then if, say, $T_i$ is determined on $\mathcal{N}_T(i):=\{j\in\mathcal{N}:L_{ij}=1\}$, we can bound \[A_{Tn}\leq\max_{i\in\mathcal{N}}\left|\mathcal{N}_T(i)\cup\left\{j\in\mathcal{N}\left|\exists k\in\mathcal{N}:L_{ik}L_{jk}=1\right.\right\}\right|,\] i.e. the size of the largest network neighborhood at a radius 2 around any node in $\mathcal{N}$.

\subsection{Examples for Assumption \ref{ass:bd_influence_ass}}

We finally turn to a discussion of primitive conditions for Assumption \ref{ass:bd_influence_ass}.

\subsubsection{Exogenous Exposure Model} Consider the exogenous exposure model $Y_i = h(D_i,\tilde{T}_i(\mathbf{D},\mathbf{L}))$, where $\tilde{T}_i(\mathbf{D},\mathbf{L})=\frac{\sum_{j\neq i}L_{ij}D_j}{\sum_{j\neq i}L_{ij}}$ and suppose that $h(d,t)$ is Lipschitz-continuous in $t$ with Lipschitz constant $0<H^*<\infty$. We also assume that the network neighborhood $\mathcal{N}_T(i):=\left\{j\in\mathcal{N}: L_{ij}=1\right\}$ has size $\varrho_n n$ for all nodes and some sequence $\varrho_n$ such that $n\varrho_n\geq\varrho>0$ for all $n$. For estimation of the direct effect, corresponding to the exposure measure $T_i(\mathbf{D},\mathbf{L}):=D_i$, $\mathcal{N}_T(i)=\{i\}$, so that $|\varphi_{ij}|\leq
H^*\dum\{L_{ij}=1\}/(n\varrho_n)$. Hence Assumption \ref{ass:bd_influence_ass} holds with $C_n = O\left(\frac{1}{n\varrho_n}\right)$ which converges to zero since $n\varrho_n$ was assumed to be bounded away from zero.

\subsubsection{Infinitesimal Shift} Consider again the model in Example \ref{ex:infinitesimal_ex}, where we consider a shift from $\mathbf{X}_0$ to $\mathbf{X}_1=\mathbf{X}_0+\delta\mathbf{D}$ where $\delta>0$ is small, and $D_1,\dots,D_n$ are assigned i.i.d. with $\mathbb{E}_{\pi_0}[D_i]=0$ and $\var_{\pi_0}(D_i)=1$. Then from the expression in (\ref{inf_eq_RF}), $\cov_{\pi_0}(Y_j(\mathbf{D}),D_i)/\delta = \mathbf{e}_j'(\mathbf{I}-\mathbf{H_Y})^{-1}\mathbf{H_X}\mathbf{e}_i+o(1)$. If  we let $|\lambda_{min}|$ denote the smallest absolute eigenvalue of $\mathbf{I}-\mathbf{H_Y}$, then
\[\left|\frac1{n^2}\sum_{i\neq j}\varphi_{ij}(t)\varphi_{ji}(t)\right| = \frac1{n^2}\left|\boldsymbol{\iota}_n'(\mathbf{I}-\mathbf{H_Y}')^{-1}
(\mathbf{I}-\mathbf{H_Y})^{-1}\boldsymbol{\iota}_n\right|\leq \frac1{n^2}\|\boldsymbol{\iota}_n\|^2|\lambda_{min}|^{-2}=\frac1n|\lambda_{min}|^{-2}
\]
Hence, for the problem of estimating direct effects, Assumption \ref{ass:bd_influence_ass} holds with $C_n=\frac1n|\lambda_{min}|^{-2}$ which converges to zero as long as $\sqrt{n}\lambda_{min}$ diverges to infinity.

\subsubsection{``Patient-Zero" Scenario} Consider the setting from Example \ref{ex:patient_zero} where we assume that the network consists of $g_n$ connected components of equal size, each of which is fully connected, and there are $s_n$ injection points in each connected component, i.e. units at which an infectious chain may start. We also assume that the vaccine is effective for each unit, $W_i=1$, and that $D_1,\dots,D_n$ are i.i.d. Bernoulli with success probability $p_0$. Then the treatment status of unit $i$ is correlated with the outcome of unit $j\neq i$ if $i$ is an injection point for the connected component containing $j$, and if $D_k=1$ for all other injection points $k$ in that same connected component. We can therefore bound
\[\left|\frac1{n^2}\sum_{i\neq j}\varphi_{ij}(t)\varphi_{ji}(t)\right|\leq \frac1{n^2}\frac{n}{g_n}s_np_0^{s_n-1}\]
where $s_np_0^{s_n-1}$ is bounded from above for any fixed assignment probability $p_0$. Hence, for the problem of estimating direct effects, Assumption \ref{ass:bd_influence_ass} holds with $C_n=\frac{1}{ng_n}$.

\section{Proofs for Results in Section \ref{sec:asy_prop}}

\flushleft\textsc{Proof of Theorem \ref{unbiased_thm}:} Since for each $i$, \[\mathbb{E}_{\pi_0}\left[\tilde{r}_{it}(\tilde{\mathbf{D}})|\mathbf{D}_{-\mathcal{N}_T(i)}\right]=\pi_{T}(\tilde{\mathbf{D}}_{\mathcal{N}_T(i)}|\mathbf{D}_{-\mathcal{N}_T(i)};t,i)\]
and $\mathbf{D}$ is independent of $\mathbf{Y}(\cdot)$, we have that $\mathbb{E}_{\pi_0}[u_i(t)]=0$ by the law of iterated expectations. Notice that under Assumptions \ref{ass:bd_outcome_ass}-\ref{ass:positivity_ass} these expectations are guaranteed to exist. Therefore, the claim follows immediately from (\ref{coa_est_error}) \qed

\flushleft\textsc{Proof of Theorem \ref{consistency_thm}:} We prove consistency by showing that $\var(\hat{V}_n(t)-V^*(t|\mathbf{Y}(\cdot),\mathbf{D}))\rightarrow0$. In what follows, we only consider the case in which $T(\mathbf{D};i)=t$ is supported by a unique configuration $\mathbf{d}_{\mathcal{N}_T(i)}$ of assignments on $\mathcal{N}_T(i)$. The general case follows from a completely parallel argument but is notationally more difficult to present.

We first bound unconditional variances $\var(u_{i}(t))$ and covariances $\cov(u_{i}(t),u_{j}(t))$ for units $i,j$. First consider the case $i=j$.
We can use the Cauchy-Schwarz inequality and Assumption \ref{ass:positivity_ass} to bound
\[\cov(u_i(t),u_i(t))=\var(Y_i\tilde{r}_{it}) \leq 4B_Y^2\var(\tilde{r}_{it})\leq 4B_Y^2B_{Tn}\]
Similarly for $i\neq j$ with $\mathcal{A}_T(i)\cap\mathcal{A}_T(j)\neq\emptyset$,
\[|\cov(u_i(t),u_j(t))|\leq4B_Y^2\sqrt{\var(\tilde{r}_{it})\var(\tilde{r}_{jt})}\leq 4B_Y^2B_{Tn}\]

Finally, consider units $i,j$ with $\mathcal{A}_T(i)\cap\mathcal{A}_T(j)=\emptyset$. For a pair of units $i,j$, we write $\mathcal{N}_T(i,j):=\mathcal{N}_T(i)\cup\mathcal{N}_T(j)$, and for any set of indices $\mathcal{B}\subset\mathcal{N}$ we let $\mathbf{D}_{\mathcal{B}}$ and $\mathbf{D}_{-\mathcal{B}}$ denote the vector of unit-specific assignments to units $k\in\mathcal{B}$, and $k\in\mathcal{B}^c$, respectively. Notice that by construction of $\tilde{r}_{it}$, $\mathbb{E}[\tilde{r}_{it}|\mathbf{D}_{-\mathcal{N}_T(i)}]=0$ a.s. Since $\mathcal{N}_T(i),\mathcal{N}_T(j)\subset\mathcal{N}_T(i,j)$, we also have $\mathbb{E}[\tilde{r}_{it}|\mathbf{D}_{-\mathcal{N}_T(i,j)}]=
\mathbb{E}[\tilde{r}_{jt}|\mathbf{D}_{-\mathcal{N}_T(i,j)}]=0$. Moreover, if $\mathcal{A}_T(i)\cap\mathcal{A}_T(j)=\emptyset$, $\tilde{r}_{it},\tilde{r}_{jt}$ are also independent conditional on $\mathbf{D}_{-\mathcal{N}_T(i,j)}$. It then follows by the ANOVA identity that  \[|\cov(u_i(t),u_j(t))|=\mathbb{E}\left[\cov(u_i(t),u_j(t)|\mathbf{D}_{-\mathcal{N}_T(i,j)})\right].\]

Writing $Y_i^*(t)=Y_i(\mathbf{d}_{\mathcal{N}_T(i)};\mathbf{D}_{-\mathcal{N}_T(i)})$, we therefore have that if $\mathcal{A}_T(i)\cap\mathcal{A}_T(j)=\emptyset$, then
\begin{eqnarray}
\nonumber |\cov(u_i(t),u_j(t))|&=&\left|\mathbb{E}\left[u_i(t)u_j(t)|\mathbf{D}_{-\mathcal{N}_T(i,j)}\right]\right|\\
\nonumber&\leq&
\left|\mathbb{E}\left[\left.\tilde{r}_{it}\tilde{r}_{jt}\left(Y_i^*(t)Y_j^*(t)
-\mathbb{E}\left[Y_i^*(t)Y_j^*(t)|\mathbf{Y}(\cdot),\mathbf{D}_{-\mathcal{N}_T(i,j)}\right]\right)\right|\mathbf{D}_{-\mathcal{N}_T(i,j)}\right]\right|\\
\nonumber&&+ \left|\mathbb{E}\left[\left.\tilde{r}_{it}\tilde{r}_{jt}\mathbb{E}\left[Y_i^*(t)Y_j^*(t)|\mathbf{Y}(\cdot),\mathbf{D}_{-\mathcal{N}_T(i,j)}\right] \right|\mathbf{D}_{-\mathcal{N}_T(i,j)}\right]\right|\\
\nonumber &=&\left|\mathbb{E}\left[\left.\tilde{r}_{it}\tilde{r}_{jt}\left(Y_i^*(t)Y_j^*(t)
-\mathbb{E}\left[Y_i^*(t)Y_j^*(t)|\mathbf{Y}(\cdot),\mathbf{D}_{-\mathcal{N}_T(i,j)}\right]\right)\right|
\mathbf{D}_{-\mathcal{N}_T(i,j)}\right]\right|\\
\nonumber&\leq&\sqrt{\var(\tilde{r}_{it})\var(\tilde{r}_{jt})}\mathbb{E}|\varphi_{ij}|\mathbb{E}|\varphi_{ji}|
\end{eqnarray}
where the first inequality uses the triangle inequality, the last equality uses the law of iterated expectations and the last step follows from the Cauchy-Schwarz inequality.

Hence the variance of the estimator in (\ref{coa_ipw_estimator}) can be bounded by
\begin{eqnarray}\nonumber \var\left(\hat{V}_n(t)-V^*(t)\right)&=&\frac1{n^2}\sum_{i=1}^n\var(Y_i\tilde{r}_{it})+\frac1{n^2}\sum_{i=1}^n\sum_{j\neq i}\cov(Y_i\tilde{r}_{it},Y_j\tilde{r}_{jt})\\
\nonumber &\leq&\frac{4B_Y^2}{n^2}B_{Tn}(1+nA_{Tn}) + B_{Tn}C_n^2
\end{eqnarray}
which converges to zero if $\frac{B_{Tn}A_{Tn}}{n}\rightarrow0$ and $B_{Tn}C_n^2\rightarrow0$. The claim of the theorem then follows from Chebyshev's inequality\qed

\flushleft\textsc{Proof of Theorem \ref{normality_thm}:}
Write $\mathbf{X}_i=(X_{it})_{t}$, where
\[X_{it}\equiv u_i(t):=\sum_{\tilde{\mathbf{d}}\in\mathcal{D}^{|\mathcal{N}_T(i)|}}
\left(\tilde{r}_{it}\left(\tilde{\mathbf{d}}\right)
-\pi_{T}(\tilde{\mathbf{d}}|\mathbf{D}_{-\mathcal{N}_T(i)};t,i)\right)
Y_i(\tilde{\mathbf{d}};\mathbf{D}_{-\mathcal{N}_T(i)}).\]
By the law of iterated expectations,
\begin{eqnarray}\nonumber\mathbb{E}\left[X_{it}|\mathbf{Y}(\cdot),\mathbf{D}_{-\mathcal{N}_T(i)}\right]&=&
\sum_{\tilde{\mathbf{d}}\in\mathcal{D}^{|\mathcal{N}_T(i)|}}\mathbb{E}\left[\left.\left(\tilde{r}_{it}\left(\tilde{\mathbf{d}}\right)
-\pi_{T}(\tilde{\mathbf{d}}|\mathbf{D}_{-\mathcal{N}_T(i)};t,i)\right)
Y_i(\tilde{\mathbf{d}};\mathbf{D}_{-\mathcal{N}_T(i)})\right|\mathbf{Y}(\cdot),\mathbf{D}_{-\mathcal{N}_T(i)}\right]\\
\nonumber&=&\sum_{\tilde{\mathbf{d}}\in\mathcal{D}^{|\mathcal{N}_T(i)|}}Y_i(\tilde{\mathbf{d}};\mathbf{D}_{-\mathcal{N}_T(i)})
\mathbb{E}\left[\left.\left(\tilde{r}_{it}\left(\tilde{\mathbf{d}}\right)
-\pi_{T}(\tilde{\mathbf{d}}|\mathbf{D}_{-\mathcal{N}_T(i)};t,i)\right)
\right|\mathbf{Y}(\cdot),\mathbf{D}_{-\mathcal{N}_T(i)}\right]\\
\nonumber&=&0,\end{eqnarray}
noting that given $\mathbf{Y}(\cdot),\mathbf{D}_{-\mathcal{N}_T(i)}$, $Y_i(\tilde{\mathbf{d}};\mathbf{D}_{-\mathcal{N}_T(i)})$ is nonstochastic. In particular, $\mathbb{E}[\mathbf{X}_{i}|\mathbf{Y}(\cdot)]=0$.

We also let $\Omega_n$ be as defined in the main text, and define $\tilde{\mathbf{X}}_n:=n^{\varrho}\Omega_n^{-1/2}\mathbf{X}_i$. We then set
\[\tilde{Z}_n:=n^{\varrho}\Omega_n^{-1/2}Z_n=\frac1n\sum_{i=1}^n\tilde{\mathbf{X}}_i.\]
The main claim of the theorem is that $\tilde{Z}_n\stackrel{d}{\rightarrow}Z\sim N(0,I)$.

\subsubsection{Stein Bound} We first address the case of a single exposure level $t\in\mathcal{T}$ in which $\tilde{Z}_n$ and $Z$ are scalar-valued, and $\Omega_n$ is therefore given by the scalar $\omega_n^2\equiv\omega_n^2(t,t)$. We prove convergence using an upper bound on the Wasserstein distance $d_W(Z,Z_n)$, where for two random variables $X,W$
\[d_W(X,W):=\sup_{h\in Lip(1)}\left|\mathbb{E}[h(X)]-\mathbb{E}[h(W)]\right|\]
and $Lip(1):=\left\{h:\mathbb{R}\rightarrow\mathbb{R}:|h(w_1)-h(w_2)|\leq |w_1-w_2|\right\}$ denotes the class of Lipschitz-continuous functions with Lipschitz constant less than or equal to $1$.

Theorem 3.1 in \cite{Ros11} then states that for $Z\sim N(0,1)$ and an arbitrary random variable $W$,
\[d_W(Z,W)\leq \sup_{h\in\mathcal{H}}\mathbb{E}\left|Wh(W)-h'(W)\right|\]
for the set of test functions
\[\mathcal{H}:=\left\{h:\mathbb{R}\rightarrow\mathbb{R}:\|h\|,\|h''\|\leq2,\|h'\|\leq \sqrt{2/\pi}\right\}\]
where $\|h\|$ denotes the sup-norm of a function $h$.

We now apply that result to the random sequence $\tilde{Z}_n$, where we show that the upper bound on the right-hand side converges to zero under the conditions of this theorem. In order to analyze that expectation, we define the sigma field $\mathcal{F}_{ni}:=\sigma\left(\mathbf{Y}(\cdot),\left(D_j\right)_{j\notin \mathcal{A}_T(i)}\right)$, where $\sigma(W)$ denotes the sigma-field generated by the random variable $W$. We then define
\[\tilde{Z}_{ni}:=\mathbb{E}\left[\tilde{Z}_n\left|\mathcal{F}_{ni}\right.\right]\]

We can therefore rewrite
\begin{eqnarray}
\nonumber\mathbb{E}[\tilde{Z}_nh(\tilde{Z}_n)-h'(\tilde{Z}_n)]&=&\mathbb{E}\left[\frac1n\sum_{i=1}^n\tilde{X}_ih(\tilde{Z}_n)-h'(\tilde{Z}_n)\right]\\
\nonumber&=&\mathbb{E}\left[\frac1n\sum_{i=1}^n\tilde{X}_i(h(\tilde{Z}_n)-h(\tilde{Z}_{ni}))-h'(\tilde{Z}_n)\right]
+\mathbb{E}\left[\frac1n\sum_{i=1}^n\tilde{X}_ih(\tilde{Z}_{ni})\right]\\
\nonumber&=&\mathbb{E}\left[\frac1{2n}\sum_{i=1}^nh''(\tilde{Z}_{ni}^*)\tilde{X}_i(\tilde{Z}_n-\tilde{Z}_{ni})^2\right]\\
\nonumber&&+\mathbb{E}\left[\left(\frac1n\sum_{i=1}^n\tilde{X}_i(\tilde{Z}_n-\tilde{Z}_{ni})-1\right)h'(\tilde{Z}_n)\right]\\
\nonumber&&+\mathbb{E}\left[\frac1n\sum_{i=1}^n\tilde{X}_ih(\tilde{Z}_{ni})\right]\\
\label{app_clt_stein_bound}&=:& \mathbb{E}\left[T_1\right] + \mathbb{E}\left[T_2\right] + \mathbb{E}\left[T_3\right]
\end{eqnarray}
where the second step uses a second-order mean-value expansion of $h(\tilde{Z}_n)-h(\tilde{Z}_{ni})=h'(\tilde{Z}_n)(\tilde{Z}_n-\tilde{Z}_{ni}) + \frac12h''(\tilde{Z}_{ni}^*)(\tilde{Z}_n-\tilde{Z}_{ni})^2$ with intermediate value $\tilde{Z}_{ni}^*$ between $\tilde{Z}_{ni}$ and $\tilde{Z}_n$.

\subsubsection{Bound on $|\mathbb{E} T_2|$}

For the second term, recall that $\omega_n^2$ was assumed to be bounded away from zero by some constant $\kappa>0$. We can use Jensen's inequality and that $\|h'\|\leq \sqrt{2/\pi}$ for all test functions in $\mathcal{H}$ to bound
\begin{eqnarray}
\nonumber \left(\mathbb{E}|T_2|\right)^2&\leq&\mathbb{E}\left[T_2^2\right]=
\mathbb{E}\left[\left(\frac{n^{2\varrho}}{n}\sum_{i=1}^nX_i(Z_n-Z_{ni})-\omega_n^2\right)\frac{h'(\tilde{Z}_n)}{\omega_n^2}\right]\\
\label{ET2_bound}&\leq& \frac{2}{\pi \kappa}\mathbb{E}\left[\left(\frac{n^{2\varrho}}n\sum_{i=1}^nX_i(Z_n-Z_{ni}) - \omega_n^2\right)^2\right]
\end{eqnarray}
where $Z_n:=\frac1n\sum_{i=1}^n\mathbf{X}_i$ and $Z_{ni}:=\mathbb{E}\left[Z_n|\mathcal{F}_{ni}\right]$. In order to show that $\mathbb{E}|T_2|$ is asymptotically negligible, it therefore suffices to show that the expectation on the right-hand side converges to zero.

We first establish that $\mathbb{E}\left[\frac{n^{2\varrho}}n\sum_{i=1}^nX_i(Z_n-Z_{ni})\right] - \omega_n^2\stackrel{p}{\rightarrow}0$. Since
$Z_n-Z_{ni} = \frac1n\sum_{j=1}^n\left(X_j - \mathbb{E}[X_j|\mathcal{F}_{ni}]\right)$, we have that
\begin{eqnarray}\nonumber X_i(Z_n-Z_{ni})&=&\frac1n\sum_{j=1}^nX_i \left(X_j - \mathbb{E}[X_j|\mathcal{F}_{ni}]\right)\\
\nonumber&=&\frac1n\sum_{j\in\mathcal{A}_T(i)}
X_i \left(X_j - \mathbb{E}[X_j|\mathcal{F}_{ni}]\right)+\frac1n\sum_{j\notin\mathcal{A}_T(i)}
X_i \left(X_j - \mathbb{E}[X_j|\mathcal{F}_{ni}]\right)\\
\nonumber&=&X_iU_{i1} + X_iU_{i2}
\end{eqnarray}
where $U_{i1}:=\frac1n\sum_{j:\mathcal{A}_T(i)\cap\mathcal{A}_T(j)\neq\emptyset}(X_j-\mathbb{E}[X_j|\mathcal{F}_{ni}])$ and $U_{i2}:=
\frac1n\sum_{j:\mathcal{A}_T(i)\cap\mathcal{A}_T(j)=\emptyset}(X_j-\mathbb{E}[X_j|\mathcal{F}_{ni}])$.

Since $\mathbb{E}\left[X_i|\mathcal{F}_{ni}\right]=0$, it follows that for any pair $i,j$ of indices,
\[\mathbb{E}\left[X_i \left(X_j - \mathbb{E}[X_j|\mathcal{F}_{ni}]\right)\right] = \cov(u_i(t),u_j(t)|\mathcal{F}_{ni})\]
so that
\[\frac{n^{2\varrho}}{n^2}\sum_{i=1}^n\mathbb{E}\left[\sum_{j\in\mathcal{A}_T(i)}X_i \left(X_j - \mathbb{E}[X_j|\mathcal{F}_{ni}]\right)\right]
=\omega_n^2\]

We then denote $\omega_{ij}\equiv\omega_{ij}(t,t)$, as defined in the main text, so that after substituting $Z_n=\frac1n\sum_{i=1}^nX_i$ and multiplying out the square,
\begin{eqnarray}
\nonumber\left(\frac{n^{2\varrho}}{n^2}\sum_{i=1}^n\left(X_i(U_{i1} + U_{i2}) - \omega_{ij}\right)\right)^2 &=&\frac{n^{4\varrho}}{n^4}\sum_{i=1}^n\sum_{k=1}^n\left(X_i(U_{i1} + U_{i2}) - \omega_{is}\right)(X_k(U_{k1} + U_{k2}) - \omega_{kt})\\
\nonumber&=&\frac{n^{4\varrho}}{n^2}\sum_{i=1}^n\sum_{k=1}^n\cov(X_i(U_{i1} + U_{i2}),X_k(U_{k1} + U_{k2}))
\end{eqnarray}
We can evaluate the pairwise covariances by summing over terms in the definitions of $U_{i1},U_{i2}$.

For the following calculations, note first that potential values \[\tilde{Y}_{i,jkl}:=
Y_i\left(\mathbf{d}_{\mathcal{A}_T(i)},\mathbf{d}_{\mathcal{A}_T(j)},\mathbf{d}_{\mathcal{A}_T(k)},\mathbf{d}_{\mathcal{A}_T(l)};
\mathbf{D}_{-(\mathcal{A}_T(i)\cup\mathcal{A}_T(j)\cup\mathcal{A}_T(k)\cup\mathcal{A}_T(l))}\right)\] are measurable under both $\mathcal{F}_{ni}$ and $\mathcal{F}_{nj}$ for any fixed $\mathbf{d}_{\mathcal{B}}\in\mathcal{D}^{|\mathcal{B}|}$ for $\mathcal{B}=\mathcal{A}_T(i),\mathcal{A}_T(j),\mathcal{A}_T(k),\mathcal{A}_T(l)$. Hence, products of the form
\[\mathbb{E}\left[\tilde{Y}_{i,jkl}\tilde{Y}_{j,ikl}\tilde{Y}_{k,ijl}\tilde{Y}_{l,ijk}\tilde{r}_i(t)\tilde{r}_j(t)\tilde{r}_k(t)\tilde{r}_l(t)\right]=0\]
unless each index in $i,j,k,l$ shares a dependency neighborhood with at least one other index.

We can therefore distinguish several cases depending on the overlap in dependency neighborhoods $\mathcal{A}_T(m)$ for $m=i,j,k,l$. We say that the index $i$ is \emph{free} among the tetrad $\{i,j,k,l\}$ if $\mathcal{A}_T(i)\cap\left(\mathcal{A}_T(j)\cup\mathcal{A}_T(k)\cup\mathcal{A}_T(l)\right)=\emptyset$. Also write $\tilde{Y}_i=Y_i\left(\mathbf{d}_{\mathcal{A}_T(i)};\mathbf{D}_{-\mathcal{A}_T(i)}\right)$ and
\[\tilde{Y}_{j,i}:=
Y_i\left(\mathbf{d}_{\mathcal{A}_T(i)},\mathbf{d}_{\mathcal{A}_T(j)};\mathbf{D}_{-(\mathcal{A}_T(i)\cup\mathcal{A}_T(j))}\right)\]
and $\Delta_i\tilde{Y}_j:=\tilde{Y}_i-\tilde{Y}_{j,i}$. Then if $i$ is free among $\{i,j,k,l\}$, $\mathbb{E}\left[\tilde{Y}_i\tilde{Y}_{j,i}\tilde{Y}_{k,i}\tilde{Y}_{l,i}\tilde{r}_i(t)\tilde{r}_j(t)\tilde{r}_k(t)\tilde{r}_l(t)\right]=0$ by the same reasoning as before, so that
\begin{eqnarray}
\nonumber\mathbb{E}\left[\tilde{Y}_i\tilde{Y}_j\tilde{Y}_k\tilde{Y}_l\tilde{r}_i(t)\tilde{r}_j(t)\tilde{r}_k(t)\tilde{r}_l(t)\right]
&=&\sum_{\{j_1,j_2,j_3\}=\{j,k,l\}}\left\{\frac{}{}
\mathbb{E}\left[\tilde{Y}_i\tilde{Y}_{j_1,i}\tilde{Y}_{j_2,i}\Delta_i\tilde{Y}_{j_3}\tilde{r}_i(t)\tilde{r}_j(t)\tilde{r}_k(t)\tilde{r}_l(t)\right]
\right.\\
\nonumber&&+\mathbb{E}\left[\tilde{Y}_i\tilde{Y}_{j_1,i}\Delta_i\tilde{Y}_{j_2}\Delta_i\tilde{Y}_{j_3}\tilde{r}_i(t)\tilde{r}_j(t)\tilde{r}_k(t)\tilde{r}_l(t)\right]\\
\label{T2_prod1}&&+\left.\frac{}{}\mathbb{E}\left[\tilde{Y}_i\Delta_i\tilde{Y}_j\Delta_i\tilde{Y}_k\Delta_i\tilde{Y}_l\tilde{r}_i(t)\tilde{r}_j(t)\tilde{r}_k(t)\tilde{r}_l(t)\right]\right\}
\end{eqnarray}
Similarly, if two indices $i,j$ are both free among $\{i,j,k,l\}$, then
\begin{eqnarray}
\nonumber \mathbb{E}\left[\tilde{Y}_i\tilde{Y}_j\tilde{Y}_k\tilde{Y}_l\tilde{r}_i(t)\tilde{r}_j(t)\tilde{r}_k(t)\tilde{r}_l(t)\right]
&=&\sum_{i_1,i_2\in\{i,j\}}\sum_{\{j_1,j_2\}=\{k,l\}}\left\{\frac{}{}
\mathbb{E}\left[\tilde{Y}_{i_1}\tilde{Y}_{i_2}\Delta_{i_1}\tilde{Y}_{j_1}\Delta_{i_2}\tilde{Y}_{j_2}\tilde{r}_i(t)\tilde{r}_j(t)\tilde{r}_k(t)\tilde{r}_l(t)\right]\right.\\
\label{T2_prod2}&&+\left.\frac{}{}\mathbb{E}\left[\tilde{Y}_i\Delta_i\tilde{Y}_j\Delta_i\tilde{Y}_k\Delta_i\tilde{Y}_l\tilde{r}_i(t)\tilde{r}_j(t)\tilde{r}_k(t)\tilde{r}_l(t)\right]\right\}
\end{eqnarray}
For our calculations below, we employ weakly narrower bounds for changes $|\Delta_{p_2}\tilde{Y}_{p_1}|$ than for levels $|\tilde{Y}_{p_1}|$, so that the first term in the expressions (\ref{T2_prod1}) (\ref{T2_prod2}) will determine the leading terms for the purposes of asymptotic rates, potentially up to a multiplicative constant no larger than $4!=24$, so the second and third term will be left implicit in our calculations below.

To analyze the terms, on $U_{i1},U_{i2}$ denote
\[\upsilon_{ji} = X_j - \mathbb{E}[X_j|\mathcal{F}_{ni}]\]
If $\mathcal{A}_T(j)\cap\mathcal{A}_T(i)=\emptyset$, $\tilde{r}_{i}(\mathbf{d}_{\mathcal{N}_T(j)})$ is $\mathcal{F}_{ni}$-measurable since $\pi_T(\mathbf{d}_{\mathcal{N}_T(j)}|\mathbf{D}_{-\mathcal{N}_T(j)})$ is constant in the conditioning argument by assumption, and $\pi_0(\mathbf{d}_{\mathcal{N}_T(j)}|\mathbf{D}_{-\mathcal{N}_T(j)})\equiv\pi_0(\mathbf{d}_{\mathcal{N}_T(j)}|\mathbf{D}_{-(\mathcal{N}_T(j)\cup\mathcal{A}_T(i))})$
 whenever $\mathcal{A}_T(j)\cap\mathcal{A}_T(i)=\emptyset$. It follows that
\[\upsilon_{ji}=\sum_{\tilde{\mathbf{d}}_{\mathcal{N}_T(j)}}\left[\tilde{r}_i(\tilde{\mathbf{d}}_{\mathcal{N}_T(j)}) -
\pi_T(\tilde{\mathbf{d}}_{\mathcal{N}_T(j)};\mathbf{D}_{-\mathcal{N}_T(j)})\right]\Delta_i\tilde{Y}_j \]
 so that
\begin{eqnarray}
\label{T2_prod3} \mathbb{E}\left[\tilde{Y}_i\upsilon_{ji}\tilde{Y}_k\upsilon_{lk}\tilde{r}_i(t)\tilde{r}_j(t)\tilde{r}_k(t)\tilde{r}_l(t)\right]
&=&\mathbb{E}\left[\tilde{Y}_i\Delta_i\tilde{Y}_j\tilde{Y}_k\Delta_k\tilde{Y}_l\tilde{r}_i(t)\tilde{r}_j(t)\tilde{r}_k(t)\tilde{r}_l(t)\right]
\end{eqnarray}
 regardless of which indices are free. We can now use these expressions to bound components of the term $\mathbb{E}[T_2^2]$.

For an ordered tetrad $(i,j,k,l)$ with edge pairs $(i,j),(k,l)$, let $F_{ij,kl}^{p}$ denote an indicator variable for the event that node $p\in\{i,j,k,l\}$ is disconnected under the dependency subgraph on $\{i,j,k,l\}$ with edges restricted to $\left\{(p_1,p_2):
p_1\in\{i,j\},p_2\in\{k,l\}\right\}$. We also let $|\varphi_{i,jkl}|:=|\varphi_{ij}| + |\varphi_{ik}| + |\varphi_{il}|$. We then bound
\[|\cov(u_i(t)\upsilon_{ji},u_k(t)\upsilon_{lk})|\leq B_{Tn}^2(2B_Y)^4\left(\frac{\mathbb{E}|\varphi_{il}|}{B_Y}\right)^{F_{ij,kl}^{i}}
\left(\frac{\mathbb{E}|\varphi_{j,kl}|}{B_Y}\right)^{F_{ij,kl}^{j}}\left(\frac{\mathbb{E}|\varphi_{kj}|}{B_Y}\right)^{F_{ij,kl}^{k}}
\left(\frac{\mathbb{E}|\varphi_{l,ij}|}{B_Y}\right)^{F_{ij,kl}^{l}}\]
using the Cauchy-Schwarz inequality, so that
\begin{eqnarray}\nonumber|\cov(X_iU_{i1},X_kU_{k1})|&\leq&24\frac{B_{Tn}^2(2B_Y)^4}{n^2}\sum_{j\in\mathcal{A}_T(i)}\sum_{l\in\mathcal{A}_T(k)}
\left(\frac{\mathbb{E}|\varphi_{il}|}{2B_Y}\right)^{F_{ij,kl}^{i}}
\left(\frac{\mathbb{E}|\varphi_{j,kl}|}{2B_Y}\right)^{F_{ij,kl}^{j}}
\left(\frac{\mathbb{E}|\varphi_{kj}|}{2B_Y}\right)^{F_{ij,kl}^{k}}
\left(\frac{\mathbb{E}|\varphi_{l,ij}|}{2B_Y}\right)^{F_{ij,kl}^{l}}\end{eqnarray}
where the multiplicative factor $24$ yields a conservative bound that also accounts for asymptotically negligible higher-order terms in finite sample.

If furthermore, $\mathcal{A}_T(i)\cap\mathcal{A}_T(j)= \mathcal{A}_T(k)\cap\mathcal{A}_T(l)=\emptyset$,
\[|\cov(u_i(t)\upsilon_{ji},u_k(t)\upsilon_{lk})|\leq B_{Tn}^2(2B_Y)^2\left(\frac{\mathbb{E}|\varphi_{il}|}{B_Y}\right)^{F_{ij,kl}^{i}}
\left(\frac{\mathbb{E}|\varphi_{kj}|}{B_Y}\right)^{F_{ij,kl}^{k}}
\mathbb{E}|\varphi_{j,kl}|\mathbb{E}|\varphi_{l,ij}|\]
so that
\begin{eqnarray}\nonumber|\cov(X_iU_{i2},X_kU_{k2})|&\leq&24\frac{B_{Tn}^2(2B_Y)^2}{n^2}\sum_{j\notin\mathcal{A}_T(i)}
\sum_{l\notin\mathcal{A}_T(k)}
\left(\frac{\mathbb{E}|\varphi_{il}|}{2B_Y}\right)^{F_{ij,kl}^{i}}\left(\frac{\mathbb{E}|\varphi_{kj}|}{2B_Y}\right)^{F_{ij,kl}^{k}}
\mathbb{E}|\varphi_{j,kl}|\mathbb{E}|\varphi_{l,ij}|\end{eqnarray}
Using similar steps, we also find that
\begin{eqnarray}\nonumber|\cov(X_iU_{i1},X_kU_{k2})|
&\leq&24\frac{B_{Tn}^2(2B_Y)^3}{n^2}\sum_{j\in\mathcal{A}_T(i)}\sum_{l\notin\mathcal{A}_T(k)}
\left(\frac{\mathbb{E}|\varphi_{il}|}{2B_Y}\right)^{F_{ij,kl}^{i}}
\left(\frac{\mathbb{E}|\varphi_{j,kl}|}{2B_Y}\right)^{F_{ij,kl}^{j}}
\left(\frac{\mathbb{E}|\varphi_{kj}|}{2B_Y}\right)^{F_{ij,kl}^{k}}\mathbb{E}|\varphi_{l,ij}|
\end{eqnarray}
and an analogous expression for $\cov(X_iU_{i2},X_kU_{k1})$.

Therefore, summing over all index pairs,
\begin{eqnarray}
\nonumber\frac1{n^2}\sum_{i=1}^n\sum_{k=1}^n|\cov(X_iU_{i1},X_kU_{k1})|&\leq&24B_{Tn}^2A_{Tn}^2(2B_Y)^2\left(\frac{A_{Tn}^2(2B_Y)^2}{n^2}
+C_n^2\right)
\end{eqnarray}
Similarly,
\begin{eqnarray}
\nonumber\frac1{n^2}\sum_{i=1}^n\sum_{k=1}^n|\cov(X_iU_{i1},X_kU_{k2})|&\leq&24B_{Tn}^2\left(\frac{4A_{Tn}^3(2B_Y)^3}{n^3}
+\frac{A_{Tn}^2(2B_Y)^3C_n}{n^2}+\frac{A_{Tn}(2B_Y)C_n^2}{n}\right)
\end{eqnarray}
and
\begin{eqnarray}
\nonumber\frac1{n^2}\sum_{i=1}^n\sum_{k=1}^n|\cov(X_iU_{i2},X_kU_{k2})|&\leq&24B_{Tn}^2\left(
\frac{4C_nA_{Tn}(2B_Y)^3}{n^3}+\frac{4C_n^2A_{Tn}^2(2B_Y)^2}{n^2}+ C_n^4\right)
\end{eqnarray}

Hence all relevant terms behave at rates $\left(\frac{A_{Tn}}n\right)^qC_n^{4-q}$ for $q=0,\dots,4$, so that the leading term corresponds to the case $q=0$ or $q=4$, depending on whether $\frac{A_{Tn}}n$ or $C_n$ goes to zero at a faster rate. Substituting this into (\ref{ET2_bound}), we therefore obtain
\begin{eqnarray}
\nonumber\left(\mathbb{E}|T_2|\right)^2&\leq&\frac{2}{\pi \kappa}\mathbb{E}\left[\left(\frac{n^{2\varrho}}n\sum_{i=1}^nX_i(Z_n-Z_{ni}) - \omega_n^2\right)^2\right]\\
\label{ET2_conv}&\leq&O\left(n^{4\varrho}B_{Tn}^2\left[\frac{A_{Tn}^4(2B_Y)^4}{n^4}+C_n^4\right]\right)
\end{eqnarray}
which converges to zero under the assumptions of the theorem.

\subsubsection{Bound on $|\mathbb{E} T_1|$} We next determine an absolute bound for the expectation of
\begin{eqnarray}
\nonumber T_1&:=&\frac1{2n}\sum_{i=1}^nh''(\tilde{Z}_{ni}^*)\tilde{X}_i(\tilde{Z}_n-\tilde{Z}_{ni})^2 \\
\nonumber&=&\frac1{2n}\sum_{i=1}^nh''(\tilde{Z}_{ni}^*)\tilde{X}_i(\tilde{U}_{i1}+\tilde{U}_{i2})^2
\end{eqnarray}
where $\tilde{U}_{i1}:=\frac1n\sum_{j:\mathcal{A}_T(i)\cap\mathcal{A}_T(j)\neq\emptyset}\tilde{\upsilon}_{ji}$ and
$\tilde{U}_{i2}:=\frac1n\sum_{j:\mathcal{A}_T(i)\cap\mathcal{A}_T(j)=\emptyset}\tilde{\upsilon}_{ji}$, and we define $\tilde{\upsilon}_{ji}:=\tilde{X}_j
-\mathbb{E}\left[\left.\tilde{X}_j\right|\mathcal{F}_{ni}\right]$. We can therefore write
\[T_1 = \frac1{2n}\sum_{i=1}^nh''(\tilde{Z}_{ni}^*)
\tilde{X}_i(\tilde{U}_{i1}+\tilde{U}_{i2})^2\]

Since $h\in\mathcal{H}$ and by the triangle inequality,
\begin{eqnarray}\nonumber \mathbb{E}|T_1|&\leq& \frac1{2n}\mathbb{E}\left[\sum_{i=1}^n|h''(\tilde{Z}_{ni}^*)|
|\tilde{X}_i(\tilde{U}_{i1}+\tilde{U}_{i2})^2|\right]\\
\nonumber&\leq& \frac1n\sum_{i=1}^n\mathbb{E}\left[|\tilde{X}_i|(\tilde{U}_{i1}+\tilde{U}_{i2})^2\right]
\end{eqnarray}

To evaluate the contributions of each triplet $ijk$ to the expectation, we again consider separately different cases regarding the respective dependency neighborhoods for nodes $i,j,k$. If $\mathcal{A}_T(i)\cap(\mathcal{A}_T(j)\cup\mathcal{A}_T(k)) = \emptyset$, we can bound
\[\mathbb{E}\left[|u_i(t)\upsilon_{ji}\upsilon_{ki}|\right]\leq B_YB_{Tn}^{\frac32}\mathbb{E}|\varphi_{ij}|\mathbb{E}|\varphi_{ik}|\]
Similarly, if $\mathcal{A}_T(i)\cap\mathcal{A}_T(j) \neq \emptyset$ and $\mathcal{A}_T(i)\cap\mathcal{A}_T(k) = \emptyset$,
\[\mathbb{E}\left[|u_i(t)\upsilon_{ji}\upsilon_{ki}|\right]\leq B_Y^2B_{Tn}^{\frac32}\mathbb{E}|\varphi_{ik}|\]
and for the case $\mathcal{A}_T(i)\cap\mathcal{A}_T(j) \neq \emptyset$ and $\mathcal{A}_T(i)\cap\mathcal{A}_T(k) \neq \emptyset$,
\[\mathbb{E}\left[|u_i(t)\upsilon_{ji}\upsilon_{ki}|\right]\leq B_Y^3B_{Tn}^{\frac32}\]
Averaging over all terms, we get
\begin{eqnarray}
\nonumber \frac1n\sum_{i=1}^n\mathbb{E}\left[|X_i|(U_{i1}+U_{i2})^2\right] &\leq&\frac{B_YB_{Tn}^{\frac32}}{n^2}\left(A_{Tn}^2B_Y^2
+ nA_{Tn}B_YC_n + n^2C_n^2\right)
\end{eqnarray}
Dividing by $(\kappa/n^{2\varrho})^{3/2}$, we therefore have
\begin{eqnarray}\nonumber \mathbb{E}|T_1|&\leq&\frac{n^{3\varrho}B_YB_{Tn}^{\frac32}}{\kappa^{3/2}}\left(\frac{A_{Tn}^2B_Y^2}{n^2}
+ \frac{A_{Tn}B_YC_n}n + C_n^2\right)\\
\label{ET1_conv}&=&O\left(\frac{n^{3\varrho}B_YB_{Tn}^{\frac32}}{\kappa^{3/2}}\left(\frac{A_{Tn}^2B_Y^2}{n^2}+C_n^2\right)\right)
\end{eqnarray}
where the right-hand side converges to zero under the assumptions of the theorem.

\subsection{Bound on $|\mathbb{E}T_3|$}

 We now turn to the term $T_3$. From the definition (\ref{Omega_n_def}),
\[\omega_n^2 \equiv\omega_n^2(\mathbf{Y}(\cdot),\mathbf{D})= \frac{n^{2\varrho}}{n^2}\sum_{k=1}^n\sum_{l\in \mathcal{A}_T(k)}\cov(u_k(t_1),u_l(t_2)|\mathbf{Y}(\cdot),\mathbf{D}_{-\mathcal{A}_T(k)})\]
 where Assumption \ref{ass:asy_rate_ass} guarantees that $\omega_n^2\geq\kappa>0$ almost surely. We also define its conditional expectation given $\mathcal{F}_{ni}$,
 \[\omega_{ni}^2\equiv\omega_{ni}^2(\cdot,\mathbf{D}_{\mathcal{A}_T(i)}):=\frac{n^{2\varrho}}{n^2}\sum_{k=1}^n\sum_{l\in \mathcal{A}_T(k)}\cov(u_k(t_1),u_l(t_2)|\mathbf{Y}(\cdot),\mathbf{D}_{-\mathcal{A}_T(k)\cup\mathcal{A}_T(i)})\]
Since $\omega_{ni}^2$ is a conditional expectation of $\omega_n^2\equiv\omega_n^2(\cdot,\mathbf{D})$, it cannot take values outside the support of $\omega_n^2$. By Assumption \ref{ass:asy_rate_ass}, we must therefore also have $\omega_{ni}^2\geq\kappa$ almost surely.

Since $\mathbb{E}\left[u_i(t)|\mathbf{Y}(\cdot),\mathbf{D}_{-\mathcal{N}_T(i)}\right]=0$ and $\mathcal{N}_T(i)\subset\mathcal{A}_T(i)$, it follows that $\mathbb{E}\left[u_i(t)|\mathcal{F}_{ni}\right]=0$, so that by the law of iterated expectations
\[\mathbb{E}\left[\frac{u_i(t)h(\tilde{Z}_{ni})}{\omega_{ni}}\right]
= \mathbb{E}\left[h(\tilde{Z}_{ni})\mathbb{E}\left[\frac{u_i(t)}{\omega_{ni}}|\mathcal{F}_{ni}\right]\right]=0\]
noting that $\omega_{ni}$ is $\mathcal{F}_{ni}$-measurable.

 We can now use a mean-value expansion for the mapping $u\mapsto u^{-1/2}$ to bound
\[|\omega_n^{-1}-\omega_{ni}^{-1}|=\left|\frac{\omega_n^2-\omega_{ni}^2}{2(\bar{\omega}_{ni}^2)^{3/2}}\right|\leq\frac1{2\kappa^{3}}|\omega_n^2-\omega_{ni}^2|\]
 for an intermediate value $\bar{\omega}_{ni}^2$ in the interval between $\omega_n^2$ and $\omega_{ni}^2$.

If $\mathcal{A}_T(i)\cap(\mathcal{A}_T(k)\cup\mathcal{A}_T(l))=\emptyset$, then
\[\left|\cov(u_k(t_1),u_l(t_2)|\mathbf{Y}(\cdot),\mathbf{D}_{-\mathcal{A}_T(k)\cup\mathcal{A}_T(i)})-
\cov(u_k(t_1),u_l(t_2)|\mathbf{Y}(\cdot),\mathbf{D}_{-\mathcal{A}_T(k)})\right|\leq B_{Tn}\mathbb{E}|\varphi_{ki}|\mathbb{E}|\varphi_{li}|\]
 On the other hand, for $\mathcal{A}_T(i)\cap(\mathcal{A}_T(k)\cup\mathcal{A}_T(l))\neq\emptyset$,
\[\left|\cov(u_k(t_1),u_l(t_2)|\mathbf{Y}(\cdot),\mathbf{D}_{-\mathcal{A}_T(k)\cup\mathcal{A}_T(i)})-
\cov(u_k(t_1),u_l(t_2)|\mathbf{Y}(\cdot),\mathbf{D}_{-\mathcal{A}_T(k)})\right|\leq B_{Tn}B_Y^2\]
and for $\mathcal{A}_T(i)\cap\mathcal{A}_T(k)\neq\emptyset$ and $\mathcal{A}_T(i)\cap\mathcal{A}_T(l)=\emptyset$,
\[\left|\cov(u_k(t_1),u_l(t_2)|\mathbf{Y}(\cdot),\mathbf{D}_{-\mathcal{A}_T(k)\cup\mathcal{A}_T(i)})-
\cov(u_k(t_1),u_l(t_2)|\mathbf{Y}(\cdot),\mathbf{D}_{-\mathcal{A}_T(k)})\right|\leq B_{Tn}\mathbb{E}|\varphi_{li}|B_Y\]

Hence, we can bound
\begin{eqnarray}
\nonumber|\omega_n^2-\omega_{ni}^2|&=&\left|\frac{n^{2\varrho}}{n^2}\sum_{k=1}^n\sum_{l\in \mathcal{A}_T(k)}\left(\cov(u_k(t_1),u_l(t_2)|\mathbf{Y}(\cdot),\mathbf{D}_{-\mathcal{A}_T(k)\cup\mathcal{A}_T(i)})-
\cov(u_k(t_1),u_l(t_2)|\mathbf{Y}(\cdot),\mathbf{D}_{-\mathcal{A}_T(k)})\right)\right|\\
\nonumber&=&O\left(B_{Tn}n^{2\varrho}\left(\frac{A_{Tn}^2B_Y^2}{n^2} + 2\frac{A_{Tn}B_YC_n}{n} + C_n^2\right)\right)
\end{eqnarray}
It then follows that
\begin{eqnarray}
\nonumber|\mathbb{E}T_3|&:=&\left|\mathbb{E}\left[\tilde{X}_ih(\tilde{Z}_{ni})\right]\right|\\
\nonumber&=&n^{\varrho}\left|\mathbb{E}\left[\frac{u_i(t)}{\omega_n}h(\tilde{Z}_{ni})\right]\right|\\
\nonumber&\leq & n^{\varrho}\left|\mathbb{E}\left[\frac{u_i(t)}{\omega_{ni}}h(\tilde{Z}_{ni})\right]\right| + n^{\varrho}\left|\mathbb{E}\left[u_i(t)(\omega_n - \omega_{ni})h(\tilde{Z}_{ni})\right]\right|\\
\label{ET3_conv}&=& 0 + O\left(B_{Tn}n^{2\varrho}\left(\frac{A_{Tn}^2B_Y^2}{n^2} + C_n^2\right)\right)
\end{eqnarray}
which converges to zero under Assumptions \ref{ass:positivity_ass}-\ref{ass:bd_influence_ass} and the rates assumed in the theorem.

\subsubsection{Conclusion of proof} Combining the bounds (\ref{ET2_conv}), (\ref{ET1_conv}), and (\ref{ET3_conv}), the right-hand side of (\ref{app_clt_stein_bound}) converges to zero as $n$ grows. This establishes the conclusion for the case in which $Z$ and $\tilde{Z}_n$ are scalar-valued, and the multivariate case then follows directly from the Cram\'er-Wold device\qed

\subsection{Proof of Proposition \ref{variance_bd_prp}:} In the case in which $\mathcal{D}_i(t_1)$ and $\mathcal{D}_i(t_2)$ are singleton sets, the claim follows immediately from the proof of Proposition 2.1 in \cite{IMe18}.

For the general case, a sharp bound on the term is given by the solution $W_n^*$ of the quadratic assignment problem of minimizing
\[W_n: =\frac{n^{2\varrho}}{n^2}\sum_{i=1}^n\sum_{\tilde{\mathbf{d}}_{t_1}}
\sum_{\tilde{\mathbf{d}}_{t_2}}\cov(\tilde{r}_i(\tilde{\mathbf{d}}_{t_1}),\tilde{r}_i(\tilde{\mathbf{d}}_{t_2}))Y_i^*(\tilde{\mathbf{d}}_{t_1})Y_i^*(\tilde{\mathbf{d}}_{t_2})\] subject to the constraint that the distribution of $Y_i^*(\mathbf{d}_{t_1}):=Y_i(\mathbf{d}_{t_1};\mathbf{D}_{-\mathcal{N}_T(i)})$ matches the conditional distribution $\left(\left.Y_i(\mathbf{D}_{\mathcal{N}_T(i)};\mathbf{D}_{-\mathcal{N}_T(i)})\right|\mathbf{Y}(\cdot),\mathbf{D}_n,T_i(\mathbf{D}_n)=t_1\right)$, and the distribution of $Y_i^*(\mathbf{d}_{t_2})$ matches the conditional distribution of $Y_i(\mathbf{D}_{\mathcal{N}_T(i)};\mathbf{D}_{-\mathcal{N}_T(i)})$ given $\mathbf{Y}(\cdot),\mathbf{D}_n,T_i(\mathbf{D}_n)=t_2$.

We can then consider the solution $W_n^{\circ}$ to a relaxation of that optimization problem, where we weaken the marginal constraint to the requirement that the distribution of $Y_i^*(\mathbf{d}_{t_1})$ match the conditional distribution of $Y_i(\mathbf{D}_{\mathcal{N}_T(i)};\mathbf{D}_{-\mathcal{N}_T(i)})$ given $\mathbf{Y}(\cdot),\mathbf{D}_{-\mathcal{N}_T(i)},
T_i(\mathbf{D}_n)=t_1$, and the distribution of $Y_i^*(\mathbf{d}_{t_2})$ match the conditional distribution of $Y_i(\mathbf{D}_{\mathcal{N}_T(i)};\mathbf{D}_{-\mathcal{N}_T(i)})$ given $\mathbf{Y}(\cdot),\mathbf{D}_{-\mathcal{N}_T(i)},
T_i(\mathbf{D}_n)=t_2$. Since $W_n^{\circ}$ minimizes the same objective under fewer constraints, it must be that $W_n^{\circ}\leq W_n^*$, so that $W_n^{\circ}$ also constitutes a lower bound for $W_n$.

The claim then follows from the observation that after summing over values of $\tilde{\mathbf{d}}_{t_1}$ and $\tilde{\mathbf{d}}_{t_2}$, $W_n^L$ is the solution to a quadratic assignment problem that is formally equivalent to the benchmark case in which potential values are only indexed by exposure levels $t_1,t_2$ instead, i.e. when $\mathcal{D}_i(t_1)$ and $\mathcal{D}_i(t_2)$ are singleton sets. As before, the minimum for that problem is achieved by the isotone assignment, so that $W_n^L = W_n^{\circ}$, and $W_n^L$ is therefore indeed a lower bound \qed

\bibliographystyle{econometrica}
\bibliography{mybibnew}

\end{document}